\documentclass{article}

\usepackage{arxiv}

\usepackage[utf8]{inputenc} 
\usepackage[T1]{fontenc}    
\usepackage{hyperref}       
\usepackage{url}            
\usepackage{booktabs}       
\usepackage{amsfonts}       
\usepackage{nicefrac}       
\usepackage{microtype}      
\usepackage{amsmath} 
\usepackage{amssymb} 
\usepackage{cleveref}       
\usepackage{graphicx}
\usepackage{natbib}
\usepackage{doi}
\usepackage{subcaption} 
\usepackage{multirow}   
\usepackage{algorithm, algpseudocode}

\title{Workflow-Based Evaluation of Music Generation Systems}

\date{}

\newif\ifuniqueAffiliation
\uniqueAffiliationtrue

\ifuniqueAffiliation
\author{ \href{https://orcid.org/0000-0003-1970-5353}{\includegraphics[scale=0.06]{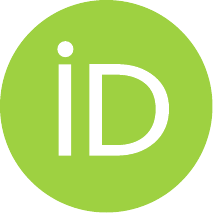}\hspace{1mm}Shayan Dadman}\thanks{Corresponding author.} \\
	Department of Computer Science\\
	UiT, The Arctic University of Tromsø\\
	Lodve Langesgate 2, 8514 Narvik, Norway \\
	\texttt{shayan.dadman@uit.no} \\
	\And
	Bernt Arild Bremdal \\
	Department of Computer Science\\
	UiT, The Arctic University of Tromsø\\
	Lodve Langesgate 2, 8514 Narvik, Norway \\
	\texttt{bernt.a.bremdal@uit.no} \\
	\AND
	Andreas Bergsland \\
	Department of Music\\
	NTNU, Norwegian University of Science and Technology\\
	Fjordgata 1, 334, Trondheim, Norway \\
	\texttt{andreas.bergsland@ntnu.no} \\
}
\else

\usepackage{authblk}

\newbox{\orcid}\sbox{\orcid}{\includegraphics[scale=0.06]{orcid.pdf}} 
\author[1]{%
	\href{https://orcid.org/0000-0003-1970-5353}{\usebox{\orcid}\hspace{1mm}Shayan Dadman\thanks{\texttt{shayan.dadman@uit.no}}}%
}
\author[1]{%
	Bernt Arild Bremdal\thanks{\texttt{bernt.a.bremdal@uit.no}}%
}
\author[2]{%
	\href{https://orcid.org/0000-0000-0000-0000}{\usebox{\orcid}\hspace{1mm}Andreas Bergsland\thanks{\texttt{andreas.bergsland@ntnu.no}}}%
}
\affil[1]{Department of Computer Science, UiT, The Arctic University of Tromsø}
\affil[2]{Department of Music, NTNU, Norwegian University of Science and Technology}
\fi

\renewcommand{\undertitle}{}

\hypersetup{
pdftitle={Workflow-Based Evaluation of Music Generation Systems},
pdfsubject={cs.HC, cs.AI, cs.SD, cs.MM},
pdfauthor={Shayan Dadman, Bernt Arild Bremdal, Andreas Bergsland},
pdfkeywords={Artificial Intelligence (AI), Music Generation Systems (MGS), Generative AI, Human-AI Co-Creation, Workflow-based Evaluation, Music production, Evaluation Framework, Creative Workflows},
}

\begin{document}
\maketitle

\begin{abstract}
	This study presents an exploratory evaluation of Music Generation Systems (MGS) within contemporary music production workflows by examining eight open-source systems. The evaluation framework combines technical insights with practical experimentation through criteria specifically designed to investigate the practical and creative affordances of the systems within the iterative, non-linear nature of music production. Employing a single-evaluator methodology as a preliminary phase, this research adopts a mixed approach utilizing qualitative methods to form hypotheses subsequently assessed through quantitative metrics. The selected systems represent architectural diversity across both symbolic and audio-based music generation approaches, spanning composition, arrangement, and sound design tasks. The investigation addresses limitations of current MGS in music production, challenges and opportunities for workflow integration, and development potential as collaborative tools while maintaining artistic authenticity. Findings reveal these systems function primarily as complementary tools enhancing rather than replacing human expertise. They exhibit limitations in maintaining thematic and structural coherence that emphasize the indispensable role of human creativity in tasks demanding emotional depth and complex decision-making. This study contributes a structured evaluation framework that considers the iterative nature of music creation. It identifies methodological refinements necessary for subsequent comprehensive evaluations and determines viable areas for AI integration as collaborative tools in creative workflows. The research provides empirically-grounded insights to guide future development in the field. Rather than claiming definitive conclusions, this work serves as a constructive contribution to the emerging discourse on MGS evaluation methodologies and their impact on music creation processes.
\end{abstract}


\section{Introduction}\label{sec:introduction}

The integration of AI with creative arts has transformed the tools available to artists and redefined human-machine collaboration, with MGS emerging as innovative technologies that can support creative processes. These systems aim to assist various aspects of music creation while expanding creators' expressive capabilities through diverse computational techniques, including deep learning, Markov models, restricted Boltzmann machines, and evolutionary algorithms \citep{civit_systematic_2022, herremans_functional_2018}. MGS can generate diverse musical styles by utilizing specific training datasets, implementing musical theory rules, or constraining the generator outputs. They can address distinct aspects of music generation \citep{herremans_functional_2018, tatar_musical_2019} including melody (note sequences fulfilling specific goals), harmony (creating harmonious music based on criteria), rhythm (producing rhythms meeting specified requirements), and timbre (manipulating tone color). The capabilities of these systems extend to generating various content types, ranging from monophonic and polyphonic melodies to single and multitrack compositions involving both MIDI and audio formats \citep{civit_systematic_2022}. Furthermore, MGS employ different generation modes to serve diverse creative needs—seeded generation (building upon existing musical content), unseeded generation (creating music from scratch), and inpainting (intelligently filling gaps in existing musical pieces). This versatility enables MGS to support a wide spectrum of creative approaches and compositional goals.

\begin{figure}[t]
  \centering
  \includegraphics[width=\linewidth]{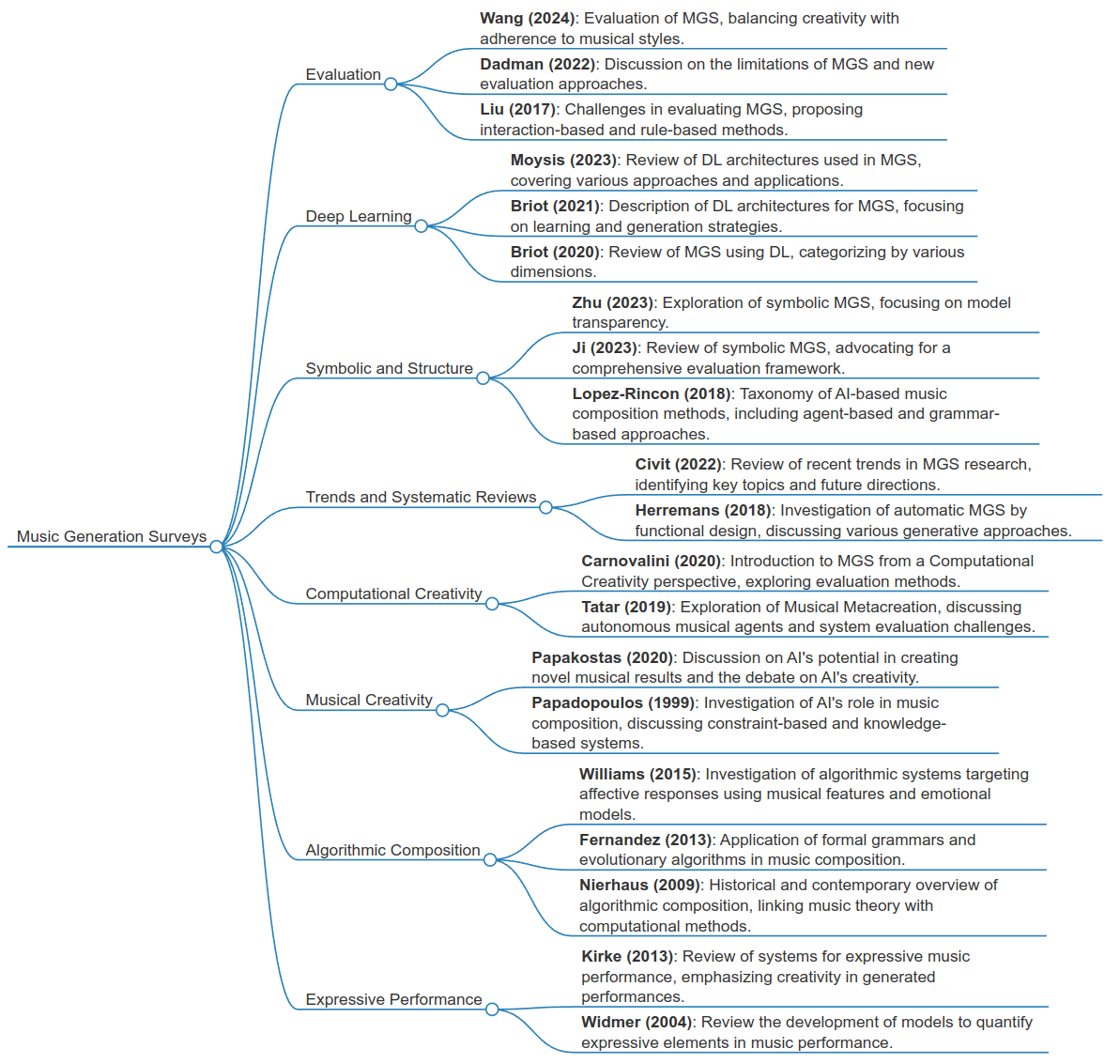}
  \caption{Categorization of surveys in MGS with brief description, grouped by primary focus areas.}
    \label{fig:mgs_review_studies}
\end{figure}

Several studies have provided overviews and analyses of MGS from different perspectives \citep{wang_review_2024,moysis_music_2023,zhu_survey_2023,ji_survey_2023,dadman_toward_2022,civit_systematic_2022,briot_artificial_2021,briot_deep_2020,carnovalini_computational_2020,papakostas_artificial_2020,tatar_musical_2019,herremans_functional_2018,lopez-rincon_algoritmic_2018,liu_computational_2017,williams_investigating_2015,fernandez_ai_2013,kirke_overview_2013,nierhaus_algorithmic_2009,widmer_computational_2004,papadopoulos_ai_1999}. Fig. \ref{fig:mgs_review_studies} briefly summarizes these studies and categorizes them based on their primary focus areas. Each category presents a distinct yet interrelated aspect of music generation that reflects the diverse methodologies and objectives pursued by researchers in the field. This categorization serves as a reference for understanding the scope and variety within AI music generation research. However, it is important to acknowledge the significant overlaps among these studies. The interdisciplinary nature of music generation often leads to research that spans multiple categories, blending techniques and theories from computational algorithms, music theory, cognitive science and artificial intelligence.

Building on this understanding, there has been a notable rise in the development of MGS, which has led to increased efforts to advance AI's creative capabilities and expand its usefulness in various musical tasks. Concurrently, the market has seen the introduction of commercial services that implement such research findings, as presented in Tab. \ref{tab:ai_music_commercial} in Appendix \ref{app:commercial_models}. These services provide a variety of tools designed to streamline the music creation process.  To boost customer appeal, they tend to place strong emphasis on creating user-friendly interfaces more than is common in open-source. Integrated solutions that address specific needs, such as royalty-free music generation or AI-assisted audio creation are also descriptive of several products. For instance, platforms like AIVA and Amper Music simplify the process of generating music. iZotope, another example from Tab. \ref{tab:ai_music_commercial} in Appendix \ref{app:commercial_models}, offers AI-powered audio plugins that focus on audio analysis and customizable settings, emphasizing the technical aspects of music production. Other noteworthy examples include Mubert, which provides personalized, royalty-free music streaming, and Brain.fm, which is designed to enhance focus and relaxation through AI-generated soundscapes.

Such commercial solutions provide easy access and ready-to-use music creation tools, especially suitable for casual creators \citep{compton_casual_2015}, as highlighted by \citet{bown_genies_2025}. However, some of these systems have encountered distinct ethical and legal challenges, as evidenced by several litigations \citep{universal2023, wired2023, musicbusiness2023}. For a comprehensive analysis and discussion on open-source and commercial systems, interested readers are referred to \citep{ma_foundation_2024, zirpoli_2023_generative, seger2023open, barnett_ethical_2023, morreale_where_2021}.

\subsection{Why Open-Source?}

While acknowledging the importance of commercial systems in the broader adoption of AI music, this study focuses, first of all, on open-source systems. Systems selected for this study has been determined primarily by accessibility. But choice has also been motivated strongly by other several factors that align with both technical and creative aspects of music creation process (more on this below). This choice supports the study's objectives as explained later.

The transparency inherent in open-source systems, as the first factor, facilitates a more profound understanding of system architectures and better control of generation processes. As \citet{bown_genies_2025} notes, indeed, the software-as-a-service nature of commercial systems creates uncertainty, as users cannot be confident that processes will remain consistent. This transparency also allows for a deeper assessment of both technical capabilities and creative affordances\footnote{'Creative affordances' refer to features that empower users to explore, experiment, and discover unexpected, novel outcomes during the creative process.}. Furthermore, the ability to examine and potentially modify model behaviors allows to investigate how these systems serve their intended creative purposes. This flexibility also enhances their practical utility in production contexts - users have more access to tinker and innovate, as noted by \citet{seger2023open}. Yet, we recognize that there is a connection between commercial viability and quality. Ease of use, for instance, is a necessity for achieving customer praise.

Additionally, the ability to modify and experiment with the system's components enables extended hands-on experimentation, which is particularly valuable for this study's approach to assessment. This alignment between open-source characteristics and the study's objectives facilitates a more complete understanding of how these systems function within contemporary music production contexts.

\subsection{Key Concepts}\label{sec:music_concepts}
 
This section describes the processes involved in contemporary music production, the evolving role of music technology, and the diverse responsibilities of music producers. Its purpose is to outline the framework for this study, setting the stage for a detailed analysis of the selected systems. This analysis aims to evaluate the effectiveness and potential of these systems within contemporary music production and to provide an informed assessment through a subjective examination of the systems discussed in this study.

\subsubsection{Contemporary Music Production}
\citet{auvinen_music_2019} defines music production as the process of creating a musical record involving a series of task sequences such as songwriting, arranging, recording, editing, mixing and mastering. This process is characterized by intricate actions and interactions between various stakeholders, including artists, producers and record companies. The central figure in this process is the producer, who links the artist, record company and audience \citep{auvinen_music_2019}. Over the past few decades, music production has become more collaborative, with producers taking on more significant roles in the creative and technical aspects of music creation \citep{zak_poetics_2001}. Contemporary music production represents a shift from traditional methods, particularly due to its integration of digital technologies. This modern approach utilizes digital audio workstations (DAWs) and a wide range of electronic music technologies with greater control over crafting sonic materials \citep{auvinen_music_2019}. Adopting DAWs, software plugins and virtual instruments has made the production process more accessible. This has reduced the dependence on expensive studio spaces and specialized industry professionals, allowing for greater creative freedom and experimentation. As a result, it enables artists and producers to create high-quality music independently.

\subsubsection{Music Technology}
According to \citet{frith_performing_1996}, music technology involves the tools and structures through which sounds are produced, reproduced and ultimately transformed into recorded music. This broad definition includes not only hardware and software but also human actions and thought processes. Over recent decades, the evolution of music technology through the integration of digital technologies has profoundly influenced every aspect of music production \citep{burgess_history_2014}. The non-linear capabilities of digital tools allow artists and producers to manipulate audio in flexible and non-sequential ways. This shift from analog to digital has facilitated a more exploratory approach to music creation, where the boundaries of sound can be continuously tested and expanded. Furthermore, the development of software plugins, virtual instruments and effects processors has expanded the palette of sounds that provide new possibilities for sonic manipulation and innovation \citep{holmes_electronic_2020}. Additionally, the integration of AI into contemporary music production highlights its potential to transform the industry through automating repetitive tasks, for instance, in the process of mixing and mastering a piece of music \citep{moffat_approaches_2019}.

\subsubsection{Music Producer}
In traditional music production, the role of producer was often viewed as an intermediary and has evolved over time, with their primary responsibilities varying depending on the specific context \citep{hennion_intermediary_1989}. However, with the rise of digital technologies, this role has expanded to include a more pronounced involvement in the creative aspects of music production \citep{burgess_art_2013}. In contemporary settings, producers are increasingly engaged in the creative process, contributing artistically and technically from the pre-production stages through to the final mixing and mastering. According to the case studies provided by \citet{auvinen_music_2019}, the role of a music producer in song arrangement and composition can vary depending on the genre and cultural context. For instance, in classical music production, a producer might focus more on the acoustics and technical settings of the recording environment, making subtle adjustments to enhance the natural sound quality \citep{auvinen_music_2019}. Conversely, in popular music, the producer's role extends into song arrangement, sound design, and even co-writing, which actively shapes the musical content during the pre-production and recording phases. These varied roles highlight the adaptability of music producers to different genres and production environments and the increasing reliance on technology as a creative tool.

\subsection{Scope of Study}\label{sec:scope_of_study}

This study presents an exploratory evaluation of MGS conducted during the period of February-March 2024. The study's evaluation framework combines technical insights with practical experimentation through evaluation criteria, specifically designed to investigate how MGS enhance creativity within the iterative, non-linear nature of music production workflows. While multiple-evaluator approaches are recognized as optimal \citep{yang_evaluation_2020}, this study employs a single-evaluator methodology as a preliminary phase to establish a foundational understanding and gather necessary insights before conducting larger-scale evaluations. In this context, the study adopts a mixed research approach by utilizing qualitative methods (subjective evaluation) to form hypotheses that can subsequently be assessed quantitatively (using Likert-scale metrics). The definitions of the evaluation criteria are informed and benefited by findings and broader discussions in prior research on human-AI co-creation and human-computer interaction.

The investigation examines eight selected open-source systems: \textit{MusicGen} \citep{copet2024simple}, \textit{M$^{2}$UGen} \citep{liu2024m2ugen}, \textit{Riffusion} \citep{Forsgren_Martiros_2022}, \textit{Magenta Studio 2.0} and \textit{Magenta DDSP-VST} \citep{magenta}, \textit{Musika} \citep{pasini2022musika}, \textit{MuseCoco} \citep{lu2023musecoco}, and \textit{MuseFormer} \citep{yu2022museformer}. While commercial systems offer valuable insights into user experience and service implementation, focusing on open-source MGS allows for deeper assessment (as noted in previous section) and supports the exploratory nature of this study.

These systems were selected based on their architectural diversity, capabilities, and creative potential. The selection represents both symbolic and audio-based music generation approaches across composition, arrangement, and sound design processes\footnote{This selection also reflects systems frequently used by artists, whose experiences and opinions are discussed in Section \ref{subsec:artists}.}. \textit{MuseCoco} and \textit{Museformer} represent advances in symbolic music generation—\textit{MuseCoco} through attribute-conditioned frameworks controlling musical parameters and \textit{Museformer} via structure-aware attention mechanisms ensuring coherence in long-form compositions. Audio-based systems like \textit{MusicGen} and \textit{M$^{2}$UGen} demonstrate text- and melody-conditioned generation capabilities, with \textit{M$^{2}$UGen} extending to multi-modal inputs including video. \textit{Magenta Studio 2.0} and \textit{DDSP-VST} integrate MIDI manipulation and timbre control within digital audio workstations (DAWs), respectively. \textit{Riffusion} (leveraging spectrogram diffusion) and \textit{Musika} (optimized for lower computational requirements) highlight accessibility considerations. Section \ref{sec:systems_overview} provides a comprehensive overview of each system's advancements and capabilites.

As \citet{ma_foundation_2024} notes, many commercial systems build upon open-source foundations, making the study's findings relevant for understanding both current capabilities and future directions. Rather than claiming definitive conclusions, this work contributes to ongoing dialogue about AI music systems evaluation methodologies and their impact on music creation process. The proposed framework, indeed, is positioned as a constructive contribution to the emerging discourse of MGS evaluation.

\subsection{Reasearch Questions and Objectives}\label{sec:research_questions}

Building upon the scope outlined above, this study examines MGS within the context of contemporary music production, specifically focusing on three tasks: composition, arrangement, and sound design. This context provides an ideal framework for evaluating current MGS capabilities, particularly in environments characterized by digital and electronic production techniques. The study employs a mixed-method evaluation approach combining qualitative observations with quantitative metrics to assess both technical capabilities and creative affordances of these systems. Within this setting, the research investigates how music producers might leverage MGS to enhance creative processes while maintaining artistic integrity and workflow efficiency.

The study is guided by three primary research questions that address limitations, integration challenges, and collaborative potential:

\begin{itemize}
  \item RQ1: What are the inherent limitations of current MGS, and how do these constraints affect their integration and utility within contemporary music production workflows?
  
  \item RQ2: What practical challenges and creative opportunities arise from embedding MGS into music production processes, particularly in terms of accessibility, usability, and workflow compatibility?
  
  \item RQ3: How can MGS be designed as collaborative tools that enhance the creative process, preserve artistic authenticity, and maintain emotional depth while adapting dynamically to evolving user preferences and production contexts?
\end{itemize}

These research questions (RQs) are designed to examine the sociotechnical ecosystem in which MGS operate, where technical capabilities, interface design, and creative workflows interact in complex ways. The proposed evaluation framework facilitates this through its mixed-method approach rather than relying on isolated technical performance measures or generic usability benchmarks (more on this in Section \ref{sec:background}). This framework is deliberately adaptable and designed to evolve alongside technological advancements and various use cases rather than providing rigid, fixed evaluation criteria. By addressing these RQs through this framework, the study aims to provide a foundation for broader discourse on how MGS can meet diverse creative expectations while acknowledging the indispensable role of human expertise in tasks demanding emotional depth and complex decision-making.

While the boundaries between composition, arrangement, and sound design often blur in practice, this study considers each domain separately for clarity. Composition encompasses the creation of initial musical ideas, including melodic, harmonic, and rhythmic elements that form a track's foundation. Arrangement involves the structural organization of musical elements over time, including formal considerations (introductions, verses, choruses) and the layering of musical textures. Sound design, particularly important in contemporary electronic music, focuses on creating and manipulating audio elements to produce timbral characteristics. For more detail on these tasks, interested readers are referred to \citet{roads2023Computer, senior2019Mixing, snoman2014Dance, holmes_electronic_2020}.

Through this investigation, this study makes the following contributions:

\begin{itemize}
  \item It proposes a mixed-method evaluation framework that balances qualitative observations with quantitative metrics (1-5 scale). This establishes a comprehensive approach for assessing the technical capabilities and creative affordances of MGS. The proposed framework is designed to evolve alongside technological advancements and adapt to diverse creative contexts and case studies.
  \item It reconceptualizes MGS as collaborative partners rather than autonomous creators. It demonstrates how these systems expand creative possibility spaces while acknowledging their limitations in maintaining thematic and structural coherence—thereby emphasizing the continued importance of human expertise for tasks requiring emotional depth and complex decision-making.
  \item The study identifies essential integration challenges, including steerability limitations, latency-quality tradeoffs, prompt engineering complexities, and others, while highlighting how open-source systems can democratize access and foster community-driven innovation.
  \item It addresses system design principles that advocate recontextualizing AI tools as extensions of familiar production metaphors rather than replacements, as exemplified by systems with real-time parametric controls and instruction-tuned capabilities.
  \item It contextualizes these technological developments within broader sociocultural frameworks, cautioning against the homogenization risks posed by over-reliance on historical data and advocating for mechanisms that encourage exploratory outputs to maintain creative diversity.
  \item The research proposes three additional evaluation dimensions—Serendipity Support, AI Assistance Balance, and Adaptation Capacity—as criteria for assessing creative exploration support and human-machine collaboration.
  \item Finally, the study provides practical insights aligned with music creators' experiences while establishing a foundation for future research through adaptive scoring mechanisms and participatory evaluation approaches.
\end{itemize}

Collectively, these contributions clarify the current state of MGS while advancing a vision for sustainable creative workflows that balance AI capabilities with human artistry.

\section{Background}\label{sec:background}

The evaluation of AI-music systems and technologies spans across different disciplines, including human-computer interaction, computational creativity, and music information retrieval, among others. This diversity of approaches has led to various methodological frameworks, each with distinct advantages and limitations.

To begin with, objective evaluation methods, as discussed by \citet{ji_survey_2023}, provide quantifiable data on technical aspects of music generation by assessing various aspects and musical elements without subjective human influence. These aspects include harmonic structure and multi-track alignment using objective measures, such as Frechet Audio Distance (FAD), BLEU scores, and genre-specific metrics (e.g., chord tone emphasis, swing deviation) \citep{gui2024Adapting, dong2020MusPy, raffel2014mir_eval}. However, these methods may not fully capture the experiential and creative nuances essential in music production \citep{deruty_development_2022}, often overlooking contextual application within music creation workflows, and miss perceptual subtleties \citep{xiong2023Comprehensive}.

While objective methods offer efficiency and reproducibility, \citet{berenzweig_large-scale_2004} argues that they present an incomplete view by overlooking creative and experiential aspects that contribute to musical meaning and engagement. Subjective evaluation allows for deeper exploration of musical elements by considering emotional responses to factors such as timbre, dynamics, harmony, and rhythm \citep{berenzweig_large-scale_2004, kasak_towards_2022}. This approach recognizes music assessment as inherently context-dependent, varying across individuals and situations—dimensions that cannot be fully addressed through objective metrics alone \citep{dadman_toward_2022}. \citet{kasak_towards_2022} emphasize that subjective methods are necessary for assessing nuances that objective measures overlook, such as the presence of unwanted artifacts and listener-specific preferences.

\citet{Linson2012CriticalII} further contrast computational analysis with human expertise using examples like \citet{schuller1958sonny} analysis of Sonny Rollins' solos. \citeauthor{schuller1958sonny}'s qualitative approach identified creative structural features—such as unexpected thematic development, phrasing choices, and long-term dependencies—that were musically significant but irreducible to rule-based frameworks. Complementing this perspective, \citet[][p. 561-563]{juslin_emotional_2008} argue that emotional responses to music are shaped by a dynamic interplay of its structural properties, personal associations, and cultural context. This, indeed, accounts for the subjective variability in how musical pieces emotionally resonate with different listeners \citep{juslin_emotional_2008}.

Moreover, subjective evaluations involving multiple participants can provide valuable insights into user experience and aesthetic reception \citep{tractinsky2000What}. However, they are inherently limited by challenges such as variability in individual musical taste, personal preferences, and expertise, which undermine consistency in subjective judgments \citep{jordanous_standardised_2012, yang_evaluation_2020, berenzweig2004LargeScale}. For example, it is reported that subjective judgments of similarity can vary across listeners and even fluctuate for the same individual depending on their mood or context \citep{juslin_emotional_2008, berenzweig2004LargeScale}. Additionally, designing effective listening experiments for diverse participant groups can introduce complexities, such as sparse data coverage, logistical limitations, and the challenge of unifying subjective opinions into a reliable evaluation framework \citep{yang_evaluation_2020, berenzweig2004LargeScale}.

Genre-specific considerations further highlight the complexity of evaluation, as musical features and their assessment can vary across different styles and users \citep{eerola2011Are}. For instance, \citet{Linson2012CriticalII} distinguish between idiomatic (e.g., jazz improvisation with quantifiable structural correlations) and non-idiomatic (e.g., freely improvised music with emergent, context-driven interactions) genres by arguing that quantitative methods inherently fail to address the latter's multivalent meanings. \citet{pressing1987micro} work reveals that while jazz solos exhibit measurable `micro-micro' and `micro-macro' correlations, free improvisations erase such patterns. This, indeed, necessitates qualitative evaluation to capture the `interweaving of social and structural factors' \citep{Linson2012CriticalII}. It also reflects \citeauthor{Linson2012CriticalII}'s discussion of \citeauthor{clarke2011Ways}'s Hendrix study, where three listeners interpreted the same arpeggiation as a military bugle call, melodic dissolution, or fingerboard traversal—divergent meanings inaccessible to quantitative analysis.

As \citet[][p. 960]{stowell_evaluation_2009} emphasize, musical interactions inherently involve `creative and affective aspects' that resist standardization and often depend on `the performer's privileged access to both the intention and the act'. This makes it challenging to distill outcomes into universally applicable quantitative metrics, especially when participant backgrounds differ. Furthermore, \citeauthor{stowell_evaluation_2009} point out that small participant populations are often unavoidable when working with specialized user groups (e.g., expert musicians). To address these challenges, they argue for evaluation approaches that can yield meaningful insights, specifically advocating for structured qualitative methods suitable for relatively small study sizes.

\citet{reimer_embracing_2021} underscore the value of exploratory evaluations as a groundwork for iterative design, where flexible methods can uncover emergent insights. They state, `Exploratory studies allow researchers to develop informed hypotheses that can be formally tested using appropriate methods with a suitable level of scientific rigor' (p. 18). However, they also advocate for adopting more structured and formal evaluative tools to address consistency and longitudinal insight gaps. Indeed, \citet{reimer_embracing_2021} emphasize the value of exploratory methods in capturing initial user perceptions and behaviors. They argue that such studies allow researchers to observe `how individuals adapt to and use technology', particularly when the goal is `to provide creative tools for creative professionals' \citep[][p. 4]{wanderley2019HCI} or non-musicians.

Building upon these methodological considerations and identified challenges in evaluation approaches, this study proposes a framework that balances quantitative and qualitative assessments with contextual relevance for contemporary music production workflows. This framework integrates systematic criteria with practical, creative contexts by addressing the limitations highlighted by previous researchers while maintaining methodological rigor. This evaluation approach emphasizes the importance of both system architecture and creative affordances to provide a structured framework that acknowledges the multidimensional nature of music generation systems. The following section outlines this methodology in detail.

\section{Evaluation Framework}\label{sec:evaluation_framework}

The evaluation framework comprises 'system-level' and 'performance' criteria derived based on previous research, which are employed to assess both system characteristics and creative affordances. This approach contextualizes creativity within the contemporary music production context, as characterized in Section \ref{sec:music_concepts}, through a systematic examination of systems' adaptability to diverse musical intentions and integration into production workflows. In doing so, it addresses concerns about standardizing subjective evaluations \citep{jordanous_standardised_2012, stowell_evaluation_2009} while emphasizing the importance of direct interaction throughout the music creation process \citep{deruty_development_2022, huang_ai_2020}.

The framework employs a systematic testing methodology comprising two phases: \textit{System Overview} and \textit{Hands-on Experimentation}. This two-phase evaluation strategy integrates both theoretical capabilities and practical performance, consistent with \citet{jordanous_standardised_2012}'s emphasis on comprehensive system assessment. The \textit{System Overview} phase utilizes the 'system-level' criteria to establish a foundational understanding of each system's architecture and potential capabilities. Conversely, the \textit{Hands-on Experimentation} phase uses the 'performance' criteria to validate these theoretical capabilities through practical music production tasks. This deliberate alignment between phases and criteria ensures methodological consistency while addressing both objective and subjective evaluation components. Sections \ref{sec:systems_overview} and \ref{sec:hands_on_experimentation} provide a detailed description of these phases.

The 'system-level' criteria analyze system architecture, features, and attributes. These criteria incorporate context-sensitive attributes, which enable structured comparisons across systems while preserving contextual nuance. Imposing fixed quantitative metrics on these attributes would result in evaluations becoming rapidly obsolete. For instance, hardware requirements pose challenges for standardized evaluations due to continuous advancements in AI technologies. These improvements—stemming from innovations in computational infrastructure and the distinct needs during model training, inference, and fine-tuning—make such fixed assessments ephemeral. \footnote{Additionally, the growing reliance on cloud computing, distributed resources and high-performance systems complicates such evaluations, which can make them temporary and susceptible to misinterpretation.} The 'performance' criteria confine practical aspects and creative affordances. They adopt a quantitative approach, where each criterion is scored on a 1-5 scale based on a standardized scoring rubric, detailed in Appendix \ref{app:evaluation_criteria}. Appendix \ref{app:evaluation_criteria} also elaborates on the rationale behind choosing this scoring metric, particularly 1-5 as the scale. Section \ref{sec:criteria_definition} will describe these criteria.

Therefore, this investigation presents a mixed research methodology by combining quantitative scoring (on a 1–5 scale) and qualitative observations (notes taken during evaluation). This methodological choice is grounded in the growing recognition of mixed method research as an effective strategy \citep{el-shimy_user-driven_2016, Linson2012CriticalII, stowell2009Evaluation, johnson2004Mixed} and the understanding that using a single data source (e.g., quantitative data) cannot adequately capture the complexity of the subject under investigation \citep{bradt2021Where, Linson2012CriticalII, stowell2009Evaluation, pressing1987micro}.

As \citet{schacher2015Movement} emphasize, such an approach facilitates triangulation by enabling convergence and contradiction analysis across different data types, thereby augmenting explanations of perceptual phenomena. This is particularly valuable in music evaluation, where subjective judgments often involve competing dimensions—such as emotional resonance and technical coherence—that single-method designs struggle to capture \citep{chu2022Empirical, schacher2015Movement}. The qualitative observational component ensures that evaluator responses, including emotional reactions and contextual interpretations, are documented alongside quantitative scoring. This documentation provides enhanced understanding and greater confidence in conclusions, as \citet{Linson2012CriticalII} notes, by revealing subtle musical choices that quantitative data alone often overlook. This dual approach acknowledges the role of qualitative measures in addressing hedonic factors like emotional resonance and creative engagement that standardized metrics cannot adequately quantify, as noted by \citet{el-shimy_user-driven_2016, Linson2012CriticalII}. This position aligns with \citet{chu2022Empirical, stowell_evaluation_2009}, who contend that qualitative responses often reveal emotional impacts and creative inspirations beyond what Likert-scale metrics alone can capture.

Building on this mixed methods foundation, it is essential to emphasize the fundamentally exploratory nature of this investigation. The research aims to evaluate the systems (presented in Section \ref{sec:scope_of_study}) and the practicality and efficacy of the framework—particularly the 'system-level' and 'performance' criteria—in evaluating AI music systems within production workflows. This exploratory approach serves the study's primary objective: establishing foundational work to examine these systems' capabilities and creative affordances in production contexts. To achieve this objective, the methodology considers the contemporary music production process described in Section \ref{sec:music_concepts}, where producers function as evaluators across technical and creative domains. This parallel is particularly relevant as the evaluation methodology aims to mirror the modern producer's role in assessing and integrating new technologies while maintaining a consistent creative vision throughout the production process.

\begin{figure}[t]
  \centering
  \includegraphics[width=\linewidth]{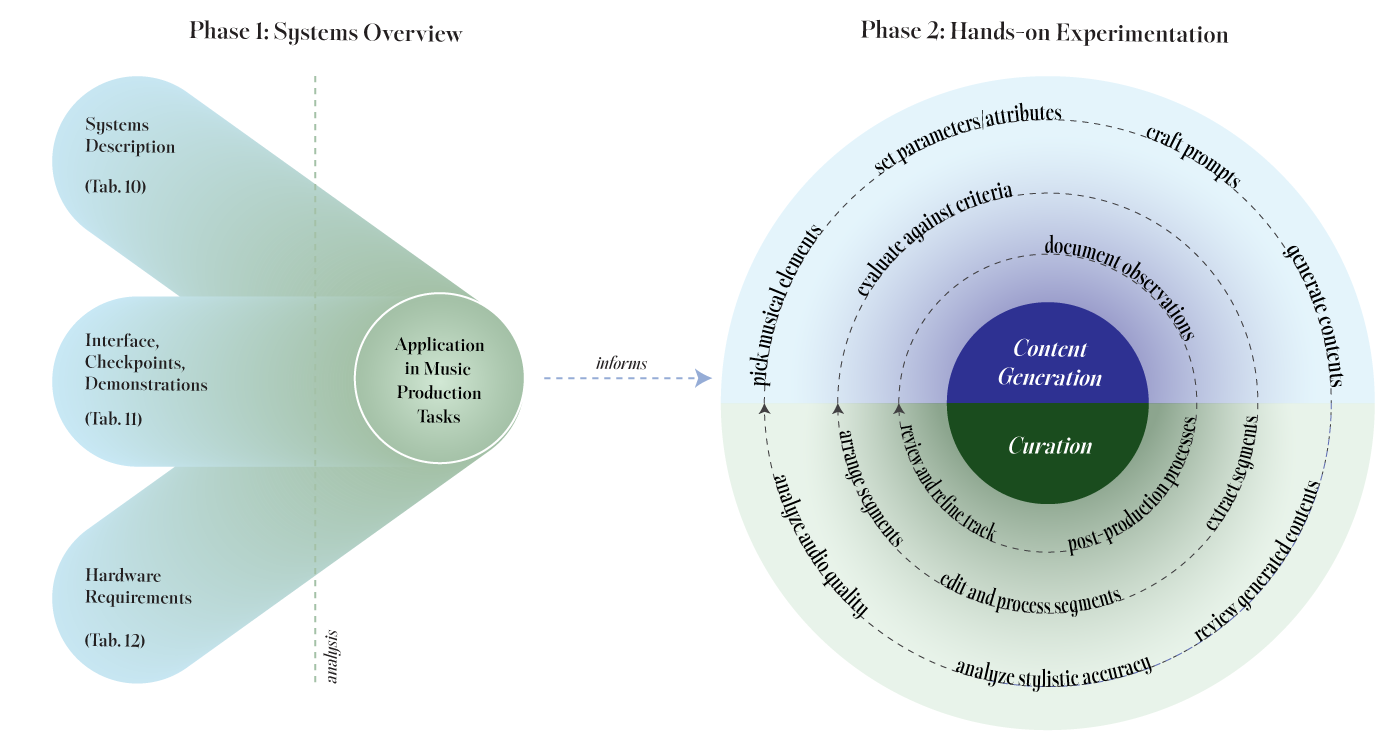}
  \caption{Diagram of a two-phase evaluation framework for MGS. Phase 1 (left) presents the \textit{Systems Overview}, comprising three analytical components: 'Systems Description' (Tab. \ref{tab:architecture_criteria}), 'Interfaces, Checkpoints, and Demonstrations' (Tab. \ref{tab:interface_criteria}), and 'Hardware Requirements' (Tab. \ref{tab:hardware_criteria}). These components collectively analyze the practical applications of the systems in production tasks. Phase 2 (right) illustrates the \textit{Hands-on Experimentation}, structured as concentric circles with Content Generation (blue) and Curation (green) at its core. This circular design represents a multi-layered workflow where various activities radiate outward in non-linear sequences. Importantly, evaluators can freely select steps from any circle at any time without adhering to a predetermined sequence or hierarchy between the concentric layers. It should be noted that these steps are non-exhaustive examples of what may be taken by an evaluator; due to the workflow-based nature of music creation, evaluators may introduce new steps that represent their own unique workflow and music creation process. This radial organization emphasizes the iterative, non-hierarchical nature of the experimentation process, allowing evaluators to navigate between steps based on their specific needs and workflows. The 'informs' connector demonstrates how insights from the systematic analysis in Phase 1 guide the experimentation in Phase 2, as elaborated in Section \ref{sec:evaluation_framework}.}
    \label{fig:eval_steps}
\end{figure}

Fig. \ref{fig:eval_steps} illustrates aspects of the evaluation framework. The following subsections detail the evaluation phases, criteria definition and the evaluation process.

\subsection{Evaluation Phases}\label{sec:evaluation_phases}

\paragraph{Phase 1: Systems Overview}

The initial phase of the evaluation employs the 'system-level' criteria to establish a foundational understanding of each system's architecture and potential capabilities. This stage focuses on the architecture and design, documented features, functionalities, and available source code provided by the system developers. The \textit{System Overview} phase relies on factual information to interpret and understand each system's technical aspects.

The analysis methodically addresses aspects such as architecture and model design, input/output modalities, conditioning mechanisms, interface availability, and hardware requirements. Each system's technical documentation is interpreted with particular attention to workflow integration possibilities, checkpoint accessibility, and demonstrations that showcase capabilities. This assessment confines both core generative frameworks and practical considerations like ease of setup and local execution options. Consequently, the \textit{System Overview} phase establishes a contextual foundation that provides comparative insights for the subsequent \textit{Hands-on Experimentation} phase.

\paragraph{Phase 2: Hands-on Experimentation}

Following the theoretical overview, the methodology transitions to the \textit{Hands-on Experimentation} phase, which applies the 'performance' criteria to validate theoretical capabilities (analyzed by 'system-level' criteria) through practical music production tasks. This phase emphasizes a cyclic 'generate-then-curate' approach \citep{deruty_development_2022, huang_ai_2020}, where iterative cycles of content generation and refinement are conducted until satisfactory results are achieved. This phase evaluates creative affordances alongside practical considerations such as usability, stylistic accuracy and creative workflows. Each criterion is systematically scored on a 1-5 scale using the standardized rubric detailed in Appendix \ref{app:evaluation_criteria}, complemented by qualitative observations that capture nuanced responses of the evaluator. This approach, as elaborated previously, acknowledges that quantitative metrics alone cannot adequately represent the complexity of creative music systems. The \textit{Hands-on Experimentation} phase comprises two complementary components: \textit{Content Generation} and \textit{Curation}, with iterative refinement central to the process.

The \textit{Content Generation} component establishes clear production goals while directly engaging with each system's parameter controls and creative capabilities. The evaluator introduces diverse musical prompts designed to test system responsiveness across various musical styles and complexities to document both the effort required for proficiency and the system's capability to adapt to musical preferences (further elaborated in Section \ref{sec:workflow}). This process examines how effectively each system balances AI assistance with user creative autonomy, including its ability to facilitate creative and unexpected discoveries while maintaining stylistic accuracy. The evaluation assesses generation speed and responsiveness to parameter adjustments, along with the systems' capacity to produce musically coherent outputs with suitable audio quality for production. These interactions simulate the initial ideation and experimentation stages of music production by providing insights into how each system responds to creative direction.

The \textit{Curation} component focuses on assessing whether the generated content should be discarded or kept and post-generation workflows by evaluating the extent to which generated content can be refined and integrated into cohesive musical compositions. This process examines content generation control, including stem separation capabilities and structural modification possibilities, in conjunction with parameter control to shape and direct the model's behavior. The evaluation also assesses DAW integration and creative workflow by considering the level of integration, capability in maintaining the creative flow, and automation features. The evaluator functions analogously to contemporary music producers—arranging, mixing, and processing the generated segments while documenting the practical challenges of integrating AI-generated content with traditional production techniques. This component yields valuable insights into each system's utility within dynamic production environments, particularly examining how the interplay between AI assistance and user autonomy influences final creative outcomes.

\subsection{Criteria Definition}\label{sec:criteria_definition}

The criteria definitions are grounded in prior research on human-AI co-creation and human-computer interaction. Studies by \citet{civit_systematic_2022, deruty_development_2022, huang_ai_2020, el-shimy_user-driven_2016} established the theoretical foundation, while additional sources referenced in Section \ref{sec:background} and Fig. \ref{fig:mgs_review_studies} contributed to the broader conceptual framework. Through comparative analysis and thematic synthesis, two categories of criteria were developed: 'system-level' criteria (presented in Tabs. \ref{tab:architecture_criteria}, \ref{tab:interface_criteria} and \ref{tab:hardware_criteria}), which describe system architecture, technical specifications, and implementation details; and 'performance' criteria (Table 4), which evaluate creative affordances and practical usability within production workflows. 

The 'system-level' criteria are based on interpretations of literature and reported results by researchers, developers, and the broader community. In contrast, the 'performance' criteria derive from the evaluator’s qualitative observations and analysis of the systems. Furthermore, the considerations provided for these criteria serve as suggestions intended to capture a broad spectrum of factors; they are subject to modification and adaptation based on the specific case study. 

Moreover, some conceptual overlap exists between these categories. These interconnections facilitate the evaluation process by capturing both discrete capabilities and emergent properties without introducing excessive complexity. Appendix \ref{app:evaluation_criteria} provides detailed tables and specifications for these criteria. The following sections outline the theoretical foundation and explain the connections between each criterion and the referenced studies.

\subsection*{System-level Criteria}

Tab. \ref{tab:architecture_criteria} analyzes architectural features and design considerations to provide a technical overview of each system. These criteria are derived from meta-analyses conducted by \citeauthor{civit_systematic_2022}, who evaluated 118 systems to identify prevalent approaches to data representation, model architecture and training methodologies. The criteria specifically respond to documented challenges in implementing end-to-end architectures, managing cultural biases in training data and balancing generative flexibility with structural coherence. Complementing this, \citeauthor{huang_ai_2020}'s empirical observations of creative teams demonstrated how architectural decisions—particularly regarding modular decomposition, generation paradigms and model steerability—directly impact creative workflows in practical contexts.

Tab. \ref{tab:interface_criteria} analyzes the accessibility, usability and practical integration of systems within creative workflows. It aligns with \citeauthor{deruty_development_2022}'s emphasis on intuitive GUI designs and \citeauthor{civit_systematic_2022}'s critique of AI systems' limited user-friendliness. The \textit{Checkpoint Accessibility and Variations} and \textit{Execution Options} criteria specifically address reproducibility and availability gaps by allowing users to explore pre-trained models without extensive computational burden, as \citeauthor{huang_ai_2020} observed. The \textit{Ease of Setup} criterion reflects \citeauthor{huang_ai_2020}'s findings that logistical complexity, setup challenges and dependency management impede adoption and disrupt creative workflows.

Tab. \ref{tab:hardware_criteria} analyzes the practical hardware considerations that determine a system's accessibility across diverse user contexts. The criteria within this table consider whether systems can operate on consumer-grade equipment or require specialized infrastructure. This directly responds to \citeauthor{huang_ai_2020}'s observation that resource-intensive models often excluded potential users and can create accessibility gaps for many AI music systems. These criteria further build upon \citet{deruty_development_2022}'s analysis of how system flexibility affects deployment across varied production contexts.

\subsection*{Performance Criteria}

Tab. \ref{tab:performance_criteria} evaluates systems based on their capacity to support creative processes and produce outputs suitable for production workflows. The criteria of \textit{Usability} and \textit{DAW Integration Capacity} reflect findings from \citeauthor{huang_ai_2020} and \citeauthor{deruty_development_2022} regarding interface and usability challenges encountered when working with AI music systems, as well as the importance of seamless integration into existing music production workflows. Specifically, \citeauthor{deruty_development_2022} highlight that contemporary popular music production is centered around DAWs, serving as essential hubs for 'in-studio composition,' where recording, editing, and mixing activities are closely intertwined with compositional processes. Complementing this perspective, \citeauthor{huang_ai_2020} document difficulties experienced by teams related to 'setup and customization issues' when using AI tools that operated independently from established production environments.

The criteria of \textit{Generation Speed} and \textit{Stylistic Accuracy} reflect \citeauthor{huang_ai_2020}'s findings concerning inefficiencies within iterative generation-curation cycles and the systems' capacity to accurately capture genre-specific conventions and styles, respectively.\footnote{As noted by \citeauthor{huang_ai_2020}, teams fine-tuned models (e.g., GPT-2 for genre-specific lyrics) to align outputs with stylistic norms; however, achieving nuanced stylistic reproduction required substantial manual intervention.}  Similarly, \citeauthor{deruty_development_2022} observed that professional artists valued AI tools that could adapt to specific musical styles while maintaining recognizable stylistic elements. This also aligns with \citeauthor{civit_systematic_2022} findings that style-specific applications are heavily influenced by training datasets, where systems trained on particular genres produce compositions closely resembling those styles. The \textit{Audio Quality} and \textit{Content Generation Control} criteria align with \citeauthor{deruty_development_2022}'s emphasis on evaluating outputs against professional production standards. These criteria also address challenges documented by \citeauthor{huang_ai_2020} regarding the maintenance of coherence and control across independently generated components, as experienced by teams working with modular AI workflows. Furthermore, they reflect \citeauthor{huang_ai_2020}'s observation that teams frequently needed to manually modify AI-generated outputs by utilizing processes such as 'stitching' multiple components together or refining individual elements to achieve stylistic alignment.

The \textit{Parameter Control} criterion reflects \citeauthor{stowell_evaluation_2009}'s research, which empirically identified the relationship between controllability and creative experience.\footnote{The concept of the \textit{Parameter Control} criterion also pertains to the system's usability and creative potential, as it encapsulates both technical precision and the user's evolving capacity to influence the system's behavior.} Their findings documented user preferences for systems that offer greater control, with one participant explicitly contrasting a system's 'randomness' with its 'controllability.' The \textit{Creative Workflow} criterion draws upon \citeauthor{deruty_development_2022}'s concept of 'workflow integration' and \citeauthor{el-shimy_user-driven_2016}'s emphasis on 'flow state' in the context of musical interfaces. Additionally, this criterion aligns with \citeauthor{huang_ai_2020}'s findings by highlighting the conversational nature of AI-assisted music creation and the importance of minimizing context-switching between creative and technical tasks. These considerations are further echoed in \citeauthor{huang_ai_2020}'s observations, where users struggled with technical complexity, steep learning curves, and the need for improved parameter control to better align system outputs with their creative objectives.

As mentioned previously, the scoring levels for these criteria employ a standardized 1-5 scoring system (Appendix \ref{app:rubric_scale}) that aims to translate these qualitative aspects into measurable dimensions. The scoring levels for each criterion are provided in Appendix \ref{app:criteria_score_levels}. This approach seeks to balance standardized evaluation as suggested by \citet{jordanous_standardised_2012} with the contextual sensitivity advocated by \citet{stowell_evaluation_2009}. These scoring levels are derived from research findings—for example, for \textit{Parameter Control} criteria, level 1 reflects \citet{huang_ai_2020}'s observed 'black box' interfaces where users struggled with unpredictable results. In contrast, level 5 incorporates their documented need for 'predictable steering mechanisms' with precise parameter adjustments. Similarly, the \textit{Creative Workflow} scores are grounded in \citeauthor{huang_ai_2020}'s findings on minimizing context-switching between creative thinking and technical troubleshooting and \citeauthor{deruty_development_2022}'s emphasis on workflow integration. Level 1 represents systems that frequently disrupt the creative flow and require a focus on technical operations, while level 5 reflects systems that seamlessly integrate into the creative process. Similarly, the \textit{Workflow Integration} scoring draws from \citeauthor{deruty_development_2022}'s production workflow analysis, with levels 3-5 reflecting progressively deeper integration with existing practices, from essential compatibility (level 3) to the seamless workflow enhancement (level 5).

For 'Stylistic Accuracy,' as another example, the progression from level 1 (failing to capture basic characteristics) to level 5 (considerable stylistic reproduction) mirrors the spectrum of capabilities observed in and expected from AI music systems as documented by \citeauthor{huang_ai_2020} and \citeauthor{deruty_development_2022}. At the lower levels, the scoring rubric addresses the fundamental challenge identified by both studies—that AI systems often struggle with basic genre fidelity by presenting either incorrect elements or significant inconsistencies. The middle tier (level 3) acknowledges systems that can handle common genres with only occasional errors. It reflects what \citeauthor{deruty_development_2022} called the 'grain' that can actually contribute positively to stylistic identity when aligned with genre expectations. The distinction between levels 4 and 5 captures the observation from \citeauthor{huang_ai_2020} that even advanced systems face a trade-off between creative exploration and stylistic consistency. The highest level is reserved for systems that achieve considerable stylistic reproduction while maintaining consistency across iterations—a balance that both studies identify as crucial for professional artistic use yet technically challenging to implement. 

\subsection{Evaluation Process}\label{subsec:evaluation_process}

The evaluation process was conducted by the first author, an AI music researcher and guitarist with formal training in electronic music production. The co-authors, who possess more than twenty years of collective experience in AI methodologies, musicology, and music production, provide additional layers of validation and methodological refinement, particularly regarding the evaluation framework. 

The evaluator documented observations and findings during both evaluation phases according to defined criteria. During the \textit{Hands-on Experimentation} phase, a more rigorous approach was taken. The evaluator systematically observed the systems' behavior, measured performance against predefined scoring criteria, documented detailed findings—including any notable deviations—and assigned final scores using a 1-5 scale. During the \textit{Hands-on Experimentation} phase, the evaluator repeated this process multiple times across various musical tasks. Fig. \ref{fig:eval_steps} demosntrates this flexible, non-linear approach to experimentation that accommodates diverse creative workflows.

The proposed evaluation framework presents a systematic examination of MGS yet involves several limitations that warrant acknowledgment. Section \ref{sec:future_directions} analyzes these considerations and methodological constraints in detail by proposing potential refinements for subsequent research endeavors.

\section{Systems Overview}\label{sec:systems_overview}

This section provides an analysis of the selected systems (presented in Section \ref{sec:scope_of_study}) using the 'system-level' criteria (Section \ref{sec:criteria_definition}). Section \ref{sec:systems_description} begins with a detailed description of the architecture and functionalities of each system, following the criteria in Tab. \ref{tab:architecture_criteria}. Tab. \ref{tab:model_specs} presents a summary of their architectural characteristics and input/output modalities.

Subsequently, Section \ref{sec:interface_demo_ckpt_analysis} analyzes the availability of interfaces, checkpoints, and demonstrations for each system according to the criteria in Tab. \ref{tab:interface_criteria}. Tab. \ref{tab:model_specs_interface} summarizes these findings, including available modes of interaction (interface types) and demonstrations of system features and best practices. This analysis highlights how the accessibility of these systems through their various interaction modes impacts user experience and facilitates practical experimentation.

Section \ref{sec:hardware_analysis} provides an overview of the hardware requirements necessary for training and inference across these systems, following the criteria in Tab. \ref{tab:hardware_criteria}. These findings help to understand the computational demands and feasibility of deploying these systems in various settings. Finally, Section \ref{sec:application_music_production_tasks} identifies suitable applications within music production tasks (Section \ref{sec:research_questions}) based on the analyzed systems' characteristics, features, and capabilities.

\subsection{Systems Description}\label{sec:systems_description}

\textbf{\textit{MusicGen}} \citep{copet2024simple}, part of the Audiocraft library\footnote{\url{https://github.com/facebookresearch/audiocraft}}, employs a transformer-based architecture to generate music from textual descriptions or melodic features. This single-stage auto-regressive Transformer model utilizes an EnCodec tokenizer \citep{defossez_high_2022} and allows for parallel prediction of codebooks\footnote{In this context, a codebook refers to a set of vectors or tokens that the model uses to efficiently encode and decode audio data. This results in more precise and controlled music generation.}. This design reduces the number of required auto-regressive steps and bypasses the necessity of self-supervised semantic representation\footnote{Self-supervised semantic representation involves deriving meaningful data representations without human annotations, which MusicGen avoids by directly generating from encoded inputs}. Additionally, it improves text conditioning through pre-trained text encoders like T5 \citep{raffel_exploring_2023} and joint text-audio representations, such as CLAP \citep{elizalde_clap_2022}. The model employs an unsupervised melody conditioning approach by leveraging chromagram-based conditioning to align the musical output closely with the given textual input \citep{copet2024simple}.

The model is trained on a diverse dataset of 20K hours of licensed music, including internal dataset, royalty-free music tracks from ShutterStock\footnote{\url{https://www.shutterstock.com/}} and Pond5\footnote{\url{https://www.pond5.com/royalty-free-music/}}, and evaluated on benchmarks like MusicCaps \citep{agostinelli2023musiclm}. According to \citet{copet2024simple} and \citet{zhu_survey_2023}, MusicGen outperforms models such as Riffusion \citep{Forsgren_Martiros_2022} and Moûsai \citep{schneider2023mousai} in text-to-music generation. It demonstrates better alignment with text descriptions and produces more consistent melodies. These improvements are measured by objective metrics like FAD \citep{kilgour_frechet_2019} and subjective assessments from listeners. However, as noted by \citet{zhu_survey_2023}, MusicGen still encounters challenges in achieving fine-grained control over music adherence and needs advancements in audio conditioning to enhance its performance.

\begin{table}[t]
    \centering
    \caption{Summary of the systems architectural characteristics and input/output modalities, organized chronologically.}\label{tab:model_specs}
    \resizebox{\textwidth}{!}{\begin{tabular}{cllp{8cm}ll}
    \toprule
    No. & Model                        &  year          & Architecture                                              & Input                 & Output   \\
    \midrule
    1   & \textit{M$^{2}$UGen}         & 2023           & Pre-trained encoders/decoders, multi-modal adapters, LLaMa 2       & Text, image, video    & Audio    \\
    2   & \textit{MusicGen}            & 2023           & Single-stage auto-regressive Transformer, EnCodec tokenizer, four parallel output streams                 & Text, audio           & Audio    \\
    3   & \textit{MuseCoco}            & 2023           & Linear Transformer, BERT$_{large}$                                   & Text                  & MIDI     \\
    4   & \textit{Magenta Studio 2.0}  & 2023           & Various deep learning architectures                                  & MIDI                  & MIDI     \\
    5   & \textit{Magenta DDSP-VST}    & 2023           & Differentiable Digital Signal Processing                  & Audio                 & Audio    \\
    6   & \textit{MuseFormer}          & 2022           & Transformer, fine- and coarse-grained attention           & MIDI                  & MIDI     \\
    7   & \textit{Musika}              & 2022           & Hierarchical autoencoder, FastGAN               & Conditioning Signals  & Audio    \\
    8   & \textit{Riffusion}           & 2022           & Latent diffusion model, variational autoencoders, U-Net, CLIP                                           & Text                  & Audio    \\
    \bottomrule  
    \end{tabular}}
  \end{table}

\textbf{\textit{MuseCoco}} \citep{lu2023musecoco}, termed as Music Composition Copilot, introduces a novel two-stage approach to generating symbolic music from text descriptions by leveraging musical attributes. The first stage, text-to-attribute understanding, utilizes a pre-trained BERT$_{large}$ \citep{devlin_bert_2019} model to extract musical attributes such as tempo, rhythm, melody, and harmony from text by achieving over 99 percent accuracy. This demonstrates its ability to comprehend and classify diverse musical attributes from text inputs. This stage is enhanced by synthesizing text-attribute pairs using ChatGPT\footnote{\url{chat.openai.com}} for refined fluency and coherence \citep{lu2023musecoco}.

The subsequent stage, attribute-to-music generation, employs a Linear Transformer \citep{katharopoulos_transformers_2020} model trained in a self-supervised manner on a large symbolic music dataset, including MMD 
\citep{zeng2021musicbert}, EMOPIA \citep{hung_emopia_2021}, MetaMIDI \citep{ens_metamidi_2021} and others. It aims to generate music that adheres to the specified attributes. In this stage, the model utilizes a REMI-like \citep{huang_pop_2020} representation for controlling music generation through prefix tokens. During the training, \textit{MuseCoco} leverages objective and subjective attributes to guide the generation process\footnote{According to \citet{lu2023musecoco}, objective attributes, such as tempo and time signature, are quantifiable and directly extracted from MIDI files. Subjective attributes, like emotion and genre, are derived from labeled datasets.}. As a result, the model achieves an average control accuracy of 80.42 percent for different attributes such as instrument, pitch range, and key, among others\footnote{For the full list, refer to \citep{lu2023musecoco}}. 

Regarding performance, \textit{MuseCoco} has outperformed baseline systems like GPT-4 \citep{openai_gpt4_2024} and BART-base \citep{wu_exploring_2023} in musicality, controllability, and overall scoring by showing 20 percent improvement in objective control accuracy \citep{lu2023musecoco}. The authors have also expanded \textit{MuseCoco} to 1.2 billion parameters, which enhances its controllability and musicality on a larger scale. However, \textit{MuseCoco} focuses primarily on symbolic music, which may limit its applicability to audio music scenarios and does not explicitly address long sequence modeling. Additionally, the reliance on a predefined set of musical attributes and template-based text synthesis may restrict its versatility.

\textbf{\textit{Riffusion}} \citep{Forsgren_Martiros_2022} utilizes a conditional diffusion model architecture that has been fine-tuned from Stable Diffusion \footnote{The model used for Riffusion is based on the Stable Diffusion v1.5 model, which is available on Huggingface \url{https://huggingface.co/stable-diffusion-v1-5/stable-diffusion-v1-5}}. This model generates audio clips from text prompts and images of spectrograms. The architecture features a variant of denoising autoencoders in combination with a diffusion process, specifically adapted to manage and interpret the complex data distribution of audio, as represented in spectrogram form\footnote{A spectrogram is a visual representation of the spectrum of frequencies in a sound or other signal as they vary with time, using color or brightness variations to indicate the amplitude of each frequency.}. This approach allows Riffusion to produce audio frequencies over time while maintaining the coherence of the output through features such as Image-to-Image transformation, looping, and interpolation mechanisms \citep{zhu_survey_2023}. 

Regarding \textit{Riffusion}'s training and evaluation results, no official report has been made publicly available by the authors as of early 2024\footnote{The assessments presented in this study are based on the version accessed at the beginning of 2024 through \url{https://github.com/riffusion/riffusion-hobby} and \url{https://www.riffusion.com/}.}. Despite this, \textit{Riffusion} has advantages, including an interface that simplifies music generation from text or image inputs and produces music with minimal noise, according to \citet{zhu_survey_2023}. However, the model provides limited user control over the final musical output. This limitation arises from its dependence on predefined text prompts and seed images, which guide the diffusion process and restrict the variety and customization of the generated music \citep{zhu_survey_2023}. 

\textbf{\textit{Musika}} \citep{pasini2022musika} is a GAN-based music generation system that can generate audio of arbitrary length, both conditionally and unconditionally. It utilizes a hierarchical autoencoder to transform audio samples into compact, lower-dimensional representations. This design aims to optimize inference speed and reduce training time by generating magnitude and phase spectrograms with a low temporal resolution. The GAN architecture used in \textit{Musika} is adapted from the FastGAN \citep{liu_towards_2021}, recognized for its quick convergence with limited data. For training, \textit{Musika} employs a diverse range of datasets. The universal autoencoder is trained using a combination of songs from the South by Southwest (SXSW) festival\footnote{\url{https://www.sxsw.com/festivals/music/}} and the LibriTTS corpus \citep{zen_libritts_2019}. For domain-specific training, the MAESTRO \citep{hawthorne_enabling_2019} dataset is utilized for piano music, while a collection of techno tracks from Jamendo\footnote{\url{https://www.jamendo.com}} is used for techno music. 

During the generation phase, \textit{Musika} can generate audio with arbitrary length alongside a global style conditioning mechanism that ensures stylistic coherence across the generated samples\footnote{In this context, arbitrary length refers to model's capability to generate audio continuously without a predetermined endpoint. This is accomplished through a latent coordinate system that allows the model to produce seamless and coherent audio segments that can be concatenated indefinitely and maintain stylistic consistency and coherence over time. This ability enables the creation of music that could theoretically extend for any desired duration.}. It can incorporate conditioning signals, such as note density and tempo. \textit{Musika} allows for full parallelization of the audio generation process. This parallelization is made possible through a latent coordinate system, which enables independent and concurrent generation of audio segments \citep{pasini2022musika}. Regarding performance evaluation, \textit{Musika} demonstrated better quality with lower FAD scores compared to similar systems, particularly in piano music generation \citep{pasini2022musika}. According to \citep{pasini2022musika}, the model's performance is further highlighted by its capacity to generate audio at speeds up to 994 times faster than real-time on a GPU and 40 times faster on a CPU. Nevertheless, \textit{Musika} faces limitations due to a lack of free-form text conditioning and relying on specific datasets for training \citep{NEURIPS2023_38b23e23}.

\textbf{\textit{M$^{2}$UGen}} \citep{liu2024m2ugen} presents a multi-modal framework designed for music understanding and generation that accepts diverse inputs such as images, videos and text. It utilizes advanced encoders like ViT \citep{dosovitskiy_image_2021} for images, ViViT \citep{arnab_vivit_2021} for videos and MERT \citep{li2024mert} as a music encoder to processes these varied inputs. The integration of these modal encoders with understanding adaptors and the LLaMA 2 model \citep{touvron_llama_2023} allows for the interpretation of multi-modal signals and input instructions to guide music generation through decoders like AudioLDM 2 \citep{liu_audioldm_2024} and \textit{MusicGen} \citep{copet2024simple}. This process involves a pipeline where each encoder extracts relevant features from its respective modality, which are then harmonized through understanding adaptors. These adaptors bridge the gap between data types by enabling the LLaMA 2 model to synthesize a coherent representation that informs the music generation process. The decoders then translate this representation into music outputs.

In terms of performance, \textit{M$^{2}$UGen} demonstrates better performance in music understanding than MU-LLaMA \citep{liu_music_2024} by leveraging additional training on the MUCaps \citep{liu2024m2ugen} dataset to enhance text-music alignment. In text-to-music generation, \textit{M$^{2}$UGen} outperforms AudioLDM 2 and \textit{MusicGen}, particularly in CLAP \citep{elizalde_clap_2022} score, which indicates enhanced relevance of generated music to input instructions. Furthermore, its ability in prompt-based music editing surpasses models like AUDIT \citep{wang_audit_2023} and InstructME \citep{han_instructme_2023} by utilizing the LLaMA 2 for prompt comprehension and the MERT for music understanding. In multi-modal music generation, \textit{M$^{2}$UGen} presented an improved performance in various related metrics for generating music based on input images and videos \citep{liu2024m2ugen}. However, as noted by \citet{li_survey_2024}, \textit{M$^{2}$UGen} applicability in music understanding tasks is limited and could be improved further using diverse training data.

\textbf{\textit{MuseFormer}} \citep{yu2022museformer} introduces an approach to symbolic music generation by addressing the challenges of long sequence and music structure modeling. The model's architecture is based on the original Transformer framework \citep{vaswani_attention_2023}, with modifications to incorporate the novel fine- and coarse-grained attention mechanisms. MuseFormer layers replace the standard self-attention module, which allows the model to process sequences by dynamically adjusting the attention focus based on the musical structure. It employs fine-grained attention to focus on structure-related bars. This enhances the learning of structure-related correlations by directly attending to tokens from these bars. In contrast, coarse-grained attention summarizes other bars to provide a broader sketch, reducing computational costs by attending only to the summarization of these bars rather than each token individually. The structure-related bars are selected through bar-pair similarity statistics to identify the bars to be repeated or varied. This dual attention system allows \textit{MuseFormer} to handle longer musical sequences.

In terms of performance, \textit{MuseFormer} was evaluated using the Lakh MIDI dataset\footnote{\url{https://colinraffel.com/projects/lmd/}}. The dataset was preprocessed and transfered into token sequences using REMI-like representation. Through objective evaluations, \textit{MuseFormer} outperformed other Transformer-based models \citep{yu2022museformer}. The objective evaluation measured the model's perplexity and similarity error across different sequence lengths. Subjective assessments further confirm these findings, with \textit{MuseFormer} receiving the highest ratings in musicality and structural coherence, both short-term and long-term. According to \citet{yu2022museformer}, the subjective evaluations involved ten participants, of whom seven had music-related backgrounds. However, as noted by \citet{yu2022museformer}, using random sampling during inference can lead to inconsistencies in generated music quality by \textit{MuseFormer}.

\textbf{\textit{Magenta}} \citep{magenta} represents a suite of music generation systems. It includes several neural network models for music generation\footnote{A complete list of Magenta models can be found in the corresponding Magenta project GitHub repository \url{https://github.com/magenta/magenta/tree/main/magenta/models}}, which can be classified into three main types: sequential models, variational autoencoders (VAEs), and neural synthesizers. Sequential models, such as MelodyRNN, ImprovRNN, and PolyphonyRNN, are trained to learn the distribution of musical patterns and structures. This enables them to generate new music by predicting the next note in a sequence. VAEs, like MusicVAE \citep{roberts_hierarchical_2019}, are probabilistic generative models that learn the probability distribution of the input dataset and can generate new music by sampling from this learned distribution. NSynth \citep{engel_neural_2017}, a neural synthesizer, uses a WaveNet-based autoencoder to generate audio with complex sound characteristics. It provides music practitioners with intuitive control over timbre and dynamics. Additionally, the Differentiable Digital Signal Processing (DDSP) \citep{engel_ddsp_2020} model is another neural synthesizer in \textit{Magenta} that combines deep learning with traditional signal processing techniques to synthesize realistic audio. It offers music practitioners advanced sound design and manipulation tools. For a comprehensive review of these models, refer to \citet{zhu_survey_2023}.

Based on these models, Magenta provides virtual studio technology (VST)\footnote{VST is a software interface developed by  Steinberg that integrates software audio synthesizers and effect plugins with DAWs. VST plugins can emulate the sounds of traditional instruments, create new sounds, or apply audio effects to recordings. They come in two main types: VST instruments,  which generate audio, and VST effects, which process audio.} plugins designed for integration with DAWs, such as Ableton Live, including \textit{Magenta Studio 2.0} and \textit{Magenta DDSP-VST}. \textit{Magenta Studio 2.0} plugin utilizes various models and is created to enhance musical creativity within the DAWs environment. It is developed using Electron\footnote{\url{https://www.electronjs.org}} for native application packaging, TensorFlow.js\footnote{\url{https://github.com/tensorflow/tfjs}} for model implementation and Max For Live\footnote{\url{https://cycling74.com/products/maxforlive}} for MIDI clip manipulation. \textit{Magenta Studio 2.0} includes various features, including 'Generate', which produces musical phrases; 'Continue', which extends existing musical inputs; and 'Interpolate', which blends two musical inputs into new compositions; 'Groove' adjusts the timing and velocity of drum inputs to mimic the feel of live drum performances; 'Drumify' generates drum accompaniments from inputs by translating rhythms into groovy drum patterns. Among these features, 'Generate' and 'Interpolate' utilize the VAE model. The 'Generate' tool uses the VAE to create entirely new 4-bar phrases without any input. 'Interpolate' uses the VAE to blend and morph between two given musical inputs to generate up to 16 new variations that combine the characteristics of the original inputs.

\citet{engel_ddsp_2020} introduces DDSP, an approach to neural audio synthesis by blending classical digital signal processing (DSP) elements with deep learning methods to create realistic musical instrument sounds. \textit{Magenta DDSP-VST} is based on DDSP, which provides a versatile and real-time neural synthesizer and audio effect plugin compatible with various DAWs. This plugin transforms voices or other sounds into musical instruments in effects mode and allows for MIDI-controlled neural synthesizers similar to traditional virtual instruments. It operates through a three-stage process: feature extraction, DSP control prediction and synthesis. Initially, it extracts pitch and volume from incoming audio using a neural network. Then, a compact recurrent neural network predicts controls for an additive harmonic synthesizer and a subtractive noise synthesizer, which are finally mixed to produce the audio output. This process ensures that the synthesized sound matches the input sound's volume and pitch contours, even if the input was not part of the training data.

For comprehensive overview of deep learning frameworks, architectures and techniques, interested readers are refered to the studies in Fig. \ref{fig:mgs_review_studies}.

\subsection{Interface, Checkpoint and Demonstration Availability}\label{sec:interface_demo_ckpt_analysis}

All of the systems reviewed provide public access to their interfaces and pre-trained model checkpoints through platforms such as GitHub, Hugging Face, and dedicated websites. Most offer demonstrations showcasing their capabilities, ranging from simple audio samples to interactive interfaces with customizable parameters. The availability of model checkpoints enables exploration without training from scratch—an advantage given the high computational demands of these systems, which will be discussed later in Section \ref{sec:hardware_analysis}.

The interfaces vary in accessibility and design approach. Web-based interfaces like those offered by \textit{MusicGen} and \textit{M$^{2}$UGen} on Hugging Face Spaces\footnote{\url{https://huggingface.co/spaces/facebook/MusicGen}\newline \url{https://huggingface.co/spaces/M2UGen/M2UGen-Demo}} provide immediate engagement, which is beneficial for quick investigation of the practical applications, generative capabilities and limitations. The community aspects of Hugging Face also facilitate knowledge sharing and collaborative improvement of these tools among users. Command-line interfaces (CLIs) and Application Programming Interfaces (APIs), while requiring more technical expertise, offer advantages in automation, batch processing, and integration into custom workflows—features valuable for research, development and production environments.

\begin{table}[t]
    \centering
    \caption{Availability of demonstrations, pre-trained model checkpoints and user interface options for the systems in this study, including Graphic User Interface (GUI), Command-Line Interface (CLI) and web-based interface.}\label{tab:model_specs_interface}
    \begin{tabular}{clccccc}
    \toprule
    No. & Model                          & GUI        & CLI        & Web-based   & Demonstrations & Checkpoints \\
    \midrule
    1   & \textit{M$^{2}$UGen}           &            & \checkmark & \checkmark  & \checkmark     & \checkmark  \\
    2   & \textit{MusicGen}              &            & \checkmark & \checkmark  & \checkmark     & \checkmark  \\
    3   & \textit{MuseCoco}              &            & \checkmark &             &                & \checkmark  \\
    4   & \textit{Magenta Studio 2.0}    & \checkmark &            &             & \checkmark     & \checkmark  \\
    5   & \textit{Magenta DDSP-VST}      & \checkmark & \checkmark & \checkmark  & \checkmark     & \checkmark  \\
    6   & \textit{MuseFormer}            &            & \checkmark &             & \checkmark     & \checkmark  \\
    7   & \textit{Musika}                &            & \checkmark & \checkmark  & \checkmark     & \checkmark  \\
    8   & \textit{Riffusion}             &            & \checkmark & \checkmark  & \checkmark     & \checkmark  \\
    \bottomrule  
    \end{tabular}
  \end{table}

For music practitioners, systems with low technical barriers facilitate rapid assessment of creative potential. \textit{MusicGen} exemplifies this approach through Hugging Face integration, easy access generation API and local implementation via its Audiocraft toolkit\footnote{\url{https://facebookresearch.github.io/audiocraft/api_docs/audiocraft/models/musicgen.html}}. It offers four model variants ranging from small (300M parameters) to large (3.3B parameters), including the melody model that accepts both text and melodic input as generation guides. \textit{Riffusion} provides a web platform\footnote{\url{https://www.riffusion.com}} with features such as stem separation, lyrics generation, and visualization capabilities. Local implementation is also available with additional functionalities for interpolation, image-to-audio and batch generation\footnote{It is also possible to use \textit{Riffusion} within AUTOMATIC1111 web UI for Stable Diffusions models through an extension provided in \url{https://github.com/enlyth/sd-webui-riffusion}}. Similarly, \textit{Musika} leverages community development to create an ecosystem of pre-trained models accessible through Hugging Face\footnote{\url{https://huggingface.co/musika}}, with implementations for both local execution (web-based) and Google Colaboratory notebooks.

\textit{Magenta} offers perhaps the most production-oriented approach by providing plugins (GUI interface) compatible with DAWs, specifically Ableton Live, for \textit{Studio 2.0} and \textit{DDSP-VST}. These plugins are complemented by additional applications and demonstrations developed and shared by the community to showcase the capabilities and creative possibilities of Magenta models\footnote{\url{https://magenta.tensorflow.org/demos/}}. In contrast, \textit{MuseCoco} and \textit{MuseFormer} provide only CLI access for sample generation, training and fine-tuning.

The systems demonstrate varying degrees of setup complexity. \textit{Magenta Studio 2.0} and \textit{DDSP-VST} offer the most streamlined experience through simple download and integration with Ableton Live. \textit{Musika} and \textit{MusicGen} present moderate difficulty as they require Python and CUDA prerequisites. Similarly, \textit{Riffusion} represents intermediate complexity with separate inference and application components\footnote{\url{https://www.reddit.com/r/riffusion/comments/zrubc9/installation_guide_for_riffusion_app_inference/}}. \textit{M$^{2}$UGen}, \textit{MuseFormer}, and \textit{MuseCoco} involve the most challenging installations as they require multiple model checkpoints (for \textit{M$^{2}$UGen}), and complex dependency management with several open issues in their GitHub repository.

\subsection{Hardware Requirements for Training and Inference}\label{sec:hardware_analysis}

Understanding hardware specifications is essential for assessing the feasibility of training and deploying AI music generation systems. Tab. \ref{tab:hardware_requirement} provides a comprehensive overview of the necessary computational resources, including NVIDIA GPU types and quantities, CPU compatibility, and memory capacity requirements. Such information allows researchers and practitioners to determine hardware prerequisites for experimental implementations and practical applications.

\begin{table*}[t]
    \centering
    \caption{Estimated hardware requirements for training and inference of the systems considered in this study based on the analysis during the \textit{Systems Overview} phase. }\label{tab:hardware_requirement}
    \resizebox{\textwidth}{!}{\begin{tabular}{cllll}
      \toprule
      No.           & Model                        & Training                          & Inference                   \\
      \midrule
      1             & \textit{M$^{2}$UGen}         & 2x NVIDIA V100 GPUs 32GB          & 1x NVIDIA V100 GPU  32GB    \\
      2             & \textit{MusicGen}            & 4-8x NVIDIA A100 GPUs 80GB        & GPU with atleast 16GB RAM   \\
      3             & \textit{MuseCoco}            & 8x NVIDIA V100 GPUs 32GB          & 1x NVIDIA V100 GPU  32GB    \\
      4             & \textit{Magenta Studio 2.0}  & N.A                               & CPU 16GB                    \\
      5             & \textit{Magenta DDSP-VST}    & 1x NVIDIA GTX 1060 6GB            & CPU 16GB                    \\
      6             & \textit{MuseFormer}          & 8x NVIDIA V100 GPUs 32GB          & 1x NVIDIA V100 GPU  32GB    \\
      7             & \textit{Musika}              & 1x NVIDIA RTX 2080 Ti 11GB        & 1x NVIDIA RTX 2080 Ti 11GB / CPU 16GB  \\
      8             & \textit{Riffusion}           & 1x NVIDIA RTX GPU 8GB             & 1x NVIDIA GTX 1060 6GB      \\
      \bottomrule
    \end{tabular}}
  \end{table*}

The training of computationally intensive systems such as \textit{MusicGen} demands substantial resources, typically necessitating 4 to 8 high-performance NVIDIA A100 GPUs, each equipped with 80GB of VRAM. These requirements present significant accessibility barriers for individual artists and smaller studios due to their prohibitive cost and resource intensity. In contrast, systems such as \textit{Musika} and \textit{Riffusion} offer better accessibility by requiring only a single high-end GPU such as the NVIDIA RTX 2080 Ti (11GB) or a standard RTX GPU (8GB), respectively.

For inference processes—utilizing trained models to generate music from inputs—the hardware requirements are generally less demanding than training but vary considerably across systems. \textit{MusicGen} requires a GPU with a minimum of 16GB of VRAM for medium-sized models (~1.5B parameters). Although this requirement is less intensive than training, it may still exceed the resources available to many potential users. Notably, \textit{Magenta Studio} and \textit{DDSP-VST} can perform inference on a CPU with 16GB of RAM, representing the most accessible option among the systems considered here.

It is important to note that this study did not utilize the exact hardware specifications listed (Section \ref{sec:hands_on_experimentation} elaborates on how these systems were accessed and utilized). Furthermore, some research groups have not reported comprehensive hardware requirements. Consequently, the information presented in Table \ref{tab:hardware_requirement} serves as an estimated guideline for the computational resources necessary to operate these systems.

\subsection{Application in Music Production Tasks}\label{sec:application_music_production_tasks}

Based on the features and capabilities of each examined system, this section analyzes their applications in music production tasks: Composition (C), Arrangement (A), and Sound Design (SD). Table \ref{tab:music_tasks_models} suggests the appropriate applications for the systems within these music production contexts.

Text-to-audio systems like \textit{MusicGen} and \textit{Riffusion} excel in early-stage composition by accelerating ideation through audio generation conditioned on textual or melodic inputs. Their capacity to produce 10-second motifs to 4-minute segments presents a new possibility for curating sample collections. However, their lack of structured output coherence limits their utility to concept development rather than full-track composition. This positions them as creative catalysts rather than substitutes for structured arrangement workflows\footnote{We refer to a deliberate, controlled approach to music creation with precise structural organization—a process MGS cannot fully replicate.}.

Symbolic generation systems like \textit{MuseCoco} and \textit{MuseFormer} address higher-level compositional challenges by balancing linguistic input interpretation with structural integrity. \textit{MuseCoco}’s attribute-based control enables targeted exploration of harmonic/melodic variations. This makes it suitable for iterative refinement of pre-existing motifs. Conversely, \textit{MuseFormer}’s architectural design allows for the generation of extended compositions requiring thematic consistency. Both systems compensate for text-to-audio tools’ structural deficiencies, but their reliance on symbolic representation (MIDI) may appeal less to creators accustomed to audio workflows.

\begin{table}[t]
    \centering
    \caption{Suggested music production tasks based on the analysis during the \textit{Systems Overview} phase of this study, including Composition (C), Arrangement (A), Sound Design (SD)}\label{tab:music_tasks_models}
    \begin{tabular}{clcccc}
    \toprule
    No.           & Model Name                              & C          & A          & SD \\
    \midrule
    1             & \textit{M$^{2}$UGen}                    & \checkmark & \checkmark & \checkmark  \\
    2             & \textit{MusicGen}                       & \checkmark & \checkmark & \checkmark  \\
    3             & \textit{MuseCoco}                       & \checkmark & \checkmark &             \\
    4             & \textit{Magenta Studo 2.0}              & \checkmark & \checkmark &             \\
    5             & \textit{Magenta DDSP-VST}               &            &            & \checkmark  \\
    6             & \textit{MuseFormer}                     & \checkmark & \checkmark &             \\
    7             & \textit{Musika}                         & \checkmark &            & \checkmark  \\
    8             & \textit{Riffusion}                      & \checkmark &            & \checkmark  \\
    \bottomrule
    \end{tabular}
    \end{table} 

For arrangement tasks, \textit{Magenta Studio 2.0} demonstrates practical utility through its phrase interpolation and continuation features, which assist in bridging compositional gaps between disparate musical ideas. However, its inability to enforce style constraints may yield outputs requiring post-generation editing, which diminishes time efficiency. This contrasts with \textit{M$^{2}$UGen} 's multimodal approach, which theoretically enables arrangement decisions informed by visual narratives but struggles with latency-induced workflow disruptions. Neither system fully resolves the core challenge of maintaining artistic intentionality during automated arrangement—a gap partially can be filled by \textit{MuseFormer} 's structural awareness and \textit{MuseCoco} 's musical attributes mapping but limited by their symbolic format constraints.

In sound design, \textit{DDSP-VST} is the most relevant tool (among the systems considered) due to its real-time timbral manipulation capabilities within standard DAW environments. Its differentiable signal processing architecture provides granular control over harmonic content, which outperforms generative systems like \textit{Musika} that prioritize musical texture creation over precise sonic sculpting. However, \textit{Musika} 's unconditional generation remains valuable for exploratory soundscape design where serendipitous discoveries outweigh deterministic outcomes. The dichotomy between these approaches underscores a fundamental tension in AI-assisted sound design: generative systems expand creative possibilities but reduce replicable precision, while DSP-based tools enhance control at the expense of autonomous creativity.

The systems collectively demonstrate the potential to augment—but not yet redefine—established production workflows. Their effectiveness correlates inversely with task complexity: it is strongest in atomic tasks like motif generation, sample collection, or timbral transformation and weakest in holistic composition requiring hierarchical structural planning. 

The following section offers a more detailed exploration of these systems by expanding on the analyses and results presented. It investigates their capabilities and uses in the process of music creation, with an emphasis on producing a complete musical composition. This hands-on assessment offers a more profound understanding of these AI music systems' performance within music production workflows rather than their theoretical potential.

\section{Hand-On Experimentation}\label{sec:hands_on_experimentation}

This section examines selected systems across various stages of music production (Section \ref{sec:research_questions}),from initial conceptualization to final compositional output. This phase investigates the systems' practical utility throughout the production workflow by evaluating their creative affordances, generative capacity for musically relevant content and their efficacy in transforming conceptual ideas into cohesive musical compositions. The evaluation follows the steps outlined in Section \ref{sec:evaluation_phases}. The complete evaluation outcomes are presented in Tab. \ref{tab:hands_on_experiment_result}.

The evaluation involves an analysis of each system's integration capabilities within established production workflows and their capability in facilitating sustained creative engagement. As detailed in Section \ref{subsec:evaluation_process} and demonstrated in Fig. \ref{fig:eval_steps}, the assessment occurs during both the \textit{Content Generation} and \textit{Curation} process concurrently, wherein diverse prompts representing specific musical concepts are provided to text-to-music generative systems.

In the following analysis, we examine the operational efficacy of these systems in generating musical content, the inherent challenges in directing them toward precise musical objectives, the qualitative aspects of their outputs, and a comprehensive assessment based on predetermined evaluation criteria. Further discussion addresses the integration methodology for incorporating generated content into final compositions through iterative \textit{Curation} process. These complementary processes of generation and curation reflect the non-linear nature of music production where creative ideation and content generation continue throughout the compositional process.

The culmination of this experimental investigation is a musical composition that will be made available on SoundCloud\footnote{The link to the SoundCloud playlist is not provided due to the peer review process.}. Regarding the use of systems, web-based interfaces were utilized for systems requiring specialized configurations or lacking local implementation capabilities, specifically \textit{MusicGen} and \textit{M$^{2}$UGen} are accessed through within Hugging Face spaces. \textit{Riffusion} was accessed through both local installation\footnote{We used the AUTOMATIC1111 web UI extension to access \textit{Riffusion} locally.} and its web interface\footnote{Last accessed on 31/08/2024.}. \textit{Musika} was also utilized through local deployment. It should be noted that \textit{MuseFormer} and \textit{MuseCoco} were excluded from the \textit{Hands-on Experimentation} phase due to persistent technical impediments regarding local inference execution and the absence of accessible web-based alternatives. The systems were evaluated in a home studio environment equipped with music production tools as detailed in Tab. \ref{tab:studio_config} in Appendix \ref{app:home_studio}. The experimentation was conducted by the first author, as described in Section \ref{subsec:evaluation_process}.

\subsection{Content Generation}

The following first delineates the workflow for the \textit{Content Generation} process with particular emphasis on prompt-based systems. This process includes defining the project's musical structure by identifying essential musical elements and formulating standardized prompt templates designed to guide the systems toward contextually appropriate musical outputs. It is important to note that music creation processes varies among different creators, with diverse approaches to conceptualization, composition, and production workflows. The workflow presented, herein, is designed not as a definitive approach, but rather to enhance the transparency and structural coherence of the evaluation process, which potentially may serve as a procedural reference for future studies.

\subsubsection{Workflow}\label{sec:workflow}

\begin{figure}[t]
  \centering
  \includegraphics[width=\linewidth]{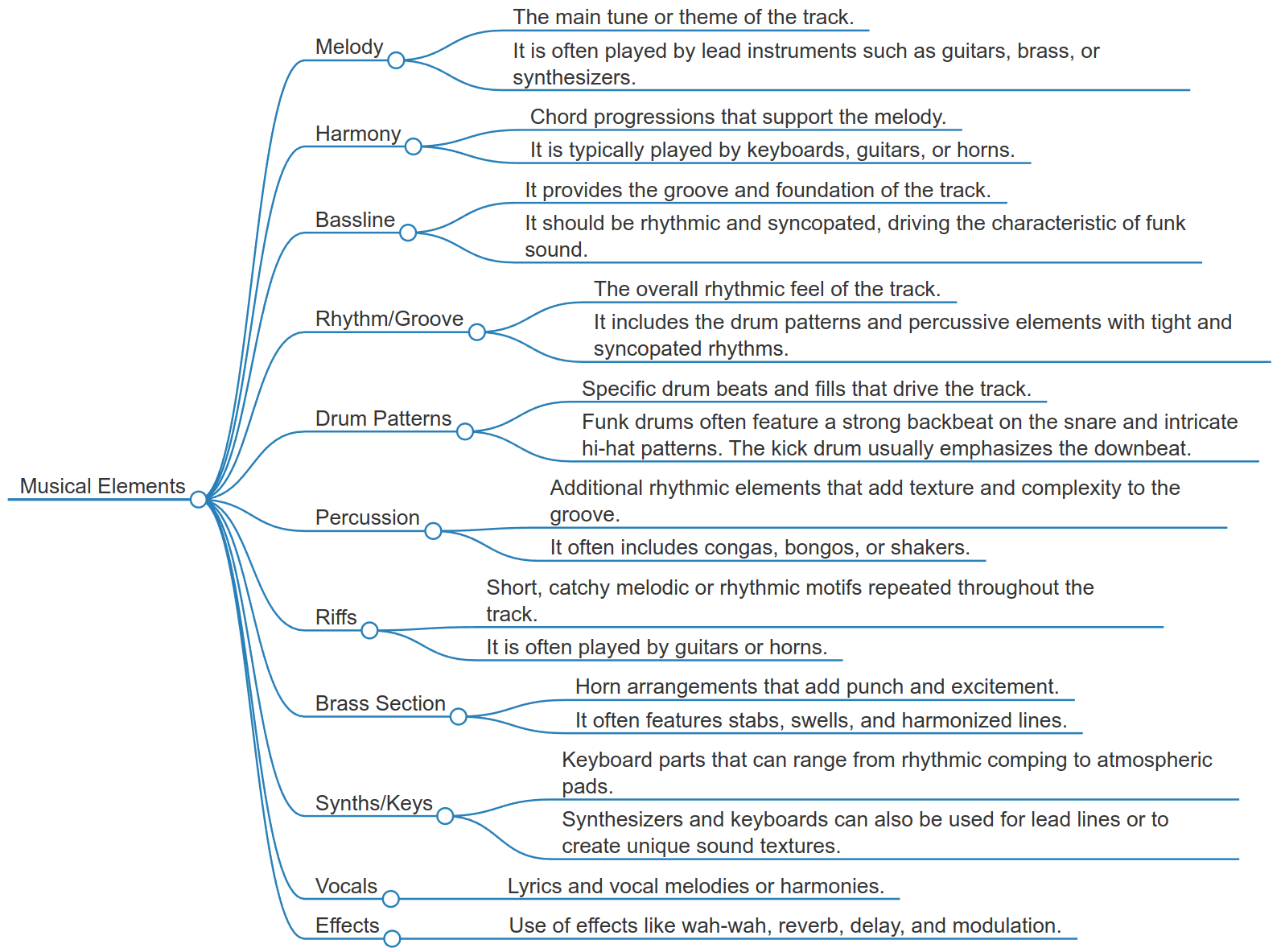}
  \caption{Overview of the musical elements used in the \textit{Content Generation} process during the \textit{Hands-on Experimentation} phase. It includes descriptions of their roles and the typical instruments involved. Each element is identified to help guide the prompt-based systems effectively, ensuring that the generated content aligns with the project's thematic and stylistic specifications. Notably, the funk-inspired characteristics of these musical elements are emphasized for further clarity.}
    \label{fig:music_elements}
\end{figure}

The initial phase of the workflow involves establishing the thematic and stylistic parameters of the final composition. This aims to align the intended creative direction and adherence to genre-specific characteristics—an essential consideration when working with prompt-based systems. In this particular context, the energetic and rhythmic qualities inherent to the funk genre serve as the primary inspiration for the final track. The compositional structure adheres to a verse-chorus format with tempo considerations ranging from 90 to 130 beats per minute (BPM). This provides sufficient flexibility for exploring the capabilities of systems, which are trained on diverse musical examples, while maintaining consistency and relevance in content generation.

To utilize the systems and produce pertinent musical content for the final composition, the musical elements presented in Fig. \ref{fig:music_elements} are considered and incorporated into the workflow. This facilitates the \textit{Content Generation} process and establishes a comparative framework for evaluating the distinctive capabilities and limitations of these systems throughout the subsequent \textit{Curation} process\footnote{These capabilities and limitations encompass each system's comprehension of musical concepts, genre-specific proficiency, instrumental representation accuracy, and interpretative fidelity to instructional parameters. For instance, some systems may excel at generating techno music but struggle with jazz, while others might be proficient at generating basslines but less effective at string instruments, which are primarily related to the system's training examples.}. The prompt creation process begins with the utilization of ChatGPT\footnote{\url{chat.openai.com}} to generate ten distinct templates for each musical element. These initial templates serve as the basis for subsequent customization and refinement of the prompts. Tab. \ref{tab:prompt_template} presents exemplary prompt templates for each musical element. The prompts are then further refined through the modification of keywords by incorporating specific musical attributes such as tempo indications, stylistic descriptors, and instrumental specifications. For other systems included in this experimentation, appropriate inputs were provided according to each system's specific requirements and parameters.

The subsequent section presents observations from the evaluator's interactions with these generative systems, focusing on their practical utility and creative affordances in generating musical content within the prescribed workflow.

\begin{table}[t]
    \caption{Examples of prompt templates for each musical element.}
    \label{tab:prompt_template}
    \centering
    \resizebox{\textwidth}{!}{\begin{tabular}{lp{12cm}}
        \toprule
        Element & Description \\
        \midrule
        Melody & Create a catchy funk melody with a syncopated rhythm and a playful, upbeat feel, suitable for a lead guitar or brass instrument. \\
        Harmony & Create a funky chord progression with syncopated rhythms and extended chords, such as 7ths and 9ths, suitable for electric piano or guitar. \\
        Bassline & Create a syncopated and punchy bassline with a strong emphasis on the off-beats, perfect for driving a funk groove. \\
        Rhythm/Groove & Create a tight, syncopated drum groove with a strong backbeat and intricate hi-hat patterns, perfect for a classic funk feel. \\
        Drum Patterns & Create a classic funk drum pattern with a strong backbeat on the snare and syncopated hi-hat rhythms, emphasizing groove and feel. \\
        Percussion & Create a lively percussion track featuring congas and bongos with syncopated rhythms that complement the main groove. \\
        Riffs & Create a catchy guitar riff with a syncopated rhythm and a bluesy feel, perfect for driving the groove of a funk track. \\
        Brass Section & Create a bold and punchy brass section riff with tight harmonies and syncopated stabs, perfect for accentuating the groove. \\
        Synths/Keys & Create a funky keyboard comping pattern with syncopated rhythms and extended chords, perfect for an electric piano or clavinet. \\
        Vocals & Create a catchy vocal hook with a rhythmic delivery and a playful, upbeat feel, perfect for a funk chorus.\\
        Effects & Create a wah-wah effect for the guitar, adding a classic funk touch with rhythmic modulation and dynamic expression. \\ 
        \bottomrule
    \end{tabular}}
  \end{table}

\subsubsection{Observations from Content Generation}\label{sec:content_generation_observations}

During experimentation, \textit{MusicGen} demonstrated capability in generating musical motifs or segments suitable for theme development. However, directing the model toward specific musical concepts proved challenging due to prompt formulation difficulties. The system occasionally produced incongruent outcomes; for instance, when prompted to generate a 'Funky bass line, 90 BPM tempo with a syncopated and rhythmic groove,' it generated a drum track instead. Similarly, when instructed to create compositions with specific instrumental elements, the outputs frequently failed to correspond to the provided descriptions.

Structural limitations were evident in the abrupt initiation and termination of compositions that resulted in the absence of proper introductions or conclusions. The \textit{MusicGen}'s inability to begin compositions on specific beats presented integration challenges for existing musical structures. Sound quality exhibited variability based on prompt specifications, though generations were generally adequate for inclusion with additional processing. An essential limitation was the inability to generate isolated instrumental tracks, necessitating source separation techniques. The web interface, while user-friendly, exhibited constraints in generation speed (approximately 200 seconds for 15-second segments) and composition length, with alternative execution methods offering improved performance at the cost of greater computational demands\footnote{The generation time also significantly varied depending on user traffic and request volume. For faster and longer generations, it was possible to use local or Google Colaboratory execution. Nevertheless, these options need higher computational resources, as noted in Tab. \ref{tab:hardware_requirement}.}.

\textit{M$^{2}$UGen}, building upon \textit{MusicGen}'s framework, inherited both its capabilities and limitations. Its chat-based interface facilitated a more conversational approach to guiding the generation process. However, \textit{M$^{2}$UGen} exhibited deficiencies in image-to-music translation, which diminished its multimodal feature's effectiveness. Other users have also reported similar issues \citep{M2UGenIssue4}. Like \textit{MusicGen}, \textit{M$^{2}$UGen} could not generate specific instrumental tracks that limited its utility when granular control over individual musical elements was required. It also shared the challenge of prompt writing with \textit{MusicGen}, where the quality of the output heavily depended on the evaluator's ability to craft effective prompts.

\textit{Musika} demonstrated particular proficiency within the techno music domain. The system generates stylistically coherent but often contextually limited compositions that lack the specific characteristics required for targeted production needs. Its practical utility is constrained by the inability to generate contextually relevant content without additional model training or fine-tuning. This requires substantial datasets and computational resources beyond this study's scope. Nevertheless, \textit{Musika}'s pre-existing checkpoints were utilized to incorporate elements into the music project despite their limited use.

\textit{Riffusion} exhibited better responsiveness and efficiency compared to other evaluated systems by offering both local and web-based interfaces with complementary advantages. The system demonstrated higher fidelity to prompt descriptions, particularly in capturing rhythmic and stylistic elements. Despite these strengths, audio quality varied significantly between interfaces, and outputs consistently required additional processing prior to integration. The local version produced notably inconsistent results, with melody-oriented prompts yielding more reliable outputs than percussion-focused ones. While prompt formulation challenges persisted, the system's accelerated generation speed facilitated more rapid iterative prompt refinement and exploration. Consistent with other evaluated systems, \textit{Riffusion}'s generations exhibited deficiencies in structural elements, particularly lacking proper introductions and conclusions.

\textit{Magenta Studio 2.0} featured an intuitive interface but produced outputs lacking the coherence and contextual appropriateness achieved by other systems. Its continuation feature generated compositions that were disconnected from the provided musical contexts. This required multiple interpolation attempts to achieve satisfactory results. Similarly, attempts to generate unconditioned percussion or melodic elements yielded outputs of insufficient quality for integration into the music project. Indeed, a significant limitation of \textit{Magenta Studio 2.0} was its inability to tailor output based on specific musical parameters or contextual inputs, which would have enhanced the relevance of generated content.

The \textit{Magenta DDSP-VST} excels in audio manipulation and transformation by incorporating various pre-built instrumental timbres, such as flute, bassoon, and trumpet. Although we did not do this, personalized sound profiles and textures can be created by training a custom DDSP model on specific examples or recording sessions, even those as short as 10 minutes\footnote{This feature allows for experimentation with new synthesis sounds. The training can be done through a Google Collaboratory Notebook, which can take between 3 to 20 hours.}. A distinctive feature of \textit{Magenta DDSP-VST} is its capability for morphing between different instrumental timbres via an XY pad interface. This provides a tactile approach to sound transformation that enables precise control over timbral and dynamic qualities to create complex harmonic textures.

\subsection{Curation}\label{sec:curation}

The final track, which emerged from the \textit{Curation} process, was created by curating, preparing, and integrating outputs from various systems. As mentioned previously (Section \ref{sec:evaluation_phases}), this process was performed simultaneously with \textit{Content Generation} throughout the experimentation phase. The following section presents the track creation process based on evaluator notes. It also highlights individual system capabilities and their collective contributions to the final composition.

\subsubsection{Observations from Curation}\label{sec:curation_observations}

All examined systems facilitated exploratory approaches to music creation to varying degrees by enabling users to draw inspiration and experiment with diverse compositional elements, sound textures, and timbral combinations. However, these systems frequently demonstrated limitations in structural coherence, often producing abrupt transitions between musical themes and textures, particularly over extended durations. This issue was most pronounced with \textit{Musika} and, to a lesser extent, with \textit{MusicGen} and \textit{Riffusion}. Consequently, these systems proved more effective for generating shorter musical segments, which could then be arranged and sequenced to form cohesive compositions.

Furthermore, \textit{Musika} appeared to be the least practical model for the production due to its unconditional generation approach, which complicated efforts to guide the model toward generating musical samples that would fulfill the project's specific requirements. By comparison, \textit{MusicGen} and \textit{Riffusion} offered improved reliability and consistency in generating usable musical samples, making them more suitable choices for this particular application.

Throughout the \textit{Curation} process, source separation algorithms such as \textit{Demucs} \citep{rouard2022hybrid} became essential tools, as most systems struggled to generate isolated instrumental tracks. The sonic characteristics of different instruments were frequently blended, resulting in outputs where individual elements lacked clear distinction. This amalgamation complicated post-production processes, particularly mixing and mastering, as the absence of clear separation obscured the timbral qualities of each instrument. Consequently, the raw audio output often fell short of the sonic characteristics typically desired in professional music production\footnote{While these outputs could potentially find application in certain experimental or lo-fi genres, they generally required post-processing to align with contemporary production standards across most mainstream musical styles.}. Therefore, once individual instruments were isolated, they underwent additional processing to achieve the desired audio quality and ensure each element contributed effectively to the overall composition. These processing steps included:

\begin{itemize}
  \item Equalization: Adjustment of the frequency spectrum to enhance clarity and balance.
  \item Compression: Normalization of dynamic range to create more compact and impactful sounds.
  \item Reverb and Delay: Application of spatial effects to simulate varied acoustic environments.
  \item Trimming: Refinement of audio clip boundaries to ensure seamless integration.
\end{itemize}

Overall, working with these systems to produce the final track proved challenging and occasionally overwhelming. Generating samples that aligned with the project's musical direction often required extended periods of experimentation. In many instances, conventional approaches—such as creating chord progressions or melodies using a MIDI editor or directly recording instrumental parts—would have been more efficient. The AI-generated samples thus functioned primarily as components within a broader production workflow, where short musical segments were generated and subsequently arranged to compose the final track.

One of the more positive aspects involved using \textit{Magenta DDSP-VST} for rendering MIDI tracks and synthesizing and manipulating sounds. This particular tool offered an intuitive interface that facilitated the exploration of sound textures by providing meaningful control over the timbral characteristics of the composition.

After preparing and arranging all generated musical segments, vocals, and sounds, post-processing was undertaken to ensure the final track was cohesive, polished, and well-balanced. This post-processing step involved several procedures to enhance audio quality and meet professional standards. The mixing process involved balancing amplitude levels, applying time-based effects such as reverb and delay, and setting appropriate stereo imaging to create spatial definitions between elements. Frequency-domain processing was applied to individual tracks to ensure spectral clarity and prevent masking artifacts. Following the mixing stage, mastering was performed to optimize the overall tonal balance, dynamic range, and loudness. 

The following summarizes the steps taken throughout the production process:

\begin{itemize}
    \item \textit{MusicGen} provided the initial building blocks for the composition—specifically a bass groove and guitar rhythms—which formed the core structure of the piece.
    
    \item A source separation algorithm \citep{rouard2022hybrid}, accessible through the Demucs online interface\footnote{\url{https://demucs.danielfrg.com}}, was employed to isolate individual instruments. The isolated stems subsequently underwent additional processing (when needed or desired), such as equalization, compression, reverb, delay, and trimming.
    
    \item \textit{Magenta Studio 2.0} was utilized to generate the drum pattern, adding a percussive layer that complemented the bass line and contributed to the overall rhythmic structure.
    
    \item Harmonic textures were generated using \textit{Riffusion}. The generated audio was then transcribed into MIDI tracks using Ableton Live's built-in functionality. The MIDI tracks were then adjusted to align with the composition's context and dynamics.
    
    \item The MIDI tracks were synthesized using \textit{Magenta DDSP-VST} to perform timbral adjustments to match the composition's aesthetic. The pre-built instruments within the plugin were utilized, while the XY pad facilitated hands-on manipulation of sound textures.
    
    \item The composition also featured a vocal track created by inputting lyrics generated by \textit{ChatGPT} into the \textit{Riffusion} web interface.
    
    \item After completing the initial production steps—comprising composition, pitch correction, and rhythmic alignment—the audio and MIDI tracks were edited and arranged to establish the intended song structure.
    
    \item Final post-production processes, including mixing, mastering, and spatial processing, were conducted to ensure a cohesive and professionally finished product.
\end{itemize}

\section{Results Analysis and Comparison}\label{sec:results_analysis}

This section analyzes to what extent the systems fulfilled the expected capabilities for specified tasks (Section \ref{sec:systems_overview}) during the \textit{Hands-on Experimentation} phase. To accomplish this, the observational notes presented previously (Section \ref{sec:hands_on_experimentation}), alongside the results from quantitative metrics shown in Tab. \ref{tab:hands_on_experiment_result}, inform the presented analysis.

\begin{table}[t]
    \centering
    \caption{Comparative overview of systems performance result based on the "performance" criteria during the \textit{Hands-on Experimentation} phase. The evaluation is based on a scale of 1-5, with 1 being the lowest and 5 being the highest, as described in Appendices \ref{app:evaluation_criteria} and \ref{app:criteria_score_levels}.}
    \label{tab:hands_on_experiment_result}
    \begin{tabular}{lcccccccc}
        \toprule
        System  & \rotatebox{90}{Usability} & \rotatebox{90}{Generation Speed} & \rotatebox{90}{Audio Quality} & \rotatebox{90}{Stylistic Accuracy} & \rotatebox{90}{Parameter Control} & \rotatebox{90}{Content Generation Control} & \rotatebox{90}{DAW Compatibility} & \rotatebox{90}{Creative Control} \\ 
        \midrule
        \textit{M$^2$UGen}          & 3  & 2  & 3     & 3  & 3  & 2  & 1  & 2 \\ 
        \textit{MusicGen}           & 3  & 2  & 3     & 3  & 3  & 2  & 1  & 2 \\ 
        \textit{Magenta Studio 2.0} & 3  & 4  & -     & 2  & 2  & 1  & 3  & 1 \\ 
        \textit{Magenta DDSP-VST}   & 4  & 5  & 4     & 4  & 4  & 4  & 3  & 4 \\ 
        \textit{Musika}             & 2  & 3  & 2     & 1  & 1  & 1  & 1  & 1 \\ 
        \textit{Riffusion}          & 3  & 4  & 3     & 3  & 3  & 3  & 1  & 3 \\ 
        \bottomrule
    \end{tabular}
  \end{table}

In composition tasks, while systems such as \textit{Riffusion} and \textit{MusicGen} demonstrated competence in generating short segments or motifs, a lack of structural coherence across outputs—characterized by abrupt transitions, incomplete introductions, and limited alignment to musical prompts—restricts their utility to sample creation rather than full-track compositions. Although these systems can facilitate the initiation of musical works, they lack proficiency in developing thematically consistent pieces that reflect specific narratives or emotional themes. 

The observational notes further highlight several mismatches between user instructions and system outputs, where prompts requesting specific instruments or grooves frequently yielded irrelevant content. This uneven performance across prompt types reveals domain-specific strengths rather than generalizable understanding of musical concepts. \textit{Riffusion} demonstrated better capability when handling prompts that included rhythmical aspects of music. For instance, it excelled when prompted to generate 'Funky bass line, 90 BPM tempo with a syncopated and rhythmic groove' and 'Drum with a deep groove, incorporating a solid backbeat with snappy snare hits and a tight kick drum pattern.' Conversely, \textit{MusicGen} exhibited relative strength with electronic textures, as it excelled at prompts like 'Digital synths with arpeggiated patterns and spacey effects.' Of particular interest is their mutual inadequacy when responding to prompts with more complex musical aspects such as 'A funky chord progression with a mix of extended chords' and 'Electric keyboards using Fender Rhodes for warm, classic funk chords and a clavinet for its distinctive percussive stabs.' These inconsistencies reveal limitations in the systems' capacity to interpret musical instructions and utilize textual inputs as reliable control mechanisms for the generation process. Section \ref{sec:challenges_prompt} will discuss this issue in detail.

Transitioning from composition to arrangement considerations, effective arrangement workflows necessitate systems that can bridge compositional gaps while maintaining stylistic integrity. \textit{Magenta Studio 2.0} partially addresses this requirement through its phrase interpolation and continuation features by facilitating a degree of structural coherence between disparate ideas. Though conceptually valuable, its inability to maintain stylistic consistency, as reflected in Tab. \ref{tab:hands_on_experiment_result}, often results in disjointed outputs that necessitate substantial editing, diminishing its effectiveness as a time-saving tool. Similarly, music arrangement represents another area of limitation for systems such as, \textit{Riffusion}, \textit{MusicGen} and \textit{M$^{2}$UGen}. While capable of generating multi-instrumental compositions, these systems lack the nuanced decision-making and control that a music producer would typically exercise during the production process. Notably, these systems fail to provide direct control over individual instruments in the generated compositions, as they are limited to textual prompt control mechanisms. These limitations underscore a fundamental challenge: preserving artistic intentionality during generation processes—a feature none of the systems effectively achieve. The observational notes consistently indicated the need for iterative refinements and user intervention to correct incoherent transitions within generated content.

When considering sound design capabilities, \textit{Magenta DDSP-VST} offers real-time, intuitive, and precise control over the timbre and texture of sounds. This system attained the highest performance ratings among all systems (Tab. \ref{tab:hands_on_experiment_result}). Similarly, \textit{MusicGen}, \textit{M$^{2}$UGen}, \textit{Riffusion}, and \textit{Musika} demonstrate capabilities in generating sounds and imitating acoustic instruments. However, these systems frequently produce complex sound layers that prove unsuitable for projects requiring specific instrument sounds or effects. This limitation necessitates additional processing steps, as elaborated in Section \ref{sec:curation_observations}, to isolate individual elements (stem separation)—a requirement that introduces complexity and potentially diminishes sound quality. Moreover, these systems encounter significant constraints in real-time interaction due to latency issues inherent in their architectures. This performance limitation substantially restricts their utility in scenarios where immediate responsiveness is essential for improvisation and interaction with musicians.

Beyond these specific functional limitations, the consistently low scores across systems for content generation control (ranging from 1-3/5) and DAW integration (1-3/5) indicate a disconnect between these technologies and professional production environments. Furthermore, the observations suggests that while current systems demonstrate proficiency in generating sonic material, they fundamentally lack the precise control mechanisms necessary for workflow integration. Sections \ref{sec:integration_mgs} and \ref{sec:collaborative_tools} will discuss this further.

\section{Discussion}\label{sec:discussion}

As observed throughout this study, the adoption of MGS can transform the role of music creators. This study delved into their potential through theoretical review and hands-on experimentation with selected systems. The evaluation framework used in this study was designed to investigate the applicability of these systems through qualitative and quantitative analysis of their performance in specified music production tasks. The findings have provided insights into the practical uses of these technologies in music production and the challenges that must be addressed to realize their full potential.

The subsequent discussion considers the creative affordances, practical value and challenges presented to address the research questions raised in Section \ref{sec:research_questions}. It aims to comprehend the potential of AI as a collaborator in the music creation process, rather than merely a tool for automation.

\subsection{Limitations of MGS in Music Production Workflows (RQ1)}\label{sec:limitations_mgs}

Through the assessments, it became evident that these systems can automate and enhance the creative process of music production to some extent. However, their integration into production workflows reveals fundamental tensions between technological sophistication and practical utility. While text-to-audio systems enable rapid musical ideation, they introduce a paradoxical relationship where accelerated content creation inversely correlates with compositional intentionality. As \citet{huang_ai_2020} observe, `ML models are not easily steerable', forcing users to generate `massive numbers of samples and curate them post-hoc' rather than directing the generative process with precision. This stochastic nature often necessitates post-generation editing to align outputs with artistic vision, which suggests their primary value lies not in autonomous creation but as catalysts for divergent thinking during creative impasses.

This challenge extends to production-integrated tools such as \textit{Magenta Studio 2.0} and \textit{DDSP-VST}, where an ergonomic divide emerges. Despite their DAW compatibility enabling workflow integration, many systems operate as opaque black boxes with conditional generation parameters that users cannot meaningfully modify. \citet{deruty_development_2022} highlight this limitation, noting that `without any visualization, the only way to navigate variations in output is by trial-and-error', which limits creative agency. Consequently, creators are restricted to superficial interactions that more closely resemble managing an unpredictable collaborator than operating a precise instrument. \textit{DDSP-VST}'s relative success with real-time parametric controls suggests that effective AI tool design requires recontextualizing—rather than replacing—existing interaction metaphors familiar to music producers.

Beyond interface considerations, performance constraints further limit the practical application of these systems. The latency-quality tradeoff observed across systems carries profound workflow implications. When generation times approach or exceed traditional composition durations, the presumed efficiency benefits become paradoxical. This necessitates a reevaluation of tool design priorities and positions these systems not as time-saving devices but as exploratory ideation tools. The cognitive burden of such approaches is substantial, as \citet{huang_ai_2020} describe how musicians must `juggle not only the creative process but also the technological processes imposed by the idiosyncrasies and lack of steerability of learning algorithms', creating parallel feedback loops of creativity and technical management that can detract from artistic focus.

These performance issues contribute to several universal limitations across all evaluated systems. Prompt formulation represents a vital bottleneck, with outputs heavily dependent on the creator's ability to craft effective prompts—a skill users may lack, leading to inconsistent results. Additionally, structural shortcomings, including the inability to isolate instrumental tracks or enforce cohesive introductions and endings, were consistently identified. These deficiencies necessitate post-processing interventions, including source separation using external algorithms like \textit{Demucs} and extensive mixing and mastering to align generated content with production standards. Such requirements diminish the systems' utility for seamless creative workflows by positioning them as supplementary tools rather than standalone solutions.

Given these constraints, these systems neither obsolete nor revolutionize traditional production practices but instead demand new hybrid competencies. Creators must now mediate between stochastic generation and intentional curation, between algorithmic suggestions and critical listening, which necessitates a redefinition of musical expertise in the AI era. These systems' value lies not in autonomous generation but in their capacity to expand creative possibility spaces when guided by users possessing both musical expertise and technical acuity. 

\subsection{Integration of MGS in Music Production Workflows (RQ2)}\label{sec:integration_mgs}

The integration of these systems into real-world music production workflows, however, presents practical challenges, particularly regarding hardware requirements. As indicated in Tab. \ref{tab:hardware_requirement}, the substantial computational resources needed for training most systems (4 to 8 GPUs) suggest that development is primarily driven by well-resourced organizations or research institutions. This creates accessibility barriers for individual producers, smaller studios, and research groups with limited resources. While inference can run on less powerful hardware, widespread adoption depends on this accessibility factor. Systems like \textit{Musika} and \textit{Magenta} offer promising alternatives through CPU compatibility and Google collaboratory notebooks as cloud-based solutions for fine-tuning and training. However, these cloud-based approaches introduce their own concerns regarding cost, data security, and vendor dependence that warrant separate investigation.

To address these accessibility challenges, online demos and web interfaces have emerged as important intermediary solutions. During the \textit{Hands-on Experimentation} phase, \textit{MusicGen}, \textit{Riffusion}, and \textit{M$^{2}$UGen}'s web interfaces allowed for system testing without substantial infrastructure investments. This approach creates a valuable feedback loop where creators can evaluate systems for specific projects while providing developers with real-world usage data. Such feedback mechanisms enable algorithmic refinements, interface improvements, and enhanced integration capabilities with existing production tools—embodying the iterative, community-driven nature of open-source development.

Beyond accessibility considerations, open-source systems offer advantages through their command-line and API interfaces, despite their resource demands. The availability of pre-trained checkpoints transforms these systems into general-purpose frameworks applicable across various domains. These systems democratize access to cutting-edge AI technologies while eliminating licensing fees, proprietary restrictions, and other integration obstacles \citep{ma_foundation_2024}. Furthermore, they enable rapid customization and foster community-driven innovation, as evidenced by developments surrounding Stable Diffusion models like ComfyUI\footnote{\url{https://github.com/comfyanonymous/ComfyUI}} and Automatic1111\footnote{\url{https://github.com/AUTOMATIC1111/stable-diffusion-webui}}. This collaborative ecosystem of tools and extensions stands in contrast to proprietary systems, which typically offer limited customization through standardized interfaces that may not accommodate diverse user needs.

The complexity of music production, which requires simultaneous management of multiple tasks as discussed in Section \ref{sec:music_concepts}, particularly benefits from open-source systems' flexibility. MusicGen exemplifies this adaptability, as its open-source nature facilitates various modifications including weight adjustments for genre-specific fine-tuning, latent space manipulation for creative exploration, and architectural optimizations for different objectives. These adaptations allow the model to address specific production challenges while expanding creative possibilities.

A notable example is instruct-MusicGen proposed by \citet{zhang_instruct-musicgen_2024}, which enhances the original model through instruction tuning that enables response to text-based editing commands. By integrating both text fusion and audio fusion modules, this approach can simultaneously process textual instructions and audio inputs. This enables various music editing capabilities, such as adding, removing, or isolating audio stems, which can potentially alleviate the corresponding shortcomings observed in this study. This approach demonstrates how open-source models can evolve to operate within DAWs and provide intelligent assistance during production by generating complementary instrumental elements (bass lines, drum patterns, harmonies) that align coherently with primary melodic content.

\subsection{Music Generation Systems as Collaborative and Creative Tools (RQ3)}\label{sec:collaborative_tools}

AI-generated music can serve as a source of inspiration, particularly during the ideation phase of composition. The ability to quickly generate ideas and explore new musical spaces can be appealing. However, there may be concerns about the authenticity and originality of AI-generated music. There is a sentiment within the music community that the human touch—characterized by intentional imperfections and unique artistic choices—is what makes music resonate on a personal level.

Despite these artistic concerns, various stakeholders approach AI music generation with different priorities. Listeners, content creators, and small businesses often value the end product—the music itself—over the methods used to create it. For these users, the ability to rapidly generate music without specialized musical knowledge presents significant advantages. Moreover, the acceptance of AI-generated music may ultimately depend on its quality and emotional impact rather than its origin. This creates opportunities for MGS in contexts where demand for new music is high and the creative process less visible, such as gaming environments, film scoring, or background music for various media

Nevertheless, the seemingly limitless possibilities for creating new musical content can paradoxically become overwhelming and counterproductive. As observed in Section \ref{sec:hands_on_experimentation}, adapting these systems to specific musical preferences presents several challenges, potentially constraining the production process through necessary limitations of sound sources and selection of viable samples. During this study's\textit{Curation} process, multiple iterations were required to obtain suitable musical content, and even when appropriate content was identified, recreating similar content to continue compositions often proved impossible. This unpredictability and lack of reproducibility frequently disrupted the creative flow, which led to frustration, extended working hours, and ultimately compromises in final compositions.

This situation suggests that the role of music creators has evolved from being solely producers to becoming arrangers of varied AI-generated music, as noted by \citep{civit_systematic_2022}. Indeed, fot MGS to be truly effective, they must generate new content while recognizing and innovating upon existing bodies of work. Consider Jazz music, where musicians rely on understanding the genre's history and standards as a foundation for improvisation. To produce authentic Jazz, the systems' training data must comprehensively cover diverse Jazz styles and encode techniques of previous masters. Additionally, AI-generated music often lacks the personal narrative and emotional journey integral to the Jazz experience, as well as the conversational interplay between instruments that requires adaptability and responsiveness. This example emphasizes the importance of positioning AI systems as enhancers of, rather than replacements for, human creativity.

In response to these limitations, collaborative approaches between humans and AI have emerged as particularly beneficial. Systems designed as creative partners and assistants to music creators \citep{dadman_toward_2022} can address many of the shortcomings of fully autonomous generation. Open-source models serve as valuable assets in developing such assistive tools. As \citet{langenkamp_how_2022} discuss, these models offer flexible and accessible platforms for innovation by enabling diverse communities to collaborate in creating and enhancing AI tools, thus incorporating broader creative perspectives into development processes. Through this collaborative ecosystem, specialized tools can emerge that aim to balance human and machine creativity while minimizing the limitations associated with autonomous generation. For instance, \citet{dadman2024crafting} proposes a framework based on multi-agent systems (MAS) that enables users to direct and refine the creative process rather than merely accepting AI-generated results. This framework involves multiple collaborative agents, with one serving as an instructor while another functions as a generator or decision-maker. This collaborative interaction can provide a more meaningful and stimulating creative process.

However, current systems often require understanding of programming, machine learning concepts, and parameter settings—potentially creating barriers for music creators who focus primarily on creative aspects rather than technical details. The successful integration of these systems into existing workflows largely depends on their compatibility with established music production software and hardware. Similar to \textit{Magenta DDSP-VST}, \textit{RAVE} by \citet{caillon2021RAVE} exemplifies another effective approach to alleviating these technical barriers. Particularly, through its MAX/MSP integration, RAVE allows creators to incorporate generative features—such as real-time timbre transfer and sound morphing—into existing patches and performance setups without requiring code-based interactions. Nevertheless, despite the interface accessibility and workflow integration, the systems' true artistic utility ultimately depends on proper training or fine-tuning to match creators' specific aesthetic preferences.

The customization process itself introduces additional challenges that can overwhelm non-technical users, as it requires managing large datasets and navigating complex model training aspects, including optimization and performance monitoring. The most important aspect of this process is assembling a dataset that accurately represents the creator's desired aesthetic, which might include their compositions or carefully curated selections. Additionally, training or fine-tuning systems demands substantial computational resources, making it a potentially prohibitive process for individual creators. In this context, intuitive interfaces become important as they democratize access to advanced AI tools by enabling creators with minimal technical expertise to leverage AI in their creative endeavors. Interface development should focus on simplifying model customization and control through clear, accessible controls and presets. Such features make it feasible for producers to employ AI tools with their own datasets—a capability essential for maintaining confidentiality and integrity of personal or proprietary musical content.

The collaborative framework highlighted by \citet{dadman2024crafting} can potentially enhance the transparency of AI operations in music creation. By allowing creators to direct and refine AI outputs, the system provides insights into decision-making processes and how inputs transform into musical elements. This transparency builds trust between creators and AI systems by ensuring creators can understand and predict technological responses to their inputs. Therefore, incorporating these principles into MGS design addresses the dual challenges of accessibility and ethical technology use. For creators concerned with originality and confidentiality, the ability to leverage AI tools without compromising these aspects represents an invaluable advancement in the field of AI-assisted music creation.

\subsection{Challenges Involved in Prompt-based Music Generation Systems}\label{sec:challenges_prompt}

As mentioned in response to RQ1, the effectiveness of systems using textual prompts hinges on users' ability to craft detailed prompts. This challenge is central to the interaction between user input and system output in prompt-based systems, often requiring trial and error in prompt design \citep{dang_how_2022}. The structure and vocabulary choices in prompts, as observed during the \textit{Hands-on Experimentation} phase, significantly influence the musical quality of the model's outputs \citep{oppenlaender_taxonomy_2023}. This dependency underscores several challenges identified by \citet{oppenlaender_taxonomy_2023, dang_how_2022, christodoulou_multimodal_2024, liu_pre-train_2021}. The following discussion will focus on two key challenges: first, the gap between the musical vocabulary and conceptual understanding of those who create the training data versus that of end-users; second, the need for extensive experimentation to determine the most effective prompts for specific models.

Regarding the first challenge, creating training data for these systems demands considerable expertise. Such datasets require annotation by individuals with a good understanding of musical concepts. For instance, in MusicCaps dataset \citep{agostinelli2023musiclm} used to evaluate \textit{MusicGen} \citep{copet2024simple}, audio files are paired with text descriptions written by ten professional musicians. These expert annotations are substantially more detailed than typical user-generated prompts, which tend to be abstract and less specific \citep{chang_open_2024}. Moreover, as \citet{christodoulou_multimodal_2024} notes, the annotation process is inherently subjective and culturally specific, reflecting human interpretations influenced by cultural contexts, individual perceptions, and domain expertise. Consequently, these annotations may not align with users' interpretations and expressions, potentially leading to a misalignment between the user's creative intentions and the model's output\footnote{This issue may affect less experienced users even more than those with extensive musical knowledge.}.

This misalignment is further complicated by the diverse linguistic practices within different music communities, each possessing unique terminologies \citep{bartleet_translating_2018}. For example, Jazz musicians employ vocabularies distinctly different from those of Hip-Hop artists. This reflects the rich histories and social contexts that have shaped these musical traditions. These linguistic nuances result in varied descriptions and interpretations of identical musical examples when presented as textual prompts. According to \citet{bartleet_translating_2018}, disparities in linguistic styles and terminologies fundamentally influence how musical concepts are understood within each community. A clear illustration of this is how 'Improvisation' in Jazz corresponds to 'Freestyling' in Hip-Hop culture. This contrast underlines the necessity for analytical approaches that consider cultural contexts rather than assuming universal frameworks for musical expression.

To address these vocabulary and interpretation challenges, \citet{christodoulou_multimodal_2024} suggests combining crowdsourcing with expert validation as a pathway toward more effective annotations. This hybrid approach utilizes crowdsourcing platforms for initial annotation tasks, followed by expert data curators who validate a subset of the results to maintain quality standards. Such methodology enhances annotation quality over time without incurring prohibitive initial costs \citep{li_learning_2022}. However, it remains essential to specify the background of data curators (experts) for the cultural reasons outlined above. Notably, such information is often absent in training examples, as seen with \textit{MusicGen} \citep{copet2024simple} and the MusicCaps dataset \citep{agostinelli2023musiclm}.

The second major challenge revolves around prompt engineering itself. As \citet{liu_pre-train_2021} explains, while various methods exist for designing effective prompts—including manual template engineering and automated template learning—the fundamental difficulty lies in crafting prompts that accurately capture and reflect the input context. This task requires deep understanding of both the model's capabilities and the nuances of music generation, making it inherently complex and necessitating several rounds of experimentation. The PAGURI study \citep{ronchini_paguri_2024} reinforces this point by demonstrating how users frequently struggle to achieve desired outputs due to discrepancies between their prompts and the model's interpretations. This iterative refinement process can be time-consuming and may not consistently produce satisfactory results, even after multiple attempts—an observation aligned with the \textit{Hands-on Experimentation} phase of this study.

To bridge these gaps between user intent and model interpretation, several promising approaches have emerged. \citet{dang_how_2022} advocates for user interfaces that assist in creating and applying prompts more effectively. They suggest that interactive tools can help users combine multiple prompts to explore various descriptions simultaneously. Such approaches enable rapid iteration and investigation of different prompt variations, as demonstrated in systems like IteraTTA \citep{yakura_iteratta_2023}. Similarly, \citet{chang_open_2024} employed instruction-tuned large language models (LLMs) to transform simple user prompts into more detailed versions. Another promising direction involves multi-agent retrieval-augmented generation (RAG) methods, where collaborative agents work together—one retrieving contextually relevant information while another generates responses based on the retrieved data, analogous to the collaborative approach discussed in Section \ref{sec:collaborative_tools}. Research by \citet{wang_survey_2024} highlights how this approach enhances divergent thinking through iterative refinement of prompts and outputs.

Complementing these technical solutions, this study emphasizes the value of user feedback mechanisms that allow models to learn from each interaction. Through this self-reinforcing cycle of learning and improvement, systems can progressively refine their responses to prompts by enhancing user satisfaction as they perceive tangible improvements in the system's outputs \citep{zeng_let_2023}. Beyond immediate practical benefits, this approach offers deeper insights into human-computer interaction by revealing how systems respond to different types of feedback and how these responses can guide the development of increasingly effective systems.

\subsection{Artists' Experience and Technical Considerations}\label{subsec:artists}

As mentioned in Section \ref{sec:evaluation_framework}, this study may be subject to certain limitations. Nonetheless, its findings and implications align with the experiences of professional music producers who have incorporated AI into their creative workflows. These real-world applications, as reported through various experiments and interviews with producers, further support the implications of this study.

For instance, Taryn Southern, during her project to produce an album entirely with AI, emphasized the necessity of retaining artistic control throughout the creative process. She stated, `It is important for me, as an artist, to be involved in every step of the creation' \citep{taryn_interviews_2024}. Similarly, Damien Roach's interaction with the \textit{Riffusion} highlighted the challenges of filtering through a vast amount of AI-generated content to find usable elements. He noted the dual nature of the outputs—both familiar and strange—which required careful selection and direction to align with his artistic vision \citep{mullen_2023}.

Moreover, the technical capabilities and aesthetic applications of various MGS like \textit{Riffusion} and \textit{Magenta} also play an important role in their adoption by music producers. For instance, some producers, including Damien, have expressed interest in the low-bitrate sound quality produced by \textit{Riffusion}, considering it an aesthetic rather than a limitation \citep{mullen_2023}. This perspective highlights the subjective nature of music production, where the perceived imperfections of AI-generated sounds can be recontextualized as desirable qualities within the creative process. However, limitations exist, as noted by Taryn, who pointed out that while tools like Amper excel at composing and producing instrumentation, they struggle with understanding complex song structures \citep{taryn_interviews_2024}. Similarly, Damien notes that while AI can generate vast amounts of content, the quality and relevance of the output can vary significantly, which necessitates a discerning (an artistic vision) and time-consuming process to identify valuable musical elements \citep{mullen_2023}.

Additionally, the effective use of AI-generated music and its application as a music technology tool depends on how well it is incorporated into the creative workflows of music producers. The practical experiences of producers such as Max Cooper demonstrate that AI models are most beneficial when they enhance rather than replace human creativity \citep{wright_2023}. Cooper's utilization of AI to propose variations and improve musical ideas based on his previous work shows the potential for AI as a dynamic assistant in music production. This method harnesses AI's computational power to enhance creativity rather than solely producing content. Furthermore, ethical considerations regarding AI in music, particularly transparency about the origins of AI-generated content, are essential for its acceptance and usability in the industry. 

Examining these insights collectively reveals that the true value of AI in music production lies in its capacity to seamlessly integrate with established human-driven creative processes by supporting rather than supplanting the creator's artistic vision.

\subsection{Final Thoughts}

This study exhibited that MGS has potential as a partner in a co-creative process rather than merely as a tool for automating tasks. When viewed through the lens of co-creation, these systems offer a unique opportunity to blend human creativity with computational power.

However, one of the critical considerations here is the balance between exploitation and exploration. Systems trained extensively on historical data (training examples) tend to exploit known patterns, styles, and structures. This can inadvertently lead to a homogenization of output that dilute the creative 'genetic material'\footnote{The 'genetic material' of music refers to the elements that define its cultural heritage and individual creativity that have evolved over time. These elements contribute to the richness and variety that characterize different musical traditions and innovations.} that makes music culturally and emotionally rich and diverse. While this data provides a foundation for understanding and learning musical conventions, it can also constrain the system's innovation ability if not paired with exploratory capabilities. This situation is similar to over-fitting in machine learning, where the model performs well on training data but struggles to generalize to new, unseen data.

To counterbalance this tendency, MGS should incorporate mechanisms that encourage exploration, resulting in unexplored, diverse and sometimes unexpected musical outputs. This exploration facilitates sustaining the creative aspect of the music creation. Indeed, these systems can stimulate divergent thinking by presenting music creators with unexpected interpretations or transformations of musical ideas. Several researchers, including \citet{doshi_generative_2024, hou_double-edged_2024, wadinambiarachchi_effects_2024, kumar_human_2024}, have noted similar perspectives. They view such systems as powerful catalysts for divergent thinking to unlock creative potential across various disciplines. In this paradigm, computational systems do not replace human creators but rather enhance and expand their creative capabilities by challenging conventional thinking patterns.

In this regard, as discussed earlier in response to the research questions, the concept of collaborative MAS framework illustrates how different AI agents can collaborate towards a common creative goal. This approach can mitigate some risks associated with over-reliance on historical data by ensuring the creative process benefits from various influences and inspirations. It also aligns with the idea of AI as a co-creator by actively participating in the creative process rather than merely executing predefined tasks. Ultimately, this draws similarities to the concept of musicking by \citet{small_musicking_1998}. Musicking, as defined by Small, is the act of engaging with music in any capacity, whether as a performer, listener, or creator. The MAS-based approach embodies this concept by encouraging a dynamic and interactive environment where AI agents and music practitioners engage in a collaborative process. This interaction mirrors the participatory essence of musicking, where the focus is on the experience and the relationships formed through the act of making music rather than solely on the final product. Furthermore, this approach also aligns with the expectations of creators as presented in Section \ref{subsec:artists}, where they anticipate AI systems to serve as collaborative partners that enhance their creative processes while respecting their artistic vision and autonomy.

\section{Future Directions}\label{sec:future_directions}

The proposed evaluation framework represents an initial exploratory study designed to enhance understanding of MGS and their integration into music creators' workflows. It acknowledges the diverse perspectives inherent in MGS assessment while adopting a mixed research approach as a pragmatic methodological stance. This methodology allows the evaluator to engage directly with the systems, documenting observations through qualitative notes while systematically applying quantitative metrics based on predefined criteria. Specifically, this study employs a single-evaluator approach to balance evaluative rigor with practical considerations, as detailed in Sections \ref{sec:background} and \ref{sec:evaluation_framework}. While this inherently limits the generalizability of findings due to reliance on a single perspective, it provides a focused lens to investigate both the practicality of the evaluation framework itself and the integration potential of MGS in production workflows.

The evaluation criteria developed for this study (Section \ref{sec:criteria_definition}) encompass system-level features and attributes, alongside the practicality and creative affordances of the systems. A central objective was to initiate dialogue regarding the evolving expectations and integration capacity of these systems within music creators' workflows. To accomplish this, we deliberately selected only open-source alternatives for evaluation. This choice allowed us to conduct and maintain our investigation with consistent depth and breadth, refining our approach throughout the review process without concerns about sudden changes or updates to the systems—a common challenge with proprietary alternatives. The framework does not aim to provide rigid, fixed evaluation criteria; rather, it demonstrates an approach that can serve as a foundation adaptable to different scenarios and case studies, maintaining relevance as technologies evolve. The necessity for such flexibility and adaptability is emphasized by several researchers \citep{young2020HCI, agres2016Evaluation, el-shimy_user-driven_2016}.

Through the combination of qualitative assessments and quantitative scoring, the single-evaluator approach functioned as an effective mechanism to examine both the utility and relevance of the established criteria by identifying specific areas for improvement. The integration of quantitative metrics with qualitative notes enhanced the depth of feedback we could elicit, as presented throughout this study and supported by \citet{el-shimy_user-driven_2016}. This methodological approach offers particular benefits when extended to studies involving multiple participants, which we intend to pursue in future research. Furthermore, the qualitative observations are guided by the specific research questions outlined in Section \ref{sec:research_questions}. These questions, which can be open-ended as noted by \citep{agres2016Evaluation}, draw upon the defined criteria and considerations to provide a structured yet flexible evaluative framework.

The criteria considerations and scoring levels, though carefully developed through iterative refinement, remain preliminary in nature. Their inherently subjective character—particularly when assessing abstract concepts like 'Creative Workflow'—poses challenges for consistency across different user groups. These dimensions vary markedly among users, with individuals possessing different levels of technical expertise or creative priorities likely to interpret and apply such criteria quite differently \citep{agres2016Evaluation, eigenfeldt2012Evaluation}. To mitigate the challenges posed by subjectivity, adaptive scoring mechanisms can be implemented to align evaluative criteria with distinct user contexts by accounting for case-specific variables. Future iterations of this framework should validate and refine scoring systems using expert panels or longitudinal studies \citep{young2020HCI}. As \citet{eigenfeldt2012Evaluation} suggest, participatory methods involving diverse demographics can effectively recalibrate subjective scoring models. Such iterative methodological approaches are essential, as \citet{agres2016Evaluation} emphasize, for ensuring the empirical robustness of creative process evaluation across diverse settings.

The framework's current emphasis on professional production standards for audio quality represents another area requiring refinement. This emphasis implicitly presumes specific aesthetic norms that may not align with all genres or creative objectives. For instance, experimental electronic genres might intentionally embrace artifacts or unconventional sound processing as valid artistic expressions rather than technical shortcomings, as highlighted in Section \ref{subsec:artists}. Research by \citet{eigenfeldt2012Evaluation} emphasizes the importance of accommodating diverse aesthetic traditions to ensure equitable evaluation. Expanding the framework to encompass such aesthetic diversity would enhance its inclusivity across various cultural and creative contexts.

Additionally, our evaluation revealed several important dimensions—'Serendipity Support,' 'AI Assistance Balance,' and 'Adaptation Capacity'—that warrant consideration as distinct evaluative criteria. Currently, these aspects are assessed indirectly within broader criteria such as 'Creative Workflow' and 'Content Generation Control.' The evaluator's notes taken during experimentation consistently highlighted these elements as factors in system usability and creative potential. Establishing these as separate criteria would enable more precise assessment of the systems' capabilities and creative affordances through quantitative metrics, while also providing structured opportunities to document participants' cognitive, perceptual, and affective responses through qualitative observations.

\section{Conclusion}\label{sec:conclusion}

This study acknowledges that MGS operate within complex sociotechnical ecosystems where technical capabilities, interface design, and creative workflows interact in complex ways. The interconnected nature of evaluation dimensions—where improvements in one area might create unexpected constraints in another—necessitates adaptive methodologies that can evolve alongside the systems they assess. Rather than presenting a definitive solution, our exploratory mixed research framework serves as a foundation for broader discourse on how MGS can meet diverse creative expectations.

Our findings reveal that MGS function primarily as complementary tools in music creation, enhancing rather than replacing human expertise. While these systems demonstrate considerable potential, they exhibit notable limitations in maintaining thematic and structural coherence throughout compositions. This emphasizes the indispensable role of human creativity in tasks demanding emotional depth and complex decision-making.

Their true value may lie in their imperfections: by generating outputs that are \textit{almost} coherent, \textit{nearly} thematic, they create a creative tension that compels artists to interrogate their own assumptions about originality, authorship, and aesthetic value. By revealing limitations in thematic coherence, MGS highlight what makes human creativity distinct: the capacity to weave fragmented ideas into narratives charged with cultural and emotional significance. This positions MGS not as competitors to human composers but as provocations—tools that force creators to articulate and defend their aesthetic choices with renewed rigor.

In this regard, the proposed evaluation framework does more than assessing systems; it maps the contours of a new creative literacy. As we observed, when users engage with systems that excel at generating variations but falter at curation, they develop hybrid skills—interpreting algorithmic outputs through the lens of their own intentionality, transforming stochastic suggestions into deliberate artistic statements. This mirrors a broader cultural shift where human expertise evolves from direct execution to strategic mediation. The observations presented by \citet{huang_ai_2020} exemplify this dynamic, where artists shape the tools themselves, turning technical limitations into sites of creative negotiation. Thereby, what emerges is not a hierarchical human-AI relationship but an ecosystem of mutual adaptation—a concept that can be reinforced by the \textit{Adaptation Capacity} metric, which aims to quantify a system's responsiveness to artistic reinvention.

In this light, the successful integration of MGS in music creation workflows hinges on careful considerations of practical and creative affordances. These elements enable music creators to preserve their unique artistic voices while leveraging the strengths of MGS. As these systems become more deeply embedded in creative processes, they should be viewed as a collaborative asset that enrich the music creation experience.

Finally, throughout the study, we have identified key components for progressing toward better integration frameworks for music generation systems. The limitations noted in our current approach provide several paths for future research. First, there is a need to extend this methodology to include proprietary systems widely adopted by music creators. We plan to conduct such studies with participants sharing comparable musical backgrounds to ensure evaluation consistency. Additionally, future work should consider adaptive scoring mechanisms that accommodate the contextual variability inherent in music production processes. Finally, we propose shifting from compatibility-based assessment toward evaluating systems' adaptive integration capacity—measuring how effectively these tools can evolve alongside creative practices rather than merely assessing their alignment with current standards.

\bibliographystyle{unsrtnat}
\bibliography{references}

\begin{thebibliography}{139}
\providecommand{\natexlab}[1]{#1}
\providecommand{\url}[1]{\texttt{#1}}
\expandafter\ifx\csname urlstyle\endcsname\relax
  \providecommand{\doi}[1]{doi: #1}\else
  \providecommand{\doi}{doi: \begingroup \urlstyle{rm}\Url}\fi

\bibitem[Civit et~al.(2022)Civit, Civit-Masot, Cuadrado, and Escalona]{civit_systematic_2022}
Miguel Civit, Javier Civit-Masot, Francisco Cuadrado, and Maria~J. Escalona.
\newblock A systematic review of artificial intelligence-based music generation: {Scope}, applications, and future trends.
\newblock \emph{Expert Systems with Applications}, 209:\penalty0 118190, December 2022.
\newblock ISSN 09574174.
\newblock \doi{10.1016/j.eswa.2022.118190}.
\newblock URL \url{https://linkinghub.elsevier.com/retrieve/pii/S0957417422013537}.

\bibitem[Herremans et~al.(2018)Herremans, Chuan, and Chew]{herremans_functional_2018}
Dorien Herremans, Ching-Hua Chuan, and Elaine Chew.
\newblock A {Functional} {Taxonomy} of {Music} {Generation} {Systems}.
\newblock \emph{ACM Computing Surveys}, 50\penalty0 (5):\penalty0 1--30, September 2018.
\newblock ISSN 0360-0300, 1557-7341.
\newblock \doi{10.1145/3108242}.
\newblock URL \url{https://dl.acm.org/doi/10.1145/3108242}.

\bibitem[Tatar and Pasquier(2019)]{tatar_musical_2019}
Kıvanç Tatar and Philippe Pasquier.
\newblock Musical agents: {A} typology and state of the art towards {Musical} {Metacreation}.
\newblock \emph{Journal of New Music Research}, 48\penalty0 (1):\penalty0 56--105, January 2019.
\newblock ISSN 0929-8215, 1744-5027.
\newblock \doi{10.1080/09298215.2018.1511736}.
\newblock URL \url{https://www.tandfonline.com/doi/full/10.1080/09298215.2018.1511736}.

\bibitem[Wang et~al.(2024)Wang, Zhao, Liu, Pang, Qin, and Wu]{wang_review_2024}
Lei Wang, Ziyi Zhao, Hanwei Liu, Junwei Pang, Yi~Qin, and Qidi Wu.
\newblock A review of intelligent music generation systems.
\newblock \emph{Neural Computing and Applications}, 36\penalty0 (12):\penalty0 6381--6401, April 2024.
\newblock ISSN 0941-0643, 1433-3058.
\newblock \doi{10.1007/s00521-024-09418-2}.
\newblock URL \url{https://link.springer.com/10.1007/s00521-024-09418-2}.

\bibitem[Moysis et~al.(2023)Moysis, Iliadis, Sotiroudis, Boursianis, Papadopoulou, Kokkinidis, Volos, Sarigiannidis, Nikolaidis, and Goudos]{moysis_music_2023}
Lazaros Moysis, Lazaros~Alexios Iliadis, Sotirios~P. Sotiroudis, Achilles~D. Boursianis, Maria~S. Papadopoulou, Konstantinos-Iraklis~D. Kokkinidis, Christos Volos, Panagiotis Sarigiannidis, Spiridon Nikolaidis, and Sotirios~K. Goudos.
\newblock Music {Deep} {Learning}: {Deep} {Learning} {Methods} for {Music} {Signal} {Processing}—{A} {Review} of the {State}-of-the-{Art}.
\newblock \emph{IEEE Access}, 11:\penalty0 17031--17052, 2023.
\newblock ISSN 2169-3536.
\newblock \doi{10.1109/ACCESS.2023.3244620}.
\newblock URL \url{https://ieeexplore.ieee.org/document/10043650/}.

\bibitem[Zhu et~al.(2023)Zhu, Baca, Rekabdar, and Rawassizadeh]{zhu_survey_2023}
Yueyue Zhu, Jared Baca, Banafsheh Rekabdar, and Reza Rawassizadeh.
\newblock A {Survey} of {AI} {Music} {Generation} {Tools} and {Models}, 2023.
\newblock URL \url{https://arxiv.org/abs/2308.12982}.

\bibitem[Ji et~al.(2023)Ji, Yang, and Luo]{ji_survey_2023}
Shulei Ji, Xinyu Yang, and Jing Luo.
\newblock A {Survey} on {Deep} {Learning} for {Symbolic} {Music} {Generation}: {Representations}, {Algorithms}, {Evaluations}, and {Challenges}.
\newblock \emph{ACM Computing Surveys}, 56\penalty0 (1):\penalty0 7:1--7:39, August 2023.
\newblock ISSN 0360-0300.
\newblock \doi{10.1145/3597493}.
\newblock URL \url{https://doi.org/10.1145/3597493}.

\bibitem[Dadman et~al.(2022)Dadman, Bremdal, Bang, and Dalmo]{dadman_toward_2022}
Shayan Dadman, Bernt~Arild Bremdal, Børre Bang, and Rune Dalmo.
\newblock Toward {Interactive} {Music} {Generation}: {A} {Position} {Paper}.
\newblock \emph{IEEE Access}, 10:\penalty0 125679--125695, 2022.
\newblock ISSN 2169-3536.
\newblock \doi{10.1109/ACCESS.2022.3225689}.
\newblock URL \url{https://ieeexplore.ieee.org/abstract/document/9966445}.

\bibitem[Briot(2021)]{briot_artificial_2021}
Jean-Pierre Briot.
\newblock From artificial neural networks to deep learning for music generation: history, concepts and trends.
\newblock \emph{Neural Computing and Applications}, 33\penalty0 (1):\penalty0 39--65, January 2021.
\newblock ISSN 1433-3058.
\newblock \doi{10.1007/s00521-020-05399-0}.
\newblock URL \url{https://doi.org/10.1007/s00521-020-05399-0}.

\bibitem[Briot et~al.(2020)Briot, Hadjeres, and Pachet]{briot_deep_2020}
Jean-Pierre Briot, Gaëtan Hadjeres, and François-David Pachet.
\newblock \emph{Deep {Learning} {Techniques} for {Music} {Generation}}.
\newblock Computational {Synthesis} and {Creative} {Systems}. Springer International Publishing, Cham, 2020.
\newblock ISBN 9783319701622 9783319701639.
\newblock \doi{10.1007/978-3-319-70163-9}.
\newblock URL \url{http://link.springer.com/10.1007/978-3-319-70163-9}.

\bibitem[Carnovalini and Rodà(2020)]{carnovalini_computational_2020}
Filippo Carnovalini and Antonio Rodà.
\newblock Computational {Creativity} and {Music} {Generation} {Systems}: {An} {Introduction} to the {State} of the {Art}.
\newblock \emph{Frontiers in Artificial Intelligence}, 3:\penalty0 14, April 2020.
\newblock ISSN 2624-8212.
\newblock \doi{10.3389/frai.2020.00014}.
\newblock URL \url{https://www.frontiersin.org/article/10.3389/frai.2020.00014/full}.

\bibitem[Kaliakatsos-Papakostas et~al.(2020)Kaliakatsos-Papakostas, Floros, and Vrahatis]{papakostas_artificial_2020}
Maximos Kaliakatsos-Papakostas, Andreas Floros, and Michael~N. Vrahatis.
\newblock Artificial intelligence methods for music generation: a review and future perspectives.
\newblock In \emph{Nature-{Inspired} {Computation} and {Swarm} {Intelligence}}, pages 217--245. Elsevier, 2020.
\newblock ISBN 9780128197141.
\newblock \doi{10.1016/B978-0-12-819714-1.00024-5}.
\newblock URL \url{https://linkinghub.elsevier.com/retrieve/pii/B9780128197141000245}.

\bibitem[Lopez-Rincon et~al.(2018)Lopez-Rincon, Starostenko, and Martín]{lopez-rincon_algoritmic_2018}
Omar Lopez-Rincon, Oleg Starostenko, and Gerardo Ayala-San Martín.
\newblock Algoritmic music composition based on artificial intelligence: {A} survey.
\newblock In \emph{2018 {International} {Conference} on {Electronics}, {Communications} and {Computers} ({CONIELECOMP})}, pages 187--193, February 2018.
\newblock \doi{10.1109/CONIELECOMP.2018.8327197}.
\newblock URL \url{https://ieeexplore.ieee.org/abstract/document/8327197/}.
\newblock ISSN: 2474-9044.

\bibitem[Liu and Ting(2017)]{liu_computational_2017}
Chien-Hung Liu and Chuan-Kang Ting.
\newblock Computational {Intelligence} in {Music} {Composition}: {A} {Survey}.
\newblock \emph{IEEE Transactions on Emerging Topics in Computational Intelligence}, 1\penalty0 (1):\penalty0 2--15, February 2017.
\newblock ISSN 2471-285X.
\newblock \doi{10.1109/TETCI.2016.2642200}.
\newblock URL \url{https://ieeexplore.ieee.org/abstract/document/7792228/}.

\bibitem[Williams et~al.(2015)Williams, Kirke, Miranda, Roesch, Daly, and Nasuto]{williams_investigating_2015}
Duncan Williams, Alexis Kirke, Eduardo~R Miranda, Etienne Roesch, Ian Daly, and Slawomir Nasuto.
\newblock Investigating affect in algorithmic composition systems.
\newblock \emph{Psychology of Music}, 43\penalty0 (6):\penalty0 831--854, November 2015.
\newblock ISSN 0305-7356, 1741-3087.
\newblock \doi{10.1177/0305735614543282}.
\newblock URL \url{http://journals.sagepub.com/doi/10.1177/0305735614543282}.

\bibitem[Fernandez and Vico(2013)]{fernandez_ai_2013}
J.~D. Fernandez and F.~Vico.
\newblock {AI} {Methods} in {Algorithmic} {Composition}: {A} {Comprehensive} {Survey}.
\newblock \emph{Journal of Artificial Intelligence Research}, 48:\penalty0 513--582, November 2013.
\newblock ISSN 1076-9757.
\newblock \doi{10.1613/jair.3908}.
\newblock URL \url{https://www.jair.org/index.php/jair/article/view/10845}.

\bibitem[Kirke and Miranda(2013)]{kirke_overview_2013}
Alexis Kirke and Eduardo~R. Miranda.
\newblock An {Overview} of {Computer} {Systems} for {Expressive} {Music} {Performance}.
\newblock In Alexis Kirke and Eduardo~R. Miranda, editors, \emph{Guide to {Computing} for {Expressive} {Music} {Performance}}, pages 1--47. Springer, London, 2013.
\newblock ISBN 9781447141235.
\newblock \doi{10.1007/978-1-4471-4123-5_1}.
\newblock URL \url{https://doi.org/10.1007/978-1-4471-4123-5_1}.

\bibitem[Nierhaus(2009)]{nierhaus_algorithmic_2009}
Gerhard Nierhaus.
\newblock \emph{Algorithmic {Composition}}.
\newblock Springer Vienna, Vienna, 2009.
\newblock ISBN 9783211755396 9783211755402.
\newblock \doi{10.1007/978-3-211-75540-2}.
\newblock URL \url{http://link.springer.com/10.1007/978-3-211-75540-2}.

\bibitem[Widmer and Goebl(2004)]{widmer_computational_2004}
Gerhard Widmer and Werner Goebl.
\newblock Computational {Models} of {Expressive} {Music} {Performance}: {The} {State} of the {Art}.
\newblock \emph{Journal of New Music Research}, 33\penalty0 (3):\penalty0 203--216, September 2004.
\newblock ISSN 0929-8215, 1744-5027.
\newblock \doi{10.1080/0929821042000317804}.
\newblock URL \url{http://www.tandfonline.com/doi/abs/10.1080/0929821042000317804}.

\bibitem[Papadopoulos and Wiggins(1999)]{papadopoulos_ai_1999}
George Papadopoulos and Geraint~A. Wiggins.
\newblock Ai methods for algorithmic composition: A survey, a critical view and future prospects.
\newblock 1999.
\newblock URL \url{https://api.semanticscholar.org/CorpusID:5055535}.

\bibitem[Compton and Mteas()]{compton_casual_2015}
Kate Compton and Michael Mteas.
\newblock Casual creators.
\newblock In \emph{International Conference on Innovative Computing and Cloud Computing}.
\newblock URL \url{https://api.semanticscholar.org/CorpusID:1305832}.

\bibitem[Bown(2025)]{bown_genies_2025}
Oliver Bown.
\newblock From genies performing magic to sages imparting wisdom: a value-centred survey of music {AI} user interfaces, creative affordances and artist objectives.
\newblock \emph{Journal of New Music Research}, pages 1--14, January 2025.
\newblock ISSN 0929-8215, 1744-5027.
\newblock \doi{10.1080/09298215.2024.2442360}.
\newblock URL \url{https://www.tandfonline.com/doi/full/10.1080/09298215.2024.2442360}.
\newblock Publisher: Informa UK Limited.

\bibitem[{Intellectual Property Helpdesk}(2023)]{universal2023}
{Intellectual Property Helpdesk}.
\newblock Universal music sues ai company anthropic for copyright infringement - levi's sues coperni for trade mark infringement.
\newblock 2023.
\newblock URL \url{https://intellectual-property-helpdesk.ec.europa.eu/news-events/news/universal-music-sues-ai-company-anthropic-copyright-infringement-levis-sues-coperni-trade-mark-2023-10-26_en}.
\newblock Accessed: 2024-11-07.

\bibitem[{Wired}(2023)]{wired2023}
{Wired}.
\newblock Us record labels sue ai music generators suno and udio for copyright infringement.
\newblock 2023.
\newblock URL \url{https://www.wired.com/story/ai-music-generators-suno-and-udio-sued-for-copyright-infringement/}.
\newblock Accessed: 2024-11-07.

\bibitem[{Music Business Worldwide}(2023)]{musicbusiness2023}
{Music Business Worldwide}.
\newblock As suno and udio admit training ai with unlicensed music, record industry says: ‘there’s nothing fair about stealing an artist’s life’s work.’.
\newblock 2023.
\newblock URL \url{https://www.musicbusinessworldwide.com/as-suno-and-udio-admit-training-ai-with-unlicensed-music-record-industry-says-theres-nothing-fair-about-stealing-an-artists-lifes-work/}.
\newblock Accessed: 2024-11-07.

\bibitem[Ma et~al.(2024)Ma, Øland, Ragni, Sette, Saitis, Donahue, Lin, Plachouras, Benetos, Shatri, Morreale, Zhang, Fazekas, Xia, Zhang, Manco, Huang, Guinot, Lin, Marinelli, Lam, Sharma, Kong, Dannenberg, Yuan, Wu, Wu, Dai, Lei, Kang, Dixon, Chen, Huang, Du, Qu, Tan, Li, Tian, Wu, Wu, Ma, and Wang]{ma_foundation_2024}
Yinghao Ma, Anders Øland, Anton Ragni, Bleiz MacSen~Del Sette, Charalampos Saitis, Chris Donahue, Chenghua Lin, Christos Plachouras, Emmanouil Benetos, Elona Shatri, Fabio Morreale, Ge~Zhang, György Fazekas, Gus Xia, Huan Zhang, Ilaria Manco, Jiawen Huang, Julien Guinot, Liwei Lin, Luca Marinelli, Max W.~Y. Lam, Megha Sharma, Qiuqiang Kong, Roger~B. Dannenberg, Ruibin Yuan, Shangda Wu, Shih-Lun Wu, Shuqi Dai, Shun Lei, Shiyin Kang, Simon Dixon, Wenhu Chen, Wenhao Huang, Xingjian Du, Xingwei Qu, Xu~Tan, Yizhi Li, Zeyue Tian, Zhiyong Wu, Zhizheng Wu, Ziyang Ma, and Ziyu Wang.
\newblock Foundation {Models} for {Music}: {A} {Survey}, September 2024.
\newblock URL \url{http://arxiv.org/abs/2408.14340}.
\newblock arXiv:2408.14340 [cs].

\bibitem[T.~Zirpoli()]{zirpoli_2023_generative}
Christopher T.~Zirpoli.
\newblock Generative artificial intelligence and copyright law.
\newblock URL \url{https://crsreports.congress.gov/product/pdf/LSB/LSB10922}.

\bibitem[Seger et~al.(2023)Seger, Dreksler, Moulange, Dardaman, Schuett, Wei, Winter, Arnold, {\'O}~h{\'E}igeartaigh, Korinek, et~al.]{seger2023open}
Elizabeth Seger, Noemi Dreksler, Richard Moulange, Emily Dardaman, Jonas Schuett, K~Wei, Christoph Winter, Mackenzie Arnold, Se{\'a}n {\'O}~h{\'E}igeartaigh, Anton Korinek, et~al.
\newblock Open-sourcing highly capable foundation models.
\newblock \emph{Research paper, Centre for the Governance of AI}, 2023.

\bibitem[Barnett(2023)]{barnett_ethical_2023}
Julia Barnett.
\newblock The {Ethical} {Implications} of {Generative} {Audio} {Models}: {A} {Systematic} {Literature} {Review}.
\newblock In \emph{Proceedings of the 2023 {AAAI}/{ACM} {Conference} on {AI}, {Ethics}, and {Society}}, pages 146--161, Montr{\textbackslash}'\{e\}al QC Canada, August 2023. ACM.
\newblock ISBN 9798400702310.
\newblock \doi{10.1145/3600211.3604686}.
\newblock URL \url{https://dl.acm.org/doi/10.1145/3600211.3604686}.

\bibitem[Morreale(2021)]{morreale_where_2021}
Fabio Morreale.
\newblock Where {Does} the {Buck} {Stop}? {Ethical} and {Political} {Issues} with {AI} in {Music} {Creation}.
\newblock \emph{Transactions of the International Society for Music Information Retrieval}, 4\penalty0 (1):\penalty0 105--113, July 2021.
\newblock ISSN 2514-3298.
\newblock \doi{10.5334/tismir.86}.
\newblock URL \url{http://transactions.ismir.net/articles/10.5334/tismir.86/}.

\bibitem[Auvinen(2019)]{auvinen_music_2019}
Tuomas Auvinen.
\newblock The {Music} producer as creative agent : studio production, technology and cultural space in the work of three {Finnish} producers.
\newblock \emph{Annales Universitatis Turkuensis. Turku: University of Turku}, January 2019.
\newblock URL \url{https://www.utupub.fi/handle/10024/146576}.

\bibitem[Zak(2001)]{zak_poetics_2001}
Albin Zak.
\newblock \emph{The {Poetics} of {Rock}: {Cutting} {Tracks}, {Making} {Records}}.
\newblock University of California Press, November 2001.
\newblock ISBN 9780520232242.
\newblock URL \url{http://www.jstor.org/stable/10.1525/j.ctt1ppbkt}.
\newblock Google-Books-ID: 5bAwDwAAQBAJ.

\bibitem[Frith(1996)]{frith_performing_1996}
Simon Frith.
\newblock \emph{Performing {Rites}: {On} the {Value} of {Popular} {Music}}.
\newblock Harvard University Press, 1996.
\newblock ISBN 9780674661967.
\newblock URL \url{https://books.google.no/books?id=BPdIfT6scIoC}.
\newblock Google-Books-ID: BPdIfT6scIoC.

\bibitem[Burgess(2014)]{burgess_history_2014}
Richard~James Burgess.
\newblock \emph{The {History} of {Music} {Production}}.
\newblock Oxford University Press, 2014.
\newblock ISBN 9780199357161.
\newblock URL \url{https://books.google.no/books?id=qMKiAwAAQBAJ}.
\newblock Google-Books-ID: ZeISDAAAQBAJ.

\bibitem[Holmes(2020)]{holmes_electronic_2020}
Thom Holmes.
\newblock \emph{Electronic and {Experimental} {Music}: {Technology}, {Music}, and {Culture}}.
\newblock Routledge, 6 edition, March 2020.
\newblock ISBN 9780429425585.
\newblock \doi{10.4324/9780429425585}.
\newblock URL \url{https://www.taylorfrancis.com/books/9780429758447}.

\bibitem[Moffat and Sandler(2019)]{moffat_approaches_2019}
David Moffat and Mark~B. Sandler.
\newblock Approaches in {Intelligent} {Music} {Production}.
\newblock \emph{Arts}, 8\penalty0 (4):\penalty0 125, December 2019.
\newblock ISSN 2076-0752.
\newblock \doi{10.3390/arts8040125}.
\newblock URL \url{https://www.mdpi.com/2076-0752/8/4/125}.

\bibitem[Hennion(1989)]{hennion_intermediary_1989}
Antoine Hennion.
\newblock An {Intermediary} {Between} {Production} and {Consumption}: {The} {Producer} of {Popular} {Music}.
\newblock \emph{Science, Technology, \& Human Values}, 14\penalty0 (4):\penalty0 400--424, October 1989.
\newblock ISSN 0162-2439, 1552-8251.
\newblock \doi{10.1177/016224398901400405}.
\newblock URL \url{http://journals.sagepub.com/doi/10.1177/016224398901400405}.

\bibitem[Burgess(2013)]{burgess_art_2013}
Richard~James Burgess.
\newblock \emph{The {Art} of {Music} {Production}: {The} {Theory} and {Practice}}.
\newblock Oxford University Press, September 2013.
\newblock ISBN 9780199359325.
\newblock URL \url{https://books.google.no/books?id=m4dNEAAAQBAJ}.
\newblock Google-Books-ID: lWEUAAAAQBAJ.

\bibitem[Yang and Lerch(2020)]{yang_evaluation_2020}
Li-Chia Yang and Alexander Lerch.
\newblock On the evaluation of generative models in music.
\newblock \emph{Neural Computing and Applications}, 32\penalty0 (9):\penalty0 4773--4784, May 2020.
\newblock ISSN 1433-3058.
\newblock \doi{10.1007/s00521-018-3849-7}.
\newblock URL \url{https://doi.org/10.1007/s00521-018-3849-7}.

\bibitem[Copet et~al.(2023)Copet, Kreuk, Gat, Remez, Kant, Synnaeve, Adi, and Défossez]{copet2024simple}
Jade Copet, Felix Kreuk, Itai Gat, Tal Remez, David Kant, Gabriel Synnaeve, Yossi Adi, and Alexandre Défossez.
\newblock Simple and {Controllable} {Music} {Generation}, 2023.
\newblock URL \url{https://arxiv.org/abs/2306.05284}.

\bibitem[Liu et~al.(2024{\natexlab{a}})Liu, Hussain, Sun, and Shan]{liu2024m2ugen}
Shansong Liu, Atin~Sakkeer Hussain, Chenshuo Sun, and Ying Shan.
\newblock M$^{2}$ugen: Multi-modal music understanding and generation with the power of large language models, 2024{\natexlab{a}}.

\bibitem[Forsgren and Martiros(2022)]{Forsgren_Martiros_2022}
Seth* Forsgren and Hayk* Martiros.
\newblock {Riffusion - Stable diffusion for real-time music generation}, 2022.
\newblock URL \url{https://riffusion.com}.

\bibitem[{Google Magenta Team}(2024)]{magenta}
{Google Magenta Team}.
\newblock Magenta: Music and art generation with machine intelligence, 2024.
\newblock URL \url{https://magenta.tensorflow.org/}.

\bibitem[Pasini and Schlüter(2022)]{pasini2022musika}
Marco Pasini and Jan Schlüter.
\newblock Musika! {Fast} {Infinite} {Waveform} {Music} {Generation}, August 2022.
\newblock URL \url{http://arxiv.org/abs/2208.08706}.
\newblock arXiv:2208.08706 [cs, eess].

\bibitem[Lu et~al.(2023)Lu, Xu, Kang, Yu, Xing, Tan, and Bian]{lu2023musecoco}
Peiling Lu, Xin Xu, Chenfei Kang, Botao Yu, Chengyi Xing, Xu~Tan, and Jiang Bian.
\newblock {MuseCoco}: {Generating} {Symbolic} {Music} from {Text}, May 2023.
\newblock URL \url{http://arxiv.org/abs/2306.00110}.
\newblock arXiv:2306.00110 [cs, eess].

\bibitem[Yu et~al.()Yu, Lu, Wang, Hu, Tan, Ye, Zhang, Qin, and Liu]{yu2022museformer}
Botao Yu, Peiling Lu, Rui Wang, Wei Hu, Xu~Tan, Wei Ye, Shikun Zhang, Tao Qin, and Tie-Yan Liu.
\newblock Museformer: {{Transformer}} with {{Fine-}} and {{Coarse-Grained Attention}} for {{Music Generation}}.
\newblock URL \url{http://arxiv.org/abs/2210.10349}.

\bibitem[Roads()]{roads2023Computer}
Curtis Roads.
\newblock \emph{The Computer Music Tutorial}.
\newblock The MIT Press, second edition edition.
\newblock ISBN 978-0-262-04491-2.

\bibitem[Senior()]{senior2019Mixing}
Mike Senior.
\newblock \emph{Mixing Secrets for the Small Studio}.
\newblock Sound on {{Sound}} Presents. Routledge/Taylor \& Francis Group, second edition edition.
\newblock ISBN 978-1-315-15001-7 978-1-351-36880-3 978-1-351-36879-7.

\bibitem[Snoman()]{snoman2014Dance}
Rick Snoman.
\newblock \emph{Dance Music Manual: Tools, Toys, and Techniques}.
\newblock Focal Press, third edition edition.
\newblock ISBN 978-0-415-82564-1.

\bibitem[Gui et~al.()Gui, Gamper, Braun, and Emmanouilidou]{gui2024Adapting}
Azalea Gui, Hannes Gamper, Sebastian Braun, and Dimitra Emmanouilidou.
\newblock Adapting {{Frechet Audio Distance}} for {{Generative Music Evaluation}}.
\newblock URL \url{http://arxiv.org/abs/2311.01616}.

\bibitem[Dong et~al.()Dong, Chen, McAuley, and Berg-Kirkpatrick]{dong2020MusPy}
Hao-Wen Dong, Ke~Chen, Julian McAuley, and Taylor Berg-Kirkpatrick.
\newblock {{MusPy}}: {{A Toolkit}} for {{Symbolic Music Generation}}.
\newblock URL \url{http://arxiv.org/abs/2008.01951}.

\bibitem[Raffel et~al.({\natexlab{a}})Raffel, McFee, Humphrey, Salamon, Nieto, Liang, Ellis, and Raffel]{raffel2014mir_eval}
Colin Raffel, Brian McFee, Eric~J Humphrey, Justin Salamon, Oriol Nieto, Dawen Liang, Daniel~PW Ellis, and C~Colin Raffel.
\newblock {{MIR}}\_{{EVAL}}: A transparent implementation of common {{MIR}} metrics.
\newblock In \emph{{{ISMIR}}}, volume~10, page 2014, {\natexlab{a}}.

\bibitem[Deruty et~al.(2022)Deruty, Grachten, Lattner, Nistal, and Aouameur]{deruty_development_2022}
Emmanuel Deruty, Maarten Grachten, Stefan Lattner, Javier Nistal, and Cyran Aouameur.
\newblock On the {Development} and {Practice} of {AI} {Technology} for {Contemporary} {Popular} {Music} {Production}.
\newblock \emph{Transactions of the International Society for Music Information Retrieval}, 5\penalty0 (1):\penalty0 35, February 2022.
\newblock ISSN 2514-3298.
\newblock \doi{10.5334/tismir.100}.
\newblock URL \url{https://transactions.ismir.net/article/10.5334/tismir.100/}.

\bibitem[Xiong et~al.()Xiong, Wang, Yu, Lin, and Wang]{xiong2023Comprehensive}
Zeyu Xiong, Weitao Wang, Jing Yu, Yue Lin, and Ziyan Wang.
\newblock A {{Comprehensive Survey}} for {{Evaluation Methodologies}} of {{AI-Generated Music}}.
\newblock URL \url{http://arxiv.org/abs/2308.13736}.

\bibitem[Berenzweig et~al.({\natexlab{a}})Berenzweig, Logan, Ellis, and Whitman]{berenzweig_large-scale_2004}
Adam Berenzweig, Beth Logan, Daniel~P.W. Ellis, and Brian Whitman.
\newblock A large-scale evaluation of acoustic and subjective music-similarity measures.
\newblock 28\penalty0 (2):\penalty0 63--76, {\natexlab{a}}.
\newblock ISSN 0148-9267, 1531-5169.
\newblock \doi{10.1162/014892604323112257}.
\newblock URL \url{https://direct.mit.edu/comj/article/28/2/63-76/93900}.

\bibitem[Kasak et~al.()Kasak, Jarina, and Ticha]{kasak_towards_2022}
Peter Kasak, Roman Jarina, and Dasa Ticha.
\newblock Towards efficiency of a subjective evaluation for music source separation.
\newblock In \emph{2022 32nd International Conference Radioelektronika ({RADIOELEKTRONIKA})}, pages 01--05. {IEEE}.
\newblock ISBN 978-1-72818-686-3.
\newblock \doi{10.1109/RADIOELEKTRONIKA54537.2022.9764949}.
\newblock URL \url{https://ieeexplore.ieee.org/document/9764949/}.

\bibitem[Linson et~al.()Linson, Dobbyn, and Laney]{Linson2012CriticalII}
Adam Linson, Chris Dobbyn, and Robin~C. Laney.
\newblock Critical issues in evaluating freely improvising interactive music systems.
\newblock In \emph{International Conference on Innovative Computing and Cloud Computing}.
\newblock URL \url{https://api.semanticscholar.org/CorpusID:95175}.

\bibitem[Schuller()]{schuller1958sonny}
Gunther Schuller.
\newblock Sonny {{Rollins}} and the challenge of thematic improvisation.
\newblock 1\penalty0 (1):\penalty0 6--11.

\bibitem[Juslin and Västfjäll(2008)]{juslin_emotional_2008}
Patrik~N. Juslin and Daniel Västfjäll.
\newblock Emotional responses to music: {The} need to consider underlying mechanisms.
\newblock \emph{Behavioral and Brain Sciences}, 31\penalty0 (5):\penalty0 559--575, October 2008.
\newblock ISSN 0140-525X, 1469-1825.
\newblock \doi{10.1017/S0140525X08005293}.
\newblock URL \url{https://www.cambridge.org/core/product/identifier/S0140525X08005293/type/journal_article}.

\bibitem[Tractinsky et~al.()Tractinsky, Katz, and Ikar]{tractinsky2000What}
N~Tractinsky, A.S Katz, and D~Ikar.
\newblock What is beautiful is usable.
\newblock 13\penalty0 (2):\penalty0 127--145.
\newblock ISSN 09535438.
\newblock \doi{10.1016/S0953-5438(00)00031-X}.
\newblock URL \url{https://academic.oup.com/iwc/article-lookup/doi/10.1016/S0953-5438(00)00031-X}.

\bibitem[Jordanous(2012)]{jordanous_standardised_2012}
Anna Jordanous.
\newblock A {Standardised} {Procedure} for {Evaluating} {Creative} {Systems}: {Computational} {Creativity} {Evaluation} {Based} on {What} it is to be {Creative}.
\newblock \emph{Cognitive Computation}, 4\penalty0 (3):\penalty0 246--279, September 2012.
\newblock ISSN 1866-9964.
\newblock \doi{10.1007/s12559-012-9156-1}.
\newblock URL \url{https://doi.org/10.1007/s12559-012-9156-1}.

\bibitem[Berenzweig et~al.({\natexlab{b}})Berenzweig, Logan, Ellis, and Whitman]{berenzweig2004LargeScale}
Adam Berenzweig, Beth Logan, Daniel~P.W. Ellis, and Brian Whitman.
\newblock A {{Large-Scale Evaluation}} of {{Acoustic}} and {{Subjective Music-Similarity Measures}}.
\newblock 28\penalty0 (2):\penalty0 63--76, {\natexlab{b}}.
\newblock ISSN 0148-9267, 1531-5169.
\newblock \doi{10.1162/014892604323112257}.
\newblock URL \url{https://direct.mit.edu/comj/article/28/2/63-76/93900}.

\bibitem[Eerola()]{eerola2011Are}
Tuomas Eerola.
\newblock Are the {{Emotions Expressed}} in {{Music Genre-specific}}? {{An Audio-based Evaluation}} of {{Datasets Spanning Classical}}, {{Film}}, {{Pop}} and {{Mixed Genres}}.
\newblock 40\penalty0 (4):\penalty0 349--366.
\newblock ISSN 0929-8215, 1744-5027.
\newblock \doi{10.1080/09298215.2011.602195}.
\newblock URL \url{http://www.tandfonline.com/doi/abs/10.1080/09298215.2011.602195}.

\bibitem[Pressing()]{pressing1987micro}
Jeff Pressing.
\newblock The micro-and macrostructural design of improvised music.
\newblock 5\penalty0 (2):\penalty0 133--172.
\newblock URL \url{https://www.jstor.org/stable/pdf/40285390.pdf}.

\bibitem[Clarke()]{clarke2011Ways}
Eric~F. Clarke.
\newblock \emph{Ways of Listening an Ecological Approach to the Perception of Musical Meaning}.
\newblock Oxford University Press.
\newblock ISBN 978-0-19-028816-7.

\bibitem[Stowell et~al.({\natexlab{a}})Stowell, Robertson, Bryan-Kinns, and Plumbley]{stowell_evaluation_2009}
D.~Stowell, A.~Robertson, N.~Bryan-Kinns, and M.D. Plumbley.
\newblock Evaluation of live human–computer music-making: Quantitative and qualitative approaches.
\newblock 67\penalty0 (11):\penalty0 960--975, {\natexlab{a}}.
\newblock ISSN 10715819.
\newblock \doi{10.1016/j.ijhcs.2009.05.007}.
\newblock URL \url{https://linkinghub.elsevier.com/retrieve/pii/S107158190900069X}.

\bibitem[Reimer and Wanderley()]{reimer_embracing_2021}
P.~J.~Charles Reimer and Marcelo~M. Wanderley.
\newblock Embracing less common evaluation strategies for studying user experience in {NIME}.
\newblock In \emph{{NIME} 2021}. {PubPub}.
\newblock \doi{10.21428/92fbeb44.807a000f}.
\newblock URL \url{https://nime.pubpub.org/pub/fidgs435}.

\bibitem[Wanderley and Mackay()]{wanderley2019HCI}
Marcelo~M. Wanderley and Wendy~E. Mackay.
\newblock {{HCI}}, {{Music}} and {{Art}}: {{An Interview}} with {{Wendy Mackay}}.
\newblock In Simon Holland, Tom Mudd, Katie Wilkie-McKenna, Andrew McPherson, and Marcelo~M. Wanderley, editors, \emph{New {{Directions}} in {{Music}} and {{Human-Computer Interaction}}}, pages 115--120. Springer International Publishing.
\newblock ISBN 978-3-319-92068-9 978-3-319-92069-6.
\newblock \doi{10.1007/978-3-319-92069-6_7}.
\newblock URL \url{http://link.springer.com/10.1007/978-3-319-92069-6_7}.

\bibitem[Huang et~al.(2020)Huang, Koops, Newton-Rex, Dinculescu, and Cai]{huang_ai_2020}
Cheng-Zhi~Anna Huang, Hendrik~Vincent Koops, Ed~Newton-Rex, Monica Dinculescu, and Carrie~J. Cai.
\newblock {AI} {Song} {Contest}: {Human}-{AI} {Co}-{Creation} in {Songwriting}, October 2020.
\newblock URL \url{http://arxiv.org/abs/2010.05388}.
\newblock arXiv:2010.05388 [cs].

\bibitem[El-Shimy and Cooperstock()]{el-shimy_user-driven_2016}
Dalia El-Shimy and Jeremy~R. Cooperstock.
\newblock User-driven techniques for the design and evaluation of new musical interfaces.
\newblock 40\penalty0 (2):\penalty0 35--46.
\newblock ISSN 0148-9267, 1531-5169.
\newblock \doi{10.1162/COMJ_a_00357}.
\newblock URL \url{https://direct.mit.edu/comj/article/40/2/35-46/94542}.

\bibitem[Stowell et~al.({\natexlab{b}})Stowell, Robertson, Bryan-Kinns, and Plumbley]{stowell2009Evaluation}
D.~Stowell, A.~Robertson, N.~Bryan-Kinns, and M.D. Plumbley.
\newblock Evaluation of live human–computer music-making: {{Quantitative}} and qualitative approaches.
\newblock 67\penalty0 (11):\penalty0 960--975, {\natexlab{b}}.
\newblock ISSN 10715819.
\newblock \doi{10.1016/j.ijhcs.2009.05.007}.
\newblock URL \url{https://linkinghub.elsevier.com/retrieve/pii/S107158190900069X}.

\bibitem[Johnson and Onwuegbuzie()]{johnson2004Mixed}
R.~Burke Johnson and Anthony~J. Onwuegbuzie.
\newblock Mixed {{Methods Research}}: {{A Research Paradigm Whose Time Has Come}}.
\newblock 33\penalty0 (7):\penalty0 14--26.
\newblock ISSN 0013-189X, 1935-102X.
\newblock \doi{10.3102/0013189X033007014}.
\newblock URL \url{https://journals.sagepub.com/doi/10.3102/0013189X033007014}.

\bibitem[Bradt()]{bradt2021Where}
Joke Bradt.
\newblock Where are the mixed methods research studies?
\newblock 30\penalty0 (4):\penalty0 311--313.
\newblock ISSN 0809-8131, 1944-8260.
\newblock \doi{10.1080/08098131.2021.1936771}.
\newblock URL \url{https://www.tandfonline.com/doi/full/10.1080/08098131.2021.1936771}.

\bibitem[Schacher et~al.()Schacher, Järveläinen, Strinning, and Neff]{schacher2015Movement}
Jan~C. Schacher, Hanna Järveläinen, Christian Strinning, and Patrick Neff.
\newblock Movement {{Perception In Music Performance}} - {{A Mixed Methods Investigation}}.
\newblock ISSN 2518-3672.
\newblock \doi{10.5281/ZENODO.851106}.
\newblock URL \url{https://zenodo.org/record/851106}.

\bibitem[Chu et~al.()Chu, Kim, Kim, Lim, Lee, Jin, Lee, Kim, and Ko]{chu2022Empirical}
Hyeshin Chu, Joohee Kim, Seongouk Kim, Hongkyu Lim, Hyunwook Lee, Seungmin Jin, Jongeun Lee, Taehwan Kim, and Sungahn Ko.
\newblock An {{Empirical Study}} on {{How People Perceive AI-generated Music}}.
\newblock In \emph{Proceedings of the 31st {{ACM International Conference}} on {{Information}} \& {{Knowledge Management}}}, pages 304--314. ACM.
\newblock ISBN 978-1-4503-9236-5.
\newblock \doi{10.1145/3511808.3557235}.
\newblock URL \url{https://dl.acm.org/doi/10.1145/3511808.3557235}.

\bibitem[Défossez et~al.()Défossez, Copet, Synnaeve, and Adi]{defossez_high_2022}
Alexandre Défossez, Jade Copet, Gabriel Synnaeve, and Yossi Adi.
\newblock High fidelity neural audio compression.
\newblock URL \url{http://arxiv.org/abs/2210.13438}.

\bibitem[Raffel et~al.({\natexlab{b}})Raffel, Shazeer, Roberts, Lee, Narang, Matena, Zhou, Li, and Liu]{raffel_exploring_2023}
Colin Raffel, Noam Shazeer, Adam Roberts, Katherine Lee, Sharan Narang, Michael Matena, Yanqi Zhou, Wei Li, and Peter~J. Liu.
\newblock Exploring the limits of transfer learning with a unified text-to-text transformer, {\natexlab{b}}.
\newblock URL \url{http://arxiv.org/abs/1910.10683}.

\bibitem[Elizalde et~al.()Elizalde, Deshmukh, Ismail, and Wang]{elizalde_clap_2022}
Benjamin Elizalde, Soham Deshmukh, Mahmoud~Al Ismail, and Huaming Wang.
\newblock {CLAP}: Learning audio concepts from natural language supervision.
\newblock URL \url{http://arxiv.org/abs/2206.04769}.

\bibitem[Agostinelli et~al.(2023)Agostinelli, Denk, Borsos, Engel, Verzetti, Caillon, Huang, Jansen, Roberts, Tagliasacchi, Sharifi, Zeghidour, and Frank]{agostinelli2023musiclm}
Andrea Agostinelli, Timo~I. Denk, Zalán Borsos, Jesse Engel, Mauro Verzetti, Antoine Caillon, Qingqing Huang, Aren Jansen, Adam Roberts, Marco Tagliasacchi, Matt Sharifi, Neil Zeghidour, and Christian Frank.
\newblock Musiclm: Generating music from text, 2023.

\bibitem[Schneider et~al.(2023)Schneider, Kamal, Jin, and Schölkopf]{schneider2023mousai}
Flavio Schneider, Ojasv Kamal, Zhijing Jin, and Bernhard Schölkopf.
\newblock Mo{\textbackslash}{\textasciicircum}usai: {Text}-to-{Music} {Generation} with {Long}-{Context} {Latent} {Diffusion}, October 2023.
\newblock URL \url{http://arxiv.org/abs/2301.11757}.
\newblock arXiv:2301.11757 [cs, eess].

\bibitem[Kilgour et~al.()Kilgour, Zuluaga, Roblek, and Sharifi]{kilgour_frechet_2019}
Kevin Kilgour, Mauricio Zuluaga, Dominik Roblek, and Matthew Sharifi.
\newblock Fr{\textbackslash}'echet audio distance: A metric for evaluating music enhancement algorithms.
\newblock URL \url{http://arxiv.org/abs/1812.08466}.

\bibitem[Devlin et~al.()Devlin, Chang, Lee, and Toutanova]{devlin_bert_2019}
Jacob Devlin, Ming-Wei Chang, Kenton Lee, and Kristina Toutanova.
\newblock {BERT}: Pre-training of deep bidirectional transformers for language understanding.
\newblock URL \url{http://arxiv.org/abs/1810.04805}.

\bibitem[Katharopoulos et~al.()Katharopoulos, Vyas, Pappas, and Fleuret]{katharopoulos_transformers_2020}
Angelos Katharopoulos, Apoorv Vyas, Nikolaos Pappas, and François Fleuret.
\newblock Transformers are {RNNs}: Fast autoregressive transformers with linear attention.
\newblock URL \url{http://arxiv.org/abs/2006.16236}.

\bibitem[Zeng et~al.(2021)Zeng, Tan, Wang, Ju, Qin, and Liu]{zeng2021musicbert}
Mingliang Zeng, Xu~Tan, Rui Wang, Zeqian Ju, Tao Qin, and Tie-Yan Liu.
\newblock {MusicBERT}: {Symbolic} {Music} {Understanding} with {Large}-{Scale} {Pre}-{Training}, June 2021.
\newblock URL \url{http://arxiv.org/abs/2106.05630}.
\newblock arXiv:2106.05630 [cs, eess].

\bibitem[Hung et~al.()Hung, Ching, Doh, Kim, Nam, and Yang]{hung_emopia_2021}
Hsiao-Tzu Hung, Joann Ching, Seungheon Doh, Nabin Kim, Juhan Nam, and Yi-Hsuan Yang.
\newblock {EMOPIA}: A multi-modal pop piano dataset for emotion recognition and emotion-based music generation.
\newblock URL \url{http://arxiv.org/abs/2108.01374}.

\bibitem[Ens and Pasquier()]{ens_metamidi_2021}
Jeff Ens and Philippe Pasquier.
\newblock Building the {MetaMIDI} dataset: Linking symbolic and audio musical data.
\newblock In \emph{{ISMIR}}, volume~22, pages 182--188.
\newblock URL \url{https://archives.ismir.net/ismir2021/paper/000022.pdf}.

\bibitem[Huang and Yang()]{huang_pop_2020}
Yu-Siang Huang and Yi-Hsuan Yang.
\newblock Pop music transformer: Beat-based modeling and generation of expressive pop piano compositions.
\newblock URL \url{http://arxiv.org/abs/2002.00212}.

\bibitem[OpenAI et~al.(2024)OpenAI, Achiam, Adler, Agarwal, Ahmad, Akkaya, Aleman, Almeida, Altenschmidt, Altman, Anadkat, Avila, Babuschkin, Balaji, Balcom, Baltescu, Bao, Bavarian, Belgum, Bello, Berdine, Bernadett-Shapiro, Berner, Bogdonoff, Boiko, Boyd, Brakman, Brockman, Brooks, Brundage, Button, Cai, Campbell, Cann, Carey, Carlson, Carmichael, Chan, Chang, Chantzis, Chen, Chen, Chen, Chen, Chen, Chess, Cho, Chu, Chung, Cummings, Currier, Dai, Decareaux, Degry, Deutsch, Deville, Dhar, Dohan, Dowling, Dunning, Ecoffet, Eleti, Eloundou, Farhi, Fedus, Felix, Fishman, Forte, Fulford, Gao, Georges, Gibson, Goel, Gogineni, Goh, Gontijo-Lopes, Gordon, Grafstein, Gray, Greene, Gross, Gu, Guo, Hallacy, Han, Harris, He, Heaton, Heidecke, Hesse, Hickey, Hickey, Hoeschele, Houghton, Hsu, Hu, Hu, Huizinga, Jain, Jain, Jang, Jiang, Jiang, Jin, Jin, Jomoto, Jonn, Jun, Kaftan, Łukasz Kaiser, Kamali, Kanitscheider, Keskar, Khan, Kilpatrick, Kim, Kim, Kim, Kirchner, Kiros, Knight, Kokotajlo, Łukasz Kondraciuk, Kondrich, Konstantinidis, Kosic, Krueger, Kuo, Lampe, Lan, Lee, Leike, Leung, Levy, Li, Lim, Lin, Lin, Litwin, Lopez, Lowe, Lue, Makanju, Malfacini, Manning, Markov, Markovski, Martin, Mayer, Mayne, McGrew, McKinney, McLeavey, McMillan, McNeil, Medina, Mehta, Menick, Metz, Mishchenko, Mishkin, Monaco, Morikawa, Mossing, Mu, Murati, Murk, Mély, Nair, Nakano, Nayak, Neelakantan, Ngo, Noh, Ouyang, O'Keefe, Pachocki, Paino, Palermo, Pantuliano, Parascandolo, Parish, Parparita, Passos, Pavlov, Peng, Perelman, de~Avila Belbute~Peres, Petrov, de~Oliveira~Pinto, Michael, Pokorny, Pokrass, Pong, Powell, Power, Power, Proehl, Puri, Radford, Rae, Ramesh, Raymond, Real, Rimbach, Ross, Rotsted, Roussez, Ryder, Saltarelli, Sanders, Santurkar, Sastry, Schmidt, Schnurr, Schulman, Selsam, Sheppard, Sherbakov, Shieh, Shoker, Shyam, Sidor, Sigler, Simens, Sitkin, Slama, Sohl, Sokolowsky, Song, Staudacher, Such, Summers, Sutskever, Tang, Tezak, Thompson, Tillet, Tootoonchian, Tseng, Tuggle, Turley, Tworek, Uribe, Vallone, Vijayvergiya, Voss, Wainwright, Wang, Wang, Wang, Ward, Wei, Weinmann, Welihinda, Welinder, Weng, Weng, Wiethoff, Willner, Winter, Wolrich, Wong, Workman, Wu, Wu, Wu, Xiao, Xu, Yoo, Yu, Yuan, Zaremba, Zellers, Zhang, Zhang, Zhao, Zheng, Zhuang, Zhuk, and Zoph]{openai_gpt4_2024}
OpenAI, Josh Achiam, Steven Adler, Sandhini Agarwal, Lama Ahmad, Ilge Akkaya, Florencia~Leoni Aleman, Diogo Almeida, Janko Altenschmidt, Sam Altman, Shyamal Anadkat, Red Avila, Igor Babuschkin, Suchir Balaji, Valerie Balcom, Paul Baltescu, Haiming Bao, Mohammad Bavarian, Jeff Belgum, Irwan Bello, Jake Berdine, Gabriel Bernadett-Shapiro, Christopher Berner, Lenny Bogdonoff, Oleg Boiko, Madelaine Boyd, Anna-Luisa Brakman, Greg Brockman, Tim Brooks, Miles Brundage, Kevin Button, Trevor Cai, Rosie Campbell, Andrew Cann, Brittany Carey, Chelsea Carlson, Rory Carmichael, Brooke Chan, Che Chang, Fotis Chantzis, Derek Chen, Sully Chen, Ruby Chen, Jason Chen, Mark Chen, Ben Chess, Chester Cho, Casey Chu, Hyung~Won Chung, Dave Cummings, Jeremiah Currier, Yunxing Dai, Cory Decareaux, Thomas Degry, Noah Deutsch, Damien Deville, Arka Dhar, David Dohan, Steve Dowling, Sheila Dunning, Adrien Ecoffet, Atty Eleti, Tyna Eloundou, David Farhi, Liam Fedus, Niko Felix, Simón~Posada Fishman, Juston Forte, Isabella Fulford, Leo Gao, Elie Georges, Christian Gibson, Vik Goel, Tarun Gogineni, Gabriel Goh, Rapha Gontijo-Lopes, Jonathan Gordon, Morgan Grafstein, Scott Gray, Ryan Greene, Joshua Gross, Shixiang~Shane Gu, Yufei Guo, Chris Hallacy, Jesse Han, Jeff Harris, Yuchen He, Mike Heaton, Johannes Heidecke, Chris Hesse, Alan Hickey, Wade Hickey, Peter Hoeschele, Brandon Houghton, Kenny Hsu, Shengli Hu, Xin Hu, Joost Huizinga, Shantanu Jain, Shawn Jain, Joanne Jang, Angela Jiang, Roger Jiang, Haozhun Jin, Denny Jin, Shino Jomoto, Billie Jonn, Heewoo Jun, Tomer Kaftan, Łukasz Kaiser, Ali Kamali, Ingmar Kanitscheider, Nitish~Shirish Keskar, Tabarak Khan, Logan Kilpatrick, Jong~Wook Kim, Christina Kim, Yongjik Kim, Jan~Hendrik Kirchner, Jamie Kiros, Matt Knight, Daniel Kokotajlo, Łukasz Kondraciuk, Andrew Kondrich, Aris Konstantinidis, Kyle Kosic, Gretchen Krueger, Vishal Kuo, Michael Lampe, Ikai Lan, Teddy Lee, Jan Leike, Jade Leung, Daniel Levy, Chak~Ming Li, Rachel Lim, Molly Lin, Stephanie Lin, Mateusz Litwin, Theresa Lopez, Ryan Lowe, Patricia Lue, Anna Makanju, Kim Malfacini, Sam Manning, Todor Markov, Yaniv Markovski, Bianca Martin, Katie Mayer, Andrew Mayne, Bob McGrew, Scott~Mayer McKinney, Christine McLeavey, Paul McMillan, Jake McNeil, David Medina, Aalok Mehta, Jacob Menick, Luke Metz, Andrey Mishchenko, Pamela Mishkin, Vinnie Monaco, Evan Morikawa, Daniel Mossing, Tong Mu, Mira Murati, Oleg Murk, David Mély, Ashvin Nair, Reiichiro Nakano, Rajeev Nayak, Arvind Neelakantan, Richard Ngo, Hyeonwoo Noh, Long Ouyang, Cullen O'Keefe, Jakub Pachocki, Alex Paino, Joe Palermo, Ashley Pantuliano, Giambattista Parascandolo, Joel Parish, Emy Parparita, Alex Passos, Mikhail Pavlov, Andrew Peng, Adam Perelman, Filipe de~Avila Belbute~Peres, Michael Petrov, Henrique~Ponde de~Oliveira~Pinto, Michael, Pokorny, Michelle Pokrass, Vitchyr~H. Pong, Tolly Powell, Alethea Power, Boris Power, Elizabeth Proehl, Raul Puri, Alec Radford, Jack Rae, Aditya Ramesh, Cameron Raymond, Francis Real, Kendra Rimbach, Carl Ross, Bob Rotsted, Henri Roussez, Nick Ryder, Mario Saltarelli, Ted Sanders, Shibani Santurkar, Girish Sastry, Heather Schmidt, David Schnurr, John Schulman, Daniel Selsam, Kyla Sheppard, Toki Sherbakov, Jessica Shieh, Sarah Shoker, Pranav Shyam, Szymon Sidor, Eric Sigler, Maddie Simens, Jordan Sitkin, Katarina Slama, Ian Sohl, Benjamin Sokolowsky, Yang Song, Natalie Staudacher, Felipe~Petroski Such, Natalie Summers, Ilya Sutskever, Jie Tang, Nikolas Tezak, Madeleine~B. Thompson, Phil Tillet, Amin Tootoonchian, Elizabeth Tseng, Preston Tuggle, Nick Turley, Jerry Tworek, Juan Felipe~Cerón Uribe, Andrea Vallone, Arun Vijayvergiya, Chelsea Voss, Carroll Wainwright, Justin~Jay Wang, Alvin Wang, Ben Wang, Jonathan Ward, Jason Wei, CJ~Weinmann, Akila Welihinda, Peter Welinder, Jiayi Weng, Lilian Weng, Matt Wiethoff, Dave Willner, Clemens Winter, Samuel Wolrich, Hannah Wong, Lauren Workman, Sherwin Wu, Jeff Wu, Michael Wu, Kai Xiao, Tao Xu, Sarah Yoo, Kevin Yu, Qiming Yuan, Wojciech Zaremba, Rowan Zellers, Chong Zhang, Marvin Zhang, Shengjia Zhao, Tianhao Zheng, Juntang Zhuang, William Zhuk, and Barret Zoph.
\newblock Gpt-4 technical report, 2024.

\bibitem[Wu and Sun()]{wu_exploring_2023}
Shangda Wu and Maosong Sun.
\newblock Exploring the efficacy of pre-trained checkpoints in text-to-music generation task.
\newblock URL \url{http://arxiv.org/abs/2211.11216}.

\bibitem[Liu et~al.()Liu, Zhu, Song, and Elgammal]{liu_towards_2021}
Bingchen Liu, Yizhe Zhu, Kunpeng Song, and Ahmed Elgammal.
\newblock Towards faster and stabilized {GAN} training for high-fidelity few-shot image synthesis.
\newblock URL \url{http://arxiv.org/abs/2101.04775}.

\bibitem[Zen et~al.()Zen, Dang, Clark, Zhang, Weiss, Jia, Chen, and Wu]{zen_libritts_2019}
Heiga Zen, Viet Dang, Rob Clark, Yu~Zhang, Ron~J. Weiss, Ye~Jia, Zhifeng Chen, and Yonghui Wu.
\newblock {LibriTTS}: A corpus derived from {LibriSpeech} for text-to-speech.
\newblock URL \url{http://arxiv.org/abs/1904.02882}.

\bibitem[Hawthorne et~al.()Hawthorne, Stasyuk, Roberts, Simon, Huang, Dieleman, Elsen, Engel, and Eck]{hawthorne_enabling_2019}
Curtis Hawthorne, Andriy Stasyuk, Adam Roberts, Ian Simon, Cheng-Zhi~Anna Huang, Sander Dieleman, Erich Elsen, Jesse Engel, and Douglas Eck.
\newblock Enabling factorized piano music modeling and generation with the {MAESTRO} dataset.
\newblock URL \url{http://arxiv.org/abs/1810.12247}.

\bibitem[Lam et~al.(2023)Lam, Tian, Li, Yin, Feng, Tu, Ji, Xia, Ma, Song, Chen, Yuping, and Wang]{NEURIPS2023_38b23e23}
Max W.~Y. Lam, Qiao Tian, Tang Li, Zongyu Yin, Siyuan Feng, Ming Tu, Yuliang Ji, Rui Xia, Mingbo Ma, Xuchen Song, Jitong Chen, Wang Yuping, and Yuxuan Wang.
\newblock Efficient neural music generation.
\newblock In A.~Oh, T.~Naumann, A.~Globerson, K.~Saenko, M.~Hardt, and S.~Levine, editors, \emph{Advances in Neural Information Processing Systems}, volume~36, pages 17450--17463. Curran Associates, Inc., 2023.
\newblock URL \url{https://proceedings.neurips.cc/paper_files/paper/2023/file/38b23e2328096520e9c889ae03e372c9-Paper-Conference.pdf}.

\bibitem[Dosovitskiy et~al.(2021)Dosovitskiy, Beyer, Kolesnikov, Weissenborn, Zhai, Unterthiner, Dehghani, Minderer, Heigold, Gelly, Uszkoreit, and Houlsby]{dosovitskiy_image_2021}
Alexey Dosovitskiy, Lucas Beyer, Alexander Kolesnikov, Dirk Weissenborn, Xiaohua Zhai, Thomas Unterthiner, Mostafa Dehghani, Matthias Minderer, Georg Heigold, Sylvain Gelly, Jakob Uszkoreit, and Neil Houlsby.
\newblock An {Image} is {Worth} 16x16 {Words}: {Transformers} for {Image} {Recognition} at {Scale}, June 2021.
\newblock URL \url{http://arxiv.org/abs/2010.11929}.
\newblock arXiv:2010.11929 [cs].

\bibitem[Arnab et~al.(2021)Arnab, Dehghani, Heigold, Sun, Lučić, and Schmid]{arnab_vivit_2021}
Anurag Arnab, Mostafa Dehghani, Georg Heigold, Chen Sun, Mario Lučić, and Cordelia Schmid.
\newblock {ViViT}: {A} {Video} {Vision} {Transformer}.
\newblock In \emph{Proceedings of the IEEE/CVF international conference on computer vision}, pages 6836--6846, 2021.
\newblock URL \url{https://openaccess.thecvf.com/content/ICCV2021/html/Arnab_ViViT_A_Video_Vision_Transformer_ICCV_2021_paper.html?ref=https://githubhelp.com}.

\bibitem[Li et~al.(2024)Li, Yuan, Zhang, Ma, Chen, Yin, Xiao, Lin, Ragni, Benetos, Gyenge, Dannenberg, Liu, Chen, Xia, Shi, Huang, Wang, Guo, and Fu]{li2024mert}
Yizhi Li, Ruibin Yuan, Ge~Zhang, Yinghao Ma, Xingran Chen, Hanzhi Yin, Chenghao Xiao, Chenghua Lin, Anton Ragni, Emmanouil Benetos, Norbert Gyenge, Roger Dannenberg, Ruibo Liu, Wenhu Chen, Gus Xia, Yemin Shi, Wenhao Huang, Zili Wang, Yike Guo, and Jie Fu.
\newblock {MERT}: {Acoustic} {Music} {Understanding} {Model} with {Large}-{Scale} {Self}-supervised {Training}, April 2024.
\newblock URL \url{http://arxiv.org/abs/2306.00107}.
\newblock arXiv:2306.00107 [cs, eess].

\bibitem[Touvron et~al.(2023)Touvron, Martin, Stone, Albert, Almahairi, Babaei, Bashlykov, Batra, Bhargava, Bhosale, Bikel, Blecher, Ferrer, Chen, Cucurull, Esiobu, Fernandes, Fu, Fu, Fuller, Gao, Goswami, Goyal, Hartshorn, Hosseini, Hou, Inan, Kardas, Kerkez, Khabsa, Kloumann, Korenev, Koura, Lachaux, Lavril, Lee, Liskovich, Lu, Mao, Martinet, Mihaylov, Mishra, Molybog, Nie, Poulton, Reizenstein, Rungta, Saladi, Schelten, Silva, Smith, Subramanian, Tan, Tang, Taylor, Williams, Kuan, Xu, Yan, Zarov, Zhang, Fan, Kambadur, Narang, Rodriguez, Stojnic, Edunov, and Scialom]{touvron_llama_2023}
Hugo Touvron, Louis Martin, Kevin Stone, Peter Albert, Amjad Almahairi, Yasmine Babaei, Nikolay Bashlykov, Soumya Batra, Prajjwal Bhargava, Shruti Bhosale, Dan Bikel, Lukas Blecher, Cristian~Canton Ferrer, Moya Chen, Guillem Cucurull, David Esiobu, Jude Fernandes, Jeremy Fu, Wenyin Fu, Brian Fuller, Cynthia Gao, Vedanuj Goswami, Naman Goyal, Anthony Hartshorn, Saghar Hosseini, Rui Hou, Hakan Inan, Marcin Kardas, Viktor Kerkez, Madian Khabsa, Isabel Kloumann, Artem Korenev, Punit~Singh Koura, Marie-Anne Lachaux, Thibaut Lavril, Jenya Lee, Diana Liskovich, Yinghai Lu, Yuning Mao, Xavier Martinet, Todor Mihaylov, Pushkar Mishra, Igor Molybog, Yixin Nie, Andrew Poulton, Jeremy Reizenstein, Rashi Rungta, Kalyan Saladi, Alan Schelten, Ruan Silva, Eric~Michael Smith, Ranjan Subramanian, Xiaoqing~Ellen Tan, Binh Tang, Ross Taylor, Adina Williams, Jian~Xiang Kuan, Puxin Xu, Zheng Yan, Iliyan Zarov, Yuchen Zhang, Angela Fan, Melanie Kambadur, Sharan Narang, Aurelien Rodriguez, Robert Stojnic, Sergey Edunov, and Thomas Scialom.
\newblock Llama 2: {Open} {Foundation} and {Fine}-{Tuned} {Chat} {Models}, July 2023.
\newblock URL \url{http://arxiv.org/abs/2307.09288}.
\newblock arXiv:2307.09288 [cs].

\bibitem[Liu et~al.(2024{\natexlab{b}})Liu, Yuan, Liu, Mei, Kong, Tian, Wang, Wang, Wang, and Plumbley]{liu_audioldm_2024}
Haohe Liu, Yi~Yuan, Xubo Liu, Xinhao Mei, Qiuqiang Kong, Qiao Tian, Yuping Wang, Wenwu Wang, Yuxuan Wang, and Mark~D. Plumbley.
\newblock {AudioLDM} 2: {Learning} {Holistic} {Audio} {Generation} with {Self}-supervised {Pretraining}, May 2024{\natexlab{b}}.
\newblock URL \url{http://arxiv.org/abs/2308.05734}.
\newblock arXiv:2308.05734 [cs, eess].

\bibitem[Liu et~al.(2024{\natexlab{c}})Liu, Hussain, Sun, and Shan]{liu_music_2024}
Shansong Liu, Atin~Sakkeer Hussain, Chenshuo Sun, and Ying Shan.
\newblock Music {Understanding} {LLaMA}: {Advancing} {Text}-to-{Music} {Generation} with {Question} {Answering} and {Captioning}.
\newblock In \emph{{ICASSP} 2024 - 2024 {IEEE} {International} {Conference} on {Acoustics}, {Speech} and {Signal} {Processing} ({ICASSP})}, pages 286--290, April 2024{\natexlab{c}}.
\newblock \doi{10.1109/ICASSP48485.2024.10447027}.
\newblock URL \url{https://ieeexplore.ieee.org/abstract/document/10447027/}.
\newblock ISSN: 2379-190X.

\bibitem[Wang et~al.(2023)Wang, Ju, Tan, He, Wu, Bian, and Zhao]{wang_audit_2023}
Yuancheng Wang, Zeqian Ju, Xu~Tan, Lei He, Zhizheng Wu, Jiang Bian, and Sheng Zhao.
\newblock {AUDIT}: {Audio} {Editing} by {Following} {Instructions} with {Latent} {Diffusion} {Models}.
\newblock \emph{Advances in Neural Information Processing Systems}, 36:\penalty0 71340--71357, December 2023.
\newblock URL \url{https://proceedings.neurips.cc/paper_files/paper/2023/hash/e1b619a9e241606a23eb21767f16cf81-Abstract-Conference.html}.

\bibitem[Han et~al.(2023)Han, Dai, Hao, He, Guo, Chen, Wang, Qian, and Song]{han_instructme_2023}
Bing Han, Junyu Dai, Weituo Hao, Xinyan He, Dong Guo, Jitong Chen, Yuxuan Wang, Yanmin Qian, and Xuchen Song.
\newblock {InstructME}: {An} {Instruction} {Guided} {Music} {Edit} {And} {Remix} {Framework} with {Latent} {Diffusion} {Models}, December 2023.
\newblock URL \url{http://arxiv.org/abs/2308.14360}.
\newblock arXiv:2308.14360 [cs, eess].

\bibitem[Li et~al.()Li, Cai, Wu, Zhang, Chen, Qi, Dong, Chen, Dong, Shi, Guo, Han, Ge, Liu, Gan, and Zhang]{li_survey_2024}
Wenjun Li, Ying Cai, Ziyang Wu, Wenyi Zhang, Yifan Chen, Rundong Qi, Mengqi Dong, Peigen Chen, Xiao Dong, Fenghao Shi, Lei Guo, Junwei Han, Bao Ge, Tianming Liu, Lin Gan, and Tuo Zhang.
\newblock A survey of foundation models for music understanding.
\newblock URL \url{http://arxiv.org/abs/2409.09601}.

\bibitem[Vaswani et~al.()Vaswani, Shazeer, Parmar, Uszkoreit, Jones, Gomez, Kaiser, and Polosukhin]{vaswani_attention_2023}
Ashish Vaswani, Noam Shazeer, Niki Parmar, Jakob Uszkoreit, Llion Jones, Aidan~N. Gomez, Lukasz Kaiser, and Illia Polosukhin.
\newblock Attention is all you need.
\newblock URL \url{http://arxiv.org/abs/1706.03762}.

\bibitem[Roberts et~al.()Roberts, Engel, Raffel, Hawthorne, and Eck]{roberts_hierarchical_2019}
Adam Roberts, Jesse Engel, Colin Raffel, Curtis Hawthorne, and Douglas Eck.
\newblock A hierarchical latent vector model for learning long-term structure in music.
\newblock URL \url{http://arxiv.org/abs/1803.05428}.

\bibitem[Engel et~al.()Engel, Resnick, Roberts, Dieleman, Eck, Simonyan, and Norouzi]{engel_neural_2017}
Jesse Engel, Cinjon Resnick, Adam Roberts, Sander Dieleman, Douglas Eck, Karen Simonyan, and Mohammad Norouzi.
\newblock Neural audio synthesis of musical notes with {WaveNet} autoencoders.
\newblock URL \url{http://arxiv.org/abs/1704.01279}.

\bibitem[Engel et~al.(2020)Engel, Hantrakul, Gu, and Roberts]{engel_ddsp_2020}
Jesse Engel, Lamtharn Hantrakul, Chenjie Gu, and Adam Roberts.
\newblock {DDSP}: {Differentiable} {Digital} {Signal} {Processing}, January 2020.
\newblock URL \url{http://arxiv.org/abs/2001.04643}.
\newblock arXiv:2001.04643 [cs, eess, stat].

\bibitem[GitHub()]{M2UGenIssue4}
GitHub.
\newblock Performance issue of m2ugen - github issue.
\newblock URL \url{https://github.com/shansongliu/M2UGen/issues/4}.
\newblock Accessed: 01 March 2024.

\bibitem[Rouard et~al.(2023)Rouard, Massa, and D{\'e}fossez]{rouard2022hybrid}
Simon Rouard, Francisco Massa, and Alexandre D{\'e}fossez.
\newblock Hybrid transformers for music source separation.
\newblock In \emph{ICASSP 23}, 2023.

\bibitem[Zhang et~al.(2024)Zhang, Ikemiya, Choi, Murata, Martínez-Ramírez, Lin, Xia, Liao, Mitsufuji, and Dixon]{zhang_instruct-musicgen_2024}
Yixiao Zhang, Yukara Ikemiya, Woosung Choi, Naoki Murata, Marco~A. Martínez-Ramírez, Liwei Lin, Gus Xia, Wei-Hsiang Liao, Yuki Mitsufuji, and Simon Dixon.
\newblock Instruct-{MusicGen}: {Unlocking} {Text}-to-{Music} {Editing} for {Music} {Language} {Models} via {Instruction} {Tuning}, May 2024.
\newblock URL \url{http://arxiv.org/abs/2405.18386}.
\newblock arXiv:2405.18386 [cs].

\bibitem[Langenkamp and Yue(2022)]{langenkamp_how_2022}
Max Langenkamp and Daniel~N. Yue.
\newblock How {Open} {Source} {Machine} {Learning} {Software} {Shapes} {AI}.
\newblock In \emph{Proceedings of the 2022 {AAAI}/{ACM} {Conference} on {AI}, {Ethics}, and {Society}}, pages 385--395, Oxford United Kingdom, July 2022. ACM.
\newblock ISBN 978-1-4503-9247-1.
\newblock \doi{10.1145/3514094.3534167}.
\newblock URL \url{https://dl.acm.org/doi/10.1145/3514094.3534167}.

\bibitem[Dadman and Bremdal(2024)]{dadman2024crafting}
Shayan Dadman and Bernt~Arild Bremdal.
\newblock Crafting {Creative} {Melodies}: {A} {User}-{Centric} {Approach} for {Symbolic} {Music} {Generation}.
\newblock \emph{Electronics}, 13\penalty0 (6):\penalty0 1116, March 2024.
\newblock ISSN 2079-9292.
\newblock \doi{10.3390/electronics13061116}.
\newblock URL \url{https://www.mdpi.com/2079-9292/13/6/1116}.

\bibitem[Caillon and Esling()]{caillon2021RAVE}
Antoine Caillon and Philippe Esling.
\newblock {{RAVE}}: {{A}} variational autoencoder for fast and high-quality neural audio synthesis.
\newblock URL \url{http://arxiv.org/abs/2111.05011}.

\bibitem[Dang et~al.(2022)Dang, Mecke, Lehmann, Goller, and Buschek]{dang_how_2022}
Hai Dang, Lukas Mecke, Florian Lehmann, Sven Goller, and Daniel Buschek.
\newblock How to {Prompt}? {Opportunities} and {Challenges} of {Zero}- and {Few}-{Shot} {Learning} for {Human}-{AI} {Interaction} in {Creative} {Applications} of {Generative} {Models}, September 2022.
\newblock URL \url{http://arxiv.org/abs/2209.01390}.
\newblock arXiv:2209.01390 [cs].

\bibitem[Oppenlaender(2023)]{oppenlaender_taxonomy_2023}
Jonas Oppenlaender.
\newblock A {Taxonomy} of {Prompt} {Modifiers} for {Text}-{To}-{Image} {Generation}.
\newblock \emph{Behaviour \& Information Technology}, pages 1--14, November 2023.
\newblock ISSN 0144-929X, 1362-3001.
\newblock \doi{10.1080/0144929X.2023.2286532}.
\newblock URL \url{http://arxiv.org/abs/2204.13988}.
\newblock arXiv:2204.13988 [cs].

\bibitem[Christodoulou et~al.(2024)Christodoulou, Lartillot, and Jensenius]{christodoulou_multimodal_2024}
Anna-Maria Christodoulou, Olivier Lartillot, and Alexander~Refsum Jensenius.
\newblock Multimodal music datasets? {Challenges} and future goals in music processing.
\newblock \emph{International Journal of Multimedia Information Retrieval}, 13\penalty0 (3):\penalty0 37, September 2024.
\newblock ISSN 2192-6611, 2192-662X.
\newblock \doi{10.1007/s13735-024-00344-6}.
\newblock URL \url{https://link.springer.com/10.1007/s13735-024-00344-6}.

\bibitem[Liu et~al.(2021)Liu, Yuan, Fu, Jiang, Hayashi, and Neubig]{liu_pre-train_2021}
Pengfei Liu, Weizhe Yuan, Jinlan Fu, Zhengbao Jiang, Hiroaki Hayashi, and Graham Neubig.
\newblock Pre-train, {Prompt}, and {Predict}: {A} {Systematic} {Survey} of {Prompting} {Methods} in {Natural} {Language} {Processing}, July 2021.
\newblock URL \url{http://arxiv.org/abs/2107.13586}.
\newblock arXiv:2107.13586 [cs].

\bibitem[Chang et~al.(2024)Chang, Srinivasan, Luthra, Lin, Nagaraja, Iandola, Liu, Ni, Zhao, Shi, and Chandra]{chang_open_2024}
Ernie Chang, Sidd Srinivasan, Mahi Luthra, Pin-Jie Lin, Varun Nagaraja, Forrest Iandola, Zechun Liu, Zhaoheng Ni, Changsheng Zhao, Yangyang Shi, and Vikas Chandra.
\newblock On the {Open} {Prompt} {Challenge} in {Conditional} {Audio} {Generation}.
\newblock In \emph{{ICASSP} 2024 - 2024 {IEEE} {International} {Conference} on {Acoustics}, {Speech} and {Signal} {Processing} ({ICASSP})}, pages 5315--5319, Seoul, Korea, Republic of, April 2024. IEEE.
\newblock ISBN 9798350344851.
\newblock \doi{10.1109/ICASSP48485.2024.10447897}.
\newblock URL \url{https://ieeexplore.ieee.org/document/10447897/}.

\bibitem[Burnard et~al.(2018)Burnard, Ross, Hassler, and Murphy]{bartleet_translating_2018}
Pamela Burnard, Valerie Ross, Laura Hassler, and Lis Murphy.
\newblock \emph{Translating {Intercultural} {Creativities} in {Community} {Music}}, volume~1.
\newblock Oxford University Press, February 2018.
\newblock \doi{10.1093/oxfordhb/9780190219505.013.6}.
\newblock URL \url{https://academic.oup.com/edited-volume/34637/chapter/295100681}.

\bibitem[Li et~al.(2022)Li, Wei, Tian, Xu, Wen, and Hu]{li_learning_2022}
Guangyao Li, Yake Wei, Yapeng Tian, Chenliang Xu, Ji-Rong Wen, and Di~Hu.
\newblock Learning to {Answer} {Questions} in {Dynamic} {Audio}-{Visual} {Scenarios}.
\newblock In \emph{2022 {IEEE}/{CVF} {Conference} on {Computer} {Vision} and {Pattern} {Recognition} ({CVPR})}, pages 19086--19096, New Orleans, LA, USA, June 2022. IEEE.
\newblock ISBN 978-1-66546-946-3.
\newblock \doi{10.1109/CVPR52688.2022.01852}.
\newblock URL \url{https://ieeexplore.ieee.org/document/9879157/}.

\bibitem[Ronchini et~al.(2024)Ronchini, Comanducci, Perego, and Antonacci]{ronchini_paguri_2024}
Francesca Ronchini, Luca Comanducci, Gabriele Perego, and Fabio Antonacci.
\newblock {PAGURI}: a user experience study of creative interaction with text-to-music models, July 2024.
\newblock URL \url{http://arxiv.org/abs/2407.04333}.
\newblock arXiv:2407.04333 [cs, eess] version: 1.

\bibitem[Yakura and Goto(2023)]{yakura_iteratta_2023}
Hiromu Yakura and Masataka Goto.
\newblock {IteraTTA}: {An} interface for exploring both text prompts and audio priors in generating music with text-to-audio models, July 2023.
\newblock URL \url{http://arxiv.org/abs/2307.13005}.
\newblock arXiv:2307.13005 [cs, eess].

\bibitem[Wang et~al.()Wang, Ma, Feng, Zhang, Yang, Zhang, Chen, Tang, Chen, Lin, Zhao, Wei, and Wen]{wang_survey_2024}
Lei Wang, Chen Ma, Xueyang Feng, Zeyu Zhang, Hao Yang, Jingsen Zhang, Zhiyuan Chen, Jiakai Tang, Xu~Chen, Yankai Lin, Wayne~Xin Zhao, Zhewei Wei, and Ji-Rong Wen.
\newblock A survey on large language model based autonomous agents.
\newblock 18\penalty0 (6):\penalty0 186345.
\newblock ISSN 2095-2228, 2095-2236.
\newblock \doi{10.1007/s11704-024-40231-1}.
\newblock URL \url{http://arxiv.org/abs/2308.11432}.

\bibitem[Zeng et~al.(2023)Zeng, Yang, Malik, Yan, Huang, and He]{zeng_let_2023}
Jingying Zeng, Jaewon Yang, Waleed Malik, Xiao Yan, Richard Huang, and Qi~He.
\newblock Let {AI} {Entertain} {You}: {Increasing} {User} {Engagement} with {Generative} {AI} and {Rejection} {Sampling}, December 2023.
\newblock URL \url{http://arxiv.org/abs/2312.12457}.
\newblock arXiv:2312.12457 [cs].

\bibitem[Taryn(2024)]{taryn_interviews_2024}
Taryn.
\newblock List of questions and answers from interviews with press.
\newblock Online, 2024.
\newblock URL \url{https://docs.google.com/document/d/1mTelMocJD788hk_x4Ce-bwnPmVXoogVG5tezB9lSirQ/edit}.
\newblock Interviews compiled and supplied by Taryn herself, including major media outlets such as Forbes, The Verge, BBC, and Fox5.

\bibitem[Mullen(2023)]{mullen_2023}
Matt Mullen.
\newblock How patten used text-to-audio ai to make an entire album: "we're at the precipice of a fundamental shift in how we think about making music".
\newblock \emph{MusicRadar}, May 2023.
\newblock URL \url{https://www.musicradar.com/news/patten-interview}.

\bibitem[Wright(2023)]{wright_2023}
Webb Wright.
\newblock Max cooper is using ai to push the frontiers of creativity and communication.
\newblock \emph{The Drum}, May 2023.
\newblock URL \url{https://www.thedrum.com/news/2023/05/31/max-cooper-ai-the-future-music-and-consciousness}.

\bibitem[Doshi and Hauser()]{doshi_generative_2024}
Anil~R. Doshi and Oliver~P. Hauser.
\newblock Generative artificial intelligence enhances creativity but reduces the diversity of novel content.
\newblock URL \url{http://arxiv.org/abs/2312.00506}.

\bibitem[Hou et~al.()Hou, Wang, Wang, Wang, and Yang]{hou_double-edged_2024}
Jinghui~(Jove) Hou, Lei Wang, Gang Wang, Harry Wang, and Shuai Yang.
\newblock The double-edged roles of generative {AI} in the creative process: Experiments on design work.
\newblock URL \url{https://papers.ssrn.com/abstract=4739471}.

\bibitem[Wadinambiarachchi et~al.()Wadinambiarachchi, Kelly, Pareek, Zhou, and Velloso]{wadinambiarachchi_effects_2024}
Samangi Wadinambiarachchi, Ryan~M. Kelly, Saumya Pareek, Qiushi Zhou, and Eduardo Velloso.
\newblock The effects of generative {AI} on design fixation and divergent thinking.
\newblock In \emph{Proceedings of the {CHI} Conference on Human Factors in Computing Systems}, pages 1--18.
\newblock \doi{10.1145/3613904.3642919}.
\newblock URL \url{http://arxiv.org/abs/2403.11164}.

\bibitem[Kumar et~al.()Kumar, Vincentius, Jordan, and Anderson]{kumar_human_2024}
Harsh Kumar, Jonathan Vincentius, Ewan Jordan, and Ashton Anderson.
\newblock Human creativity in the age of {LLMs}: Randomized experiments on divergent and convergent thinking.
\newblock URL \url{http://arxiv.org/abs/2410.03703}.

\bibitem[Small()]{small_musicking_1998}
Christopher Small.
\newblock \emph{Musicking: the meanings of performing and listening}.
\newblock Music/culture. University Press of New England.
\newblock ISBN 9780819522566 9780819522573.

\bibitem[Young and Murphy()]{young2020HCI}
Gareth~W. Young and Dave Murphy.
\newblock {{HCI Models}} for {{Digital Musical Instruments}}: {{Methodologies}} for {{Rigorous Testing}} of {{Digital Musical Instruments}}.
\newblock URL \url{http://arxiv.org/abs/2010.01328}.

\bibitem[Agres et~al.()Agres, Forth, and Wiggins]{agres2016Evaluation}
Kat Agres, Jamie Forth, and Geraint~A. Wiggins.
\newblock Evaluation of {{Musical Creativity}} and {{Musical Metacreation Systems}}.
\newblock 14\penalty0 (3):\penalty0 1--33.
\newblock ISSN 1544-3574.
\newblock \doi{10.1145/2967506}.
\newblock URL \url{https://dl.acm.org/doi/10.1145/2967506}.

\bibitem[Eigenfeldt et~al.()Eigenfeldt, Burnett, and Pasquier]{eigenfeldt2012Evaluation}
Arne Eigenfeldt, Adam Burnett, and Philippe Pasquier.
\newblock Evaluating musical metacreation in a live performance context.
\newblock Proceedings of the {{Third International Conference}} on {{Computational Creativity}}, pages 140--144.

\bibitem[Aybek and Toraman()]{aybek2022How}
Eren~Can Aybek and Cetin Toraman.
\newblock How many response categories are sufficient for {{Likert}} type scales? {{An}} empirical study based on the {{Item Response Theory}}.
\newblock 9\penalty0 (2):\penalty0 534--547.
\newblock ISSN 2148-7456.
\newblock \doi{10.21449/ijate.1132931}.
\newblock URL \url{http://dergipark.org.tr/en/doi/10.21449/ijate.1132931}.

\bibitem[Preston and Colman()]{preston2000Optimal}
Carolyn~C Preston and Andrew~M Colman.
\newblock Optimal number of response categories in rating scales: Reliability, validity, discriminating power, and respondent preferences.
\newblock 104\penalty0 (1):\penalty0 1--15.
\newblock ISSN 00016918.
\newblock \doi{10.1016/S0001-6918(99)00050-5}.
\newblock URL \url{https://linkinghub.elsevier.com/retrieve/pii/S0001691899000505}.

\bibitem[Morrison()]{morrison1972Regressions}
Donald~G. Morrison.
\newblock Regressions with {{Discrete Dependent Variables}}: {{The Effect}} on {{R}} 2.
\newblock 9\penalty0 (3):\penalty0 338.
\newblock ISSN 00222437.
\newblock \doi{10.2307/3149551}.
\newblock URL \url{https://www.jstor.org/stable/3149551?origin=crossref}.

\bibitem[Sauro()]{sauro2019ThreePoint}
Jeff Sauro.
\newblock Is a {{Three-Point Scale Good Enough}}?
\newblock URL \url{https://measuringu.com/three-points/}.

\bibitem[Wakita et~al.()Wakita, Ueshima, and Noguchi]{wakita2012Psychological}
Takafumi Wakita, Natsumi Ueshima, and Hiroyuki Noguchi.
\newblock Psychological {{Distance Between Categories}} in the {{Likert Scale}}: {{Comparing Different Numbers}} of {{Options}}.
\newblock 72\penalty0 (4):\penalty0 533--546.
\newblock ISSN 0013-1644, 1552-3888.
\newblock \doi{10.1177/0013164411431162}.
\newblock URL \url{https://journals.sagepub.com/doi/10.1177/0013164411431162}.

\end{thebibliography}

\newpage
\appendix

\section{Commerical AI Music Generation Platforms}\label{app:commercial_models}
Tab. \ref{tab:ai_music_commercial} presents a wide array of AI music generation platforms and tools for commercial purposes. It provides a list that features the key characteristics of each and includes their website URLs for further information. The information presented is solely based on the product descriptions available on the respective websites at the time of this study.

\begin{table}[h]
  \centering
  \caption{List of commerical AI music generation platforms and tools.}
  \label{tab:ai_music_commercial}
  \resizebox{\textwidth}{!}{\begin{tabular}{clp{7cm}p{4cm}}
  \toprule
  No.             & Platform Name             & Key Features                                                                  & Website URL              \\
  \midrule
  1               & AIVA                      & Generates variations of songs, deep learning algorithms, MIDI editor          & \url{https://aiva.ai}             \\
  2               & Amper Music               & Cloud-based, wide range of samples and instruments, part of Shutterstock      & \url{https://www.ampermusic.com}  \\
  3               & AudioCipher               & Text-to-MIDI VST plugin, musical cryptogram for chord/melody generation       & \url{https://audiocipher.com}     \\
  4               & iZotope                   & AI-powered audio plugins, audio analysis, custom settings                     & \url{https://www.izotope.com}     \\
  5               & StableAudio               & Text-to-audio generator, diffusion model, high-quality instrumental audio     & \url{https://stability.ai/news/stable-audio-using-ai-to-generate-music} \\
  6               & Ecrett Music              & Generates music clips for scenes/emotions                                     & \url{https://ecrettmusic.com}     \\
  7               & Soundful                  & Brand-specific, studio-quality, royalty-free music                            & \url{https://soundful.com}        \\
  8               & Flow Machines             & AI-powered composition system, iPad app for experimentation                   & \url{https://www.sony.com/en/SonyInfo/design/stories/flow-machines/} \\
  9               & Mubert                    & Personalized, royalty-free music streaming platform                           & \url{https://mubert.com}          \\
  10              & Fadr                      & Automated mastering, stem extraction, remixing and composition tools          & \url{https://fadr.com}            \\
  11              & Boomy AI                  & Automated music creation and distribution                                     & \url{https://boomy.com}           \\
  12              & Beatoven AI               & AI composer for royalty-free soundtracks                                      & \url{https://beatoven.ai}         \\
  14              & WavTool AI                & Browser-based music studio, AI-generated beats and melodies                   & \url{https://wavtool.com}         \\
  15              & Amadeus Code              & AI songwriting assistant, generates melodic hooks and song ideas              & \url{https://amadeuscode.com}     \\
  16              & Suno AI                   & Generates adaptive music and soundscapes for games and interactive media      & \url{https://suno.com}            \\
  17              & WarpSound                 & Combines art and music generation, virtual artists, NFTs                      & \url{https://www.warpsound.ai}    \\
  18              & Audio Design Desk         & AI-assisted audio creation tool, sound effects and music for content creators & \url{https://add.app}             \\
  19              & Bandlab SongStarter       & AI-generated beats and melodies, text-to-MIDI, collaboration features         & \url{https://www.bandlab.com/songstarter} \\
  20              & RipX DAW Pro by Hit'n'Mix & AI-powered stem separation, remixing, sample creation, DAW functionalities    & \url{https://hitnmix.com}         \\
  21              & Brain.fm                  & Enhances focus, relaxation, sleep, personalized soundscapes                   & \url{https://www.brain.fm}        \\
  31              & Voice Swap                & Voice cloning, library of licensed artist voices                              & \url{https://www.voice-swap.ai}   \\
  22              & Splash                    & Assists music production, text-to-singing, melody generation                  & \url{https://www.splashmusic.com} \\
  23              & Aimi                      & Studio-quality, royalty-free music generation                                 & \url{https://www.aimi.fm}         \\
  24              & HookGen                   & Generates hooks and melodies, variety of musical elements                     & \url{https://hookgen.com}         \\
  25              & Chord AI                  & Real-time chord recognition, audio to MIDI conversion                         & \url{https://www.chordai.net}     \\
  26              & Cassette AI               & Uses Latent Diffusion models to generate music patterns                       & \url{https://cassetteai.com}      \\
  27              & Vocaloid                  & Singing synthesis software, realistic voices from text, multiple voicebanks   & \url{https://www.vocaloid.com}    \\
  28              & MelodyStudio              & AI-powered songwriting tool, generates melody ideas from user input           & \url{https://melodystudio.net}    \\
  29              & EvokeMusic                & Generates royalty-free music with customization for mood, genre, instruments  & \url{https://evokemusic.ai}       \\
  30              & Kits AI                   & Provides voice tools including voice cloning and licensed artist voices       & \url{https://www.kits.ai}         \\
  31              & Musicfy                   & Generation of music through voice or text inputs, custom AI voice models      & \url{https://musicfy.lol}         \\
  32              & Songburst                 & AI-powered music creation tool                                                & \url{https://www.songburst.ai/}   \\
  33              & Two Shot                  & Personalized voice models, copyright-free compositions, rich sample library   & \url{https://twoshot.ai/}         \\
  \bottomrule  
\end{tabular}}
\end{table}

\clearpage
\section{Home Studio Configuration}\label{app:home_studio}

Tab. \ref{tab:studio_config} presents the hardware configuration utilized during the \textit{Hands-on Experimentation} phase of this study. The experimental setup consisted of a Lenovo Legion 9 16IRX8 laptop as the primary computing device, interfaced with a Native Instruments KOMPLETE Audio 6 sound card for audio processing. Audio monitoring was facilitated through KRK RP8 RoKit Classic studio monitors for reference listening, complemented by Beyerdynamic DT 990 Pro headphones for audio analysis. Musical input and parameter control (when needed) were achieved using an Arturia Keystep 32 MIDI controller, while Ableton Live 11 served as the DAWs for all audio processing, recording, and experimental procedures.

\begin{table}[h]
    \centering
    \caption{Hardware configuration used for the hands-on experimentation in this study.
    }\label{tab:studio_config}
    \begin{tabular}{ll}
    \toprule
    Equipment Type                  & Model \\
    \midrule
    Laptop                          & Lenovo Legion 9 16IRX8                  \\ 
    Sound Card                      & Native Instruments KOMPLETE Audio 6     \\ 
    Studio Monitors                 & KRK RP8 RoKit Classic                   \\ 
    Headphones                      & Beyerdynamic DT 990 Pro                 \\ 
    MIDI Controller                 & Arturia Keystep 32                      \\ 
    Digital Audio Workstation       & Ableton Live 11                         \\ 
    \bottomrule
  \end{tabular}
  \end{table}

\clearpage
\section{Methodological Justification for Scale Design in Performance Assessment}\label{app:rubric_scale}
The choice of a 5-point scale in exploratory evaluation studies is supported because it balances reliability, usability, and analytical clarity. Compared to 3-point scales, it offers better sensitivity to capture nuanced differences in performance or perceptions while avoiding the cognitive overload and inconsistency often associated with 10-point scales \citep{aybek2022How, preston2000Optimal}. Psychometric studies indicate that 5-point scales achieve strong inter-rater reliability and reduce errors related to forced choices or ambiguous distinctions \citep{morrison1972Regressions}. Additionally, they align with human cognitive limits by providing sufficient differentiation without overwhelming raters, which is particularly important in exploratory contexts where consistency and actionable feedback are essential\footnote{As noted by \citet{sauro2019ThreePoint}, a two-point scale conveys only a single dimension of information (for example, a binary option such as yes/no or agree/disagree). In contrast, a three-point scale conveys two dimensions by introducing both directional bias and neutrality. Although a four-point scale captures the intensity of directional opinion, it omits a neutral option. Therefore, from a theoretical standpoint, a five-point scale is preferable because it provides three distinct dimensions: the direction of opinion (positive or negative), the intensity of that opinion, and a neutral midpoint. }. The prevalence of 5-point scales in social science research also enhances comparability across studies by making them a practical and effective choice for evaluations requiring both granularity and interpretability \citep{wakita2012Psychological}.

\begin{table}[h]
  \centering
  \caption{Standardized scoring rubric for formative \& summative criteria (1--5 scale)}
  \label{tab:scoring_rubric}
  \begin{tabular}{ll}
    \toprule
    Score & Description \\ 
    \midrule
    1 & Fails to meet basic expectations \\ 
    2 & Meets minimum requirements with limitations \\ 
    3 & Satisfies acceptable standards; needs partial improvements \\ 
    4 & Exceeds expectations in most areas; minor issues remain \\ 
    5 & Fully meets/surpasses all expectations \\ 
    \bottomrule
  \end{tabular}
\end{table}

Therefore, to apply this 5-point scale, Table \ref{tab:scoring_rubric} is created as a clear and consistent guideline for criterion definition. This standardized scoring rubric defines performance levels from 1 to 5. The scoring levels progress from 'Fails to meet basic expectations' (1) to 'Fully meets/surpasses all expectations' (5), with intermediate levels capturing nuanced performance gradations. The rubric's design reflects the scale's cognitive and measurement principles discussed earlier. The central score of 3 represents an acceptable standard that requires partial improvements and provides a neutral reference point. Scores 2 and 4 offer intermediate assessments by capturing performance that either marginally meets minimum requirements or approaches excellence with minor limitations.

\clearpage
\section{Evaluation Framework Criteria}\label{app:evaluation_criteria}
This appendix presents the evaluation criteria, utilized in Section \ref{sec:criteria_definition}, through four tables that systematically categorize assessment criteria for MGS. Tab. \ref{tab:architecture_criteria} establishes six system-level descriptive criteria examining architecture design, input/output modalities, conditioning mechanisms, data methodologies, evaluation metrics, and technical limitations. Tab. \ref{tab:interface_criteria} outlines five interface-focused criteria addressing interaction methods, model checkpoint access, demonstration effectiveness, deployment flexibility, and setup complexity. Tab. \ref{tab:hardware_criteria} specifies four hardware-oriented criteria evaluating GPU requirements, memory demands, accessibility constraints, and operational flexibility across computing environments. Tab. \ref{tab:performance_criteria} delineates eigth performance-focused criteria for hands-on experimentation, assessing usability, generation efficiency, output quality, stylistic fidelity, parametric control, content modification capabilities, digital audio workstation integration, and creative workflow support.

\begin{table}[h]
    \centering
    \caption{This table presents architecture and design aspects of the "descriptive" criteria utilized during the \textit{Systems Overview} phase. The \textit{Criterion} column specifies the key areas of evaluation. The \textit{Description} column provides an overview of each criterion's purpose. The \textit{Considerations} column lists specific elements or features considered to evaluate within each criterion.}
    \label{tab:architecture_criteria}
    \resizebox{\textwidth}{!}{%
    \begin{tabular}{p{2cm}p{6cm}p{8cm}}
        \toprule
        Criterion & Description & Considerations \\ 
        \midrule
        Architecture\newline and Model\newline Design
        & Analyzes the fundamental generative approach and structural elements of the system.
        & Underlying generative framework and design elements (e.g., transformer, diffusion, GANs, autoencoders); sequential vs. parallel generation capabilities; handling of temporal and hierarchical musical structures. \\
        
        \midrule
        Input\newline and Output\newline Modalities
        & The specific types of inputs the system accepts and the formats of musical output it generates.
        & Types of inputs supported (e.g., text prompts, audio samples, MIDI sequences, melodic features, chord progressions); output representation formats (e.g., symbolic MIDI, audio waveforms, score notation); multi-modal capabilities. \\
        
        \midrule
        Conditioning\newline and Control\newline Mechanisms
        & Techniques used to guide generation via conditioning signals, such as text descriptions, melodic/musical attributes, or style cues.
        & Use of text-to-music alignment mechanisms for natural language steering (e.g., T5, BERT, CLAP); attribute (e.g., tempo, key, style) or melody conditioning (e.g., chromagram-based); joint text–audio representations. \\
        
        \midrule
        Data\newline and Training\newline Methodologies
        & Analyzes the dataset selection, and learning approaches that shape the system's musical knowledge, biases, and generative capabilities.
        & Diversity of training materials; scale of datasets; approaches to data curation; licensed or publicly available data; training paradigms employed. \\
        
        \midrule
        Evaluation\newline Metrics and\newline Performance\newline Assessments
        & Analyzes the evaluation methods used to measure system performance across technical, musical, and creative dimensions.
        & Use of objective metrics (e.g., Fréchet Audio Distance, perplexity); use of subjective evaluations (e.g., human listener ratings); comparative assessment against existing systems or human compositions; measures of musical coherence and stylistic consistency; approaches to measuring alignment with user intent. \\
        
        \midrule
        Limitations\newline and\newline Challenges
        & Documents the technical constraints, performance limitations and practical barriers.
        & Technical performance issues (e.g., audio artifacts, coherence problems over longer sequences); control limitations (e.g., difficulty steering generation, lack of fine-grained control); practical application barriers (e.g., workflow integration challenges, real-time performance constraints). \\
  
        \bottomrule
    \end{tabular}
    }
  \end{table}
  
  \begin{table}[t]
    \centering
    \caption{This table presents interface aspects of the "descriptive" criteria utilized during the \textit{Systems Overview} phase. The \textit{Criterion} column specifies the key areas of evaluation. The \textit{Description} column provides an overview of each criterion's purpose. The \textit{Considerations} column lists specific elements or features considered to evaluate within each criterion.}
    \label{tab:interface_criteria}
    \resizebox{\textwidth}{!}{%
    \begin{tabular}{p{2cm}p{6cm}p{8cm}}
        \toprule
        Criterion & Description & Considerations \\ 
        \midrule
        Interface Availability
        & Analyzes the range and effectiveness of interaction methods offered by the system for user engagement.
        & Available interaction modes (e.g., GUI, web-based, CLI); accessibility for non-technical users; comparison of interface options across systems; user experience; Compatibility with domain-specific tools (e.g., VST plugins for DAWs). \\
        
        \midrule
        Checkpoint\newline Accessibility and\newline variations
        & Documents the availability and configuration range of pre-trained checkpoints to enable reproducibility and adaptability to different user needs and technical constraints.
        & Access to model weights and parameters; versioning of checkpoints for comparative evaluation; documentation of training conditions; scalability across computational resources. \\
        
        \midrule
        Demonstrations
        & Analyzes how effectively the system incorporates into established creative practices and existing technological ecosystems.
        & Effectiveness in communicating system capabilities and limitations; diversity of demonstrated outputs; showcasing of originality, value, and domain competence of outputs; showcasing of potential use cases that aid adoption and exploration of creative possibilities. \\
        
        \midrule
        Execution\newline Options
        & Analyzes the flexibility of deployment across different computational environments.
        & Support for both high-performance (GPU) and accessible (CPU) environments; considerations of latency and real-time performance capabilities; options for offline and online deployment. \\
        
        \midrule
        Ease of Setup
        & Analyzes the technical barriers to system deployment and configuration.
        & Simplicity of installation process; dependency management; balance between immediate engagement and technical depth. \\
  
        \bottomrule
    \end{tabular}
    }
  \end{table}
  
  \begin{table}[t]
    \centering
    \caption{This table presents hardware aspects of the "descriptive" criteria utilized during the \textit{Systems Overview} phase. The \textit{Criterion} column specifies the key areas of evaluation. The \textit{Description} column provides an overview of each criterion's purpose. The \textit{Considerations} column lists specific elements or features considered to evaluate within each criterion.}
    \label{tab:hardware_criteria}
    \resizebox{\textwidth}{!}{%
    \begin{tabular}{p{2cm}p{6cm}p{8cm}}
        \toprule
        Criterion & Description & Considerations \\ 
        \midrule
        GPU Type and\newline Quantity
        & Evaluate the type and number of GPUs required for training and inference.
        & Consider the scalability and cost implications of using high-end GPUs versus more accessible configurations  (e.g., 4–8 NVIDIA A100 GPUs, RTX 2080 Ti). \\
        
        \midrule
        Memory\newline Capacity
        & Assess the memory requirements for training and inference.
        & Examine the accessibility of systems based on memory demands and compatibility with lower-tier hardware (e.g., 24GB VRAM, 16GB RAM). \\
        
        \midrule
        Accessibility
        & Analyze the feasibility of hardware setups.
        & Consider the impact of hardware accessibility on adoption by smaller studios, independent users, or researchers. \\
        
        \midrule
        Hardware\newline Flexibility
        & Determine whether systems can operate without GPUs.
        & Evaluate flexibility in hardware requirements to accommodate diverse user needs and resource constraints. \\
        
        \bottomrule
    \end{tabular}
    }
  \end{table}
  
  \begin{table}[h]
    \centering
    \caption{This table presents "performance" criteria utilized during the \textit{Hands-on Experimentation} phase. The \textit{Criterion} column specifies the key areas of evaluation. The \textit{Description} column provides an overview of each criterion's purpose. The \textit{Considerations} column lists specific elements or features considered to evaluate within each criterion.}
    \label{tab:performance_criteria}
    \resizebox{\textwidth}{!}{%
    \begin{tabular}{p{2cm}p{6cm}p{8cm}}
        \toprule
        Criterion & Description & Considerations \\ 
        \midrule
        Usability
        & Evaluates the system's intuitiveness, accessibility, and ease of use.
        & Interface clarity and intuitiveness; discoverability of features and functions; contextual help and documentation quality; error handling with solution suggestions. \\
  
        \midrule
        Generation\newline Speed
        & Measures the system’s efficiency in producing musical outputs.
        & Considers generation speed in relation to output length; responsiveness to parameter changes; latency impact on creative flow. \\
  
        \midrule
        Audio\newline Quality
        & Evaluates the clarity and professional standard of the generated audio.
        & Sound quality; absence of artifacts; requiring post-processing for studio production use. \\
        
        \midrule
        Stylistic\newline Accuracy
        & Evaluates the system's ability to replicate and adapt to various musical genres and styles.
        & Capturing key stylistic elements; consistency across iterations; resemblance to the genre/style of the given input (e.g., prompt, attribute, melody); ability to span a wide range of genres with accuracy. \\
        
        \midrule
        Parameter\newline Control
        & Evaluates the precision and granularity of user control over system parameters.
        & Range of control options; precision and reliability of parameter adjustments; predictability of results; ability to shape and direct the model's behavior effectively. \\
        
        \midrule
        Content\newline Generation\newline Control
        & Assesses how easily the generated content can be control and modified.
        & Considers capability for individual stem separation; possibilities for structural modifications; arrangement flexibility; support for non-destructive editing. \\
        
        \midrule
        DAW\newline Integration\newline Capacity
        & Evaluates how well the system integrates with DAWs.
        & Level of integration with DAWs; support for plugin formats; session persistence and recall; automation capabilities; ease of workflow within production environments. \\

        \midrule
        Creative\newline Workflows
        & Assesses how effectively the system supports and maintains the user's state of flow during the creative process.
        & Balance between technical operation and creative focus; ability to support iterative refinement; alignment with natural creative rhythm; capacity to maintain immersive workflows. \\
        
        \bottomrule
      \end{tabular}
      }
  \end{table}

\clearpage
\section{Performance Criteria Score Levels}\label{app:criteria_score_levels}

This appendix presents the scoring-levels tables for the 'peformance' criteria (Tab. \ref{tab:performance_criteria}) used in the evaluation of the MGS, described in Section \ref{sec:criteria_definition}. Each table outlines the scoring levels from 1 to 5 and considerations for each level. The tables are organized by criterion, and the scoring levels are based on the rubric scale presented in Appendix \ref{app:rubric_scale}.

\begin{table}[h]
    \centering
    \caption{Score levels for \textit{Usability} criterion in 'performance' criteria.}
    \label{tab:usability_criteria_score_levels}
    \resizebox{\textwidth}{!}{%
    \begin{tabular}{lp{15cm}}
        \toprule
        Score  & Considerations \\ 
        \midrule
        1
        & System requires technical expertise; interface is confusing with poor documentation; frequent errors with unhelpful messages; inaccessible to most users.\\
        
        \midrule
        2
        & Interface is functional but unintuitive; requires significant learning time; documentation exists but is incomplete; error messages are generic; limited accessibility features. \\
        
        \midrule
        3
        & Moderately intuitive interface with adequate documentation; occasional navigation challenges; basic error handling; standard accessibility features. Comparable to 'Moderate' ease of use in the assessment table.\\
        
        \midrule
        4
        & Clear, well-organized interface with comprehensive documentation; intuitive navigation; helpful error messages; good accessibility features. Comparable to 'High' ease of use in the assessment table.\\
        
        \midrule
        5
        & Exceptionally intuitive interface requiring minimal learning; excellent documentation with tutorials; proactive error prevention; comprehensive accessibility features.\\
        
        \bottomrule
    \end{tabular}
    }
  \end{table}
  
  \begin{table}[h]
    \centering
    \caption{Score levels for \textit{Generation Speed} criterion in 'performance' criteria.}
    \label{tab:generation_speed_criteria_score_levels}
    \resizebox{\textwidth}{!}{%
    \begin{tabular}{lp{15cm}}
        \toprule
        Score  & Considerations \\ 
        \midrule
        1
        & Extremely slow generation (>10 minutes for short segments); hinders creative exploration; unresponsive to parameter changes.\\
        
        \midrule
        2
        & Slow generation with long waiting periods; limits iterative processes; delayed response to parameter adjustments. \\
        
        \midrule
        3
        & Moderate generation times that are acceptable but noticeable; adequate for most workflows; reasonable responsiveness.\\
        
        \midrule
        4
        & Fast generation with minimal waiting; supports rapid iteration; quick response to parameter changes.\\
        
        \midrule
        5
        & Near-instant generation; ideal for real-time applications; immediate parameter response.\\
        
        \bottomrule
    \end{tabular}
    }
  \end{table}
  
  \begin{table}[h]
    \centering
    \caption{Score levels for \textit{Audio Quality} criterion in 'performance' criteria.}
    \label{tab:audio_quality_criteria_score_levels}
    \resizebox{\textwidth}{!}{%
    \begin{tabular}{lp{15cm}}
        \toprule
        Score  & Considerations \\ 
        \midrule
        1
        & Poor audio quality with significant artifacts; requires extensive post-processing.\\
        
        \midrule
        2
        & Basic audio quality with noticeable artifacts; requires considerable post-processing. \\
        
        \midrule
        3
        & Acceptable audio quality with some artifacts;  needs moderate post-processing.\\
        
        \midrule
        4
        & High audio quality with minor artifacts; requires minimal post-processing.\\
        
        \midrule
        5
        & Professional-grade audio quality; no post-processing needed.\\
        
        \bottomrule
    \end{tabular}
    }
  \end{table}
  
  \begin{table}[h]
    \centering
    \caption{Score levels for \textit{Stylistic Accuracy} criterion in 'performance' criteria.}
    \label{tab:stylistic_accuracy_criteria_score_levels}
    \resizebox{\textwidth}{!}{%
    \begin{tabular}{lp{15cm}}
        \toprule
        Score  & Considerations \\ 
        \midrule
        1
        & Fails to capture the basic characteristics of genres and styles; key elements are either incorrect or missing; there is no resemblance to the provided input.\\
        
        \midrule
        2
        & Present a limited range of genres and styles; exhibits significant inconsistencies; demonstrates little resemblance to the provided input. \\
        
        \midrule
        3
        & Presents common genres and styles with only occasional inaccuracies; demonstrates consistency across several iterations; often produces an output that resembles the provided input.\\
        
        \midrule
        4
        & Presents a wide range of genres and styles despite minor issues; exhibits good consistency; in most cases, the output resembles the provided input.\\
        
        \midrule
        5
        & Exhibits considerable stylistic reproduction across most genres; demonstrates consistency most of the time; produces an output that closely resembles the provided input.\\
        
        \bottomrule
    \end{tabular}
    }
  \end{table}
  
  \begin{table}[h]
    \centering
    \caption{Score levels for \textit{Parameter Control} criterion in 'performance' criteria.}
    \label{tab:parameter_control_criteria_score_levels}
    \resizebox{\textwidth}{!}{%
    \begin{tabular}{lp{15cm}}
        \toprule
        Score  & Considerations \\ 
        \midrule
        1
        & Limited control options; primarily randomized output; few adjustable parameters; unpredictable results.\\
        
        \midrule
        2
        & Basic control with general parameters; limited precision; inconsistent parameter response; minimal capabilities to control model's behavior. \\
        
        \midrule
        3
        & Moderate control with standard parameters; reasonable precision; generally predictable responses; adequate capabilities to control model's behavior.\\
        
        \midrule
        4
        & Responsive control with detailed parameters; good precision; reliable parameter response; good capabilities to control model's behavior.\\
        
        \midrule
        5
        & Exhaustive parameter control; high precision and predictable responses; well-defined capabilities to control model's behavior.\\
        
        \bottomrule
    \end{tabular}
    }
  \end{table}
  
  \begin{table}[h]
    \centering
    \caption{Score levels for \textit{Content Generation Control} criterion in 'performance' criteria.}
    \label{tab:content_editability_criteria_score_levels}
    \resizebox{\textwidth}{!}{%
    \begin{tabular}{lp{15cm}}
        \toprule
        Score  & Considerations \\ 
        \midrule
        1
        & Generated output is essentially fixed or hard to control; no stem separation or single track generation; modifications to input yield irrelevant results and may cause severe artifacts.\\
        
        \midrule
        2
        & Limited control over the generation content; poor quality stem separation (if available) or single track generation; quality and content loss when modified. \\
        
        \midrule
        3
        & Moderate control over the generated content; functional stem separation or single track generation; reasonable modification capability. \\
        
        \midrule
        4
        & High control over the generated content; clean stem separation or single track generation; good modification capabilities with minimal artifacts.\\
        
        \midrule
        5
        & High control over the generated content; perfect stem separation or single track generation; flexible modification capabilities without quality loss.\\
        
        \bottomrule
    \end{tabular}
    }
  \end{table}
  
  \begin{table}[h]
    \centering
    \caption{Score levels for \textit{DAW Integration Capacity} criterion in 'performance' criteria.}
    \label{tab:daw_compatibility_criteria_score_levels}
    \resizebox{\textwidth}{!}{%
    \begin{tabular}{lp{15cm}}
        \toprule
        Score  & Considerations \\ 
        \midrule
        1
        & No DAW integration; operates completely outside production environments; no plugin options.\\
        
        \midrule
        2
        & Minimal DAW interaction; limited to basic file import/export; no direct integration; cumbersome workflow. \\
        
        \midrule
        3
        & Functional DAW compatibility; works as plugin in major DAWs (e.g. Ableton Live, Logic pro); limited plugin formats (e.g. only VST); adequate workflow.\\
        
        \midrule
        4
        & Strong DAW integration; different plugin formats; supports automation; good session persistence.\\
        
        \midrule
        5
        & Complete DAW integration; full plugin support; comprehensive automation capabilities; perfect session persistence and recall.\\
        
        \bottomrule
    \end{tabular}
    }
  \end{table}

  \begin{table}[h]
    \centering
    \caption{Score levels for \textit{Creative Workflow} criterion in 'performance' criteria.}
    \label{tab:creative_workflow_criteria_score_levels}
    \resizebox{\textwidth}{!}{%
    \begin{tabular}{lp{15cm}}
        \toprule
        Score  & Considerations \\ 
        \midrule
        1
        & Frequently interrupts workflow; requires focus on technical aspects; impedes creative momentum; creates noticeable frustration.\\
        
        \midrule
        2
        & Periodically disrupts creative flow; technical operations often divert creative focus; maintaining creative momentum requires effort; limited iterative capabilities.\\
        
        \midrule
        3
        & Sometimes interrupts workflow; balances technical and creative needs adequately; allows maintaining flow with some adjustment; supports basic iterative refinement.\\
        
        \midrule
        4
        & Generally maintains workflow; emphasizes creative focus over technical operation; aligns well with creative rhythm; effectively reduces context-switching.\\
        
        \midrule
        5
        & Smoothly integrates with creative process; supports flow with minimal effort; accommodates natural rhythm of creation; users commonly become immersed while creating.\\
        
        \bottomrule
    \end{tabular}
    }
  \end{table}


\end{document}